\documentclass[11pt,a4paper]{article}
\pdfoutput=1
\usepackage{jheppub}
\usepackage{amsthm,amsbsy,amsfonts,mathrsfs,enumerate,float,wrapfig,amsmath}
\usepackage[utf8]{inputenc}
\usepackage{subfigure}
\usepackage{youngtab}
\usepackage{soul}
\usepackage{bm}

\newcommand{\nn}{\nonumber}
\newcommand{\be}{\begin{equation}}
\newcommand{\ee}{\end{equation}}
\newcommand{\bF}{\mathbf{F}}

\newcommand{\AS}{\mathbf{AS}}

\usepackage{tikz}
\usetikzlibrary{positioning}
\usetikzlibrary{arrows}
\title{Rank-3 antisymmetric matter on 5-brane webs}

\author[a]{Hirotaka Hayashi,}
\author[b]{Sung-Soo Kim,}
\author[c]{Kimyeong Lee,}
\author[d]{and Futoshi Yagi}

\affiliation[a]{Department of Physics, School of Science, Tokai University,\\ 4-1-1 Kitakaname, Hiratsuka-shi, Kanagawa 259-1292, Japan}
\affiliation[b]{School of Physics, University of Electronic Science and Technology of China, \\
No.4, Section 2, North Jianshe Road, Chengdu, Sichuan 610054, China}
\affiliation[c]{School of Physics, Korea Institute for Advanced Study, \\
85 Hoegi-ro Dongdaemun-gu, Seoul 02455, Korea}
\affiliation[d]{School of Mathematics, Southwest Jiaotong University,\\ 
West zone, High-tech district, Chengdu, Sichuan 611756, China}

\emailAdd{h.hayashi@tokai.ac.jp}
\emailAdd{sungsoo.kim@uestc.edu.cn}
\emailAdd{klee@kias.re.kr}
\emailAdd{futoshi\_yagi@swjtu.edu.cn}

\abstract{
We discuss Type IIB 5-brane configurations for 5d $\mathcal{N}=1$ gauge theories with hypermultiplets in the rank-3 antisymmetric representation and with various other hypermultiplets, which flow 
from a UV fixed point at the infinite coupling. We propose 5-brane web diagrams for the theories of $SU(6)$ and  $Sp(3)$ gauge groups with rank-3 antisymmetric matter and check our proposed 5-brane webs against several consistency conditions implied from the one-loop corrected prepotential.  
Using the obtained 5-brane webs for rank-3 antisymmetric matter, we apply the topological vertex method to compute the partition function for one of these $SU(6)$ gauge theories. 
}

\begin{document}
\preprint{
\begin{flushright}
\tt 
KIAS-P19004\\
\end{flushright}
}

\maketitle


\section{Introduction}\label{sec:introduction}

String theory is a useful tool to study various aspects of supersymmetric field theories. For example, 5-brane web diagrams in type IIB string theory proposed in \cite{Aharony:1997ju, Aharony:1997bh} can realize ultraviolet (UV) complete five-dimensional (5d) theories with eight supercharges. By using 5-brane web diagrams, it is possible to explicitly see non-perturbative features of 5d theories such as dualities. We can also compute 5d Nekrasov partitions by applying the topological vertex \cite{Aganagic:2003db, Iqbal:2007ii} to 5-brane webs, which makes use of a chain of string dualities between 5-brane webs in type IIB string theory and non-compact Calabi-Yau threefolds in M-theory \cite{Leung:1997tw}.

Since 5-brane web diagram is a powerful tool to study 5d theories, it is important to see how large class of 5d theories 5-brane web diagram can realize. Original 5-brane web diagrams basically yield 5d $SU(N)$ gauge theories with hypermultiplets in the fundamental or bi-fundamental representation. 
The class of gauge theories realized on 5-brane webs can be further expanded by introducing an orientifold or 7-branes created by decomposing an orientifold 7-plane \cite{Sen:1996vd}. An orientifold can change the gauge group into $SO(N)$ or $Sp(N)$ \cite{Brunner:1997gk, Bergman:2015dpa}, or it can also introduce different representations such as the symmetric or antisymmetric representation of $SU(N)$ or $Sp(N)$ \cite{Benini:2009gi,Bergman:2013aca, Bergman:2014kza, Bergman:2015dpa}. 
Recently it has been noticed that 5-brane web diagrams can provide more exotic theories which are typically not realized by brane configurations. An O5-plane may introduce the spinor representation of $SO(N)\; (7 \leq N \leq 12)$ gauge theories \cite{Zafrir:2015ftn} or it can even yield  $G_2$ gauge theories with hypermultiplets in the fundamental representation \cite{Hayashi:2018bkd}. In particular, as for 5d rank 2 theories, the authors showed in \cite{Hayashi:2018lyv} various 5-brane realizations of all the rank 2 theories which are geometrically constructed in \cite{Jefferson:2018irk}. 


It is then natural to ask if it is possible to still expand the class of 5d gauge theories which 5-brane web diagrams can construct. In this paper we argue that 5-brane web diagrams may yield further new type of gauge theories which are $SU(6)$ or $Sp(3)$ gauge theories with half-hypermultiplets in the rank-3 antisymmetric representation. The strategy to obtain the rank-3 antisymmetric representation of $SU(6)$ is to make use of a 5-brane web diagram for the $SO(12)$ gauge theory with a half-hypermultiplet in the conjugate spinor representation. Since the decomposition of the conjugate spinor representation under $SU(6) \times U(1)$ includes the rank-3 antisymmetric representation of the $SU(6)$ which is not charged under the $U(1)$, decoupling the degrees of freedom associated to the $U(1)$ should yield a 5-brane diagram of the $SU(6)$ gauge theory with a half-hypermultiplet in the rank-3 antisymmetric representation. A similar method was used to obtain the four-dimensional (4d) Seiberg-Witten curve for the $SU(6)$ gauge theory with a hypermultiplet in the rank-3 antisymmetric representation in \cite{Tachikawa:2011yr}. The extension of the construction can introduce more half-hypermultiplets in the rank-3 antisymmetric representation until four half-hypermultiplets or two hypermultiplets in the rank-3 antisymmetric representation. For related work, see~\cite{Ohmori:2018ona}.

Since 5d gauge theories with rank-3 antisymmetric matter are realized using 5-brane webs, it is also possible to compute the 5d Nekrasov partition functions. As an illustration, we explicitly compute the Nekrasov partition function for an $SU(6)$ gauge theory with a half-hypermultiplet in the rank-3 antisymmetric representation. 

We can then introduce matter in the fundamental representation and the rank-2 antisymmetric representation to 5-brane webs in addition to rank-3 antisymmetric matter. We can realize many of the $SU(6)$ gauge theories with rank-3 antisymmetric representation matter that have a six-dimensional (6d) UV completion in the list in \cite{Jefferson:2017ahm} which were obtained from the analysis of effective prepotentials. Moreover a Higgsing associated to a hypermultiplet in the rank-2 antisymmetric representation of marginal $SU(6)$ gauge theories with rank-2 and rank-3 antisymmetric matter yields 5-brane diagrams for $Sp(3)$ gauge theories with matter in the rank-3 antisymmetric representation which also have a 6d UV completion. From the construction of the 5-brane webs we also find dualities and propose explicit 6d theories for some of the marginal theories. 

The organization of the paper is as follows. In section \ref{sec:SU6TSA}, we propose 5-brane web diagrams of $SU(6)$ gauge theories with half-hypermultiplets in the rank-3 antisymmetric representation. From the obtained diagram we compute the Nekrasov partition function for an $SU(6)$ gauge theory with a half-hypermultiplet in the rank-3 antisymmetric representation. We extend the construction of 5-brane webs in section \ref{sec:SU6} by adding hypermultiplets in other representations. In particular we propose 5-brane webs for 5d $SU(6)$ gauge theories with rank-3 antisymmetric matter that have a 6d UV completion. In section \ref{sec:Sp3}, we Higgs the diagrams obtained in section \ref{sec:SU6} to construct 5-brane webs for marginal $Sp(3)$ gauge theories with rank-3 antisymmetric matter. Finally we find explicit 6d UV complete theories for some of the marginal $SU(6)$ gauge theories from the 5-brane webs and discuss dualities involving marginal $SU(6)$ gauge theories with a half-hypermultiplets in the rank-3 antisymmetric representation in section \ref{sec:6d}.





%

\bigskip

\section{$SU(6)$ gauge theories with rank-3 antisymmetric matter}\label{sec:SU6TSA}


In this section, we propose 5-brane webs for $SU(6)$ gauge theories with half-hypermultiplets in the rank three antisymmetric representation. UV complete 5d $SU(6)$ gauge theories can have at most two hypermultiplets in the rank-3 
antisymmetric representation \cite{Jefferson:2017ahm}. We will obtain brane webs with all possible number of massless rank-3 
antisymmetric half-hypermultiplets in this section.

\subsection{Decoupling from $SO(12)$ gauge theory with conjugate spinor matter}

One way to obtain the rank-3 
antisymmetric representation of $SU(6)$ is using the decomposition of the spinor or the conjugate spinor representation of $SO(12)$ under $SU(6) \times U(1)$. 
We here consider the decomposition from the conjugate spinor for later use. In this case, 
the decomposition of the conjugate spinor representation under the $SU(6) \times U(1)$ is given by
\begin{align}
SO(12)~~ &\supset~~ SU(6) \times U(1)\cr
{\bf 32}' ~~&= ~~{\bf 20}_0 \oplus {\bf 6}_{-2} \oplus \bar{{\bf 6}}_{2},
\end{align}
where the subscript stands for the $U(1)$ charge\footnote{We note that the decomposition of the spinor representation of $SO(12)$ under $SU(6)\times U(1)$ is given by ${\bf 32}= {\bf 1}_3 +{\bf 1}_{-3}+ {\bf 15}_{-1}+ \overline{\bf 15}_{1}$, where all the $SU(6)$ representations are charged under $U(1)$.}. The twenty dimensional representation ${\bf 20}_0$ is the rank-3 antisymmetric representation of $SU(6)$. The 5d $\mathcal{N}=1$ $SO(12)$ gauge theory with a conjugate spinor also contains a vector multiplet in the adjoint representation of $SO(12)$. The decomposition of the adjoint representation of $SO(12)$ under the $SU(6) \times U(1)$ is given by
\begin{align}
SO(12)~~ &\supset~~ SU(6) \times U(1)\cr
{\bf 66} ~~&= ~~{\bf 1}_0 \oplus {\bf 15}_2 + \overline{{\bf 15}}_{-2} + {\bf 35}_0.
\end{align}
Since ${\bf 6}_{-2}, \bar{{\bf 6}}_{2}, {\bf 15}_2$ and $\overline{{\bf 15}}_{-2}$ are charged under the $U(1)$ of the $SU(6) \times U(1)$, the fields in those representations acquire large mass when we give a large vev to the Coulomb branch modulus for the $U(1)${, while the singlet ${\bf 1}_{0}$ becomes non-dynamical (which will be more clear when we discuss its brane realization).}
 Therefore, when the vev for the Coulomb branch modulus of the $U(1)$ in the $SO(12)$ gauge theory 
becomes infinitely large, the low energy effective field theory should be described by the $SU(6)$ gauge theory with a hypermultiplet in the rank-3 antisymmetric representation. This method was made use of to obtain the Seiberg-Witten curve for the 4d $SU(6)$ gauge theory with rank-3 antisymmetric matter in \cite{Tachikawa:2011yr}.

We can apply this procedure to a 5-brane web for the $SO(12)$ gauge theory with a conjugate spinor for obtaining a brane web for the $SU(6)$ gauge theory with rank-3 antisymmetric matter. In order to simplify the discussion, we start from the 5d $SO(12)$ gauge theory with a half-hypermultiplet in the conjugate spinor representation. Then the decoupling procedure will lead to an $SU(6)$ gauge theory with a half-hypermultiplet in the rank-3 antisymmetric representation at low energies.

\begin{figure}
\centering
\includegraphics[width=10cm]{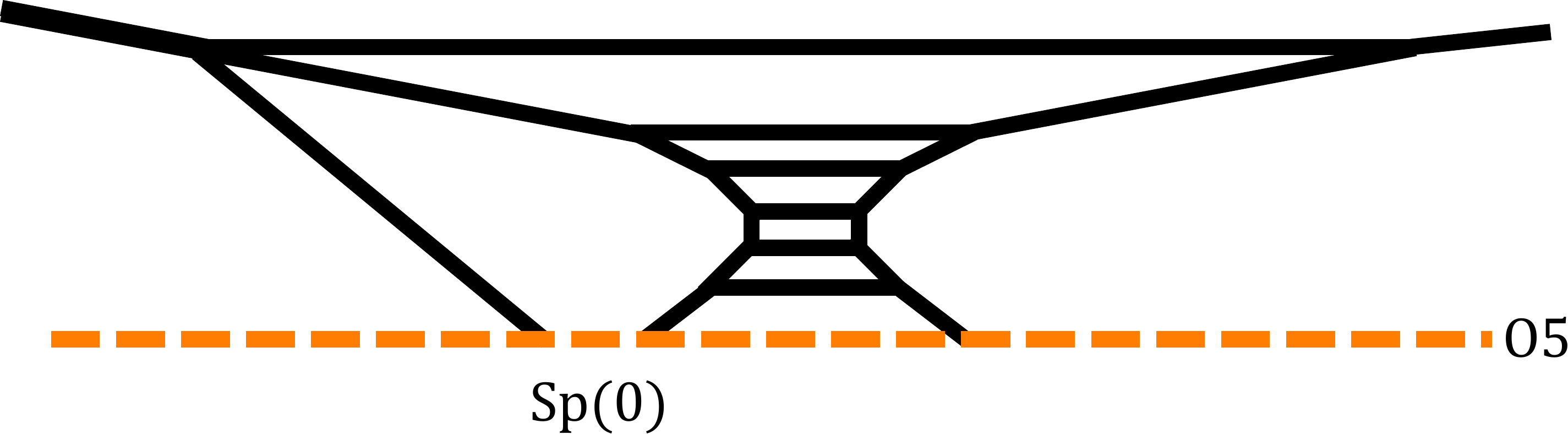} 
\caption{A 5-brane diagram which realizes the $SO(12)$ gauge theory with a half-hypermultiplet in the conjugate spinor representation. {In the left upper conner, two 5-branes of the charge $(-3,1)$ should be understood as they are bound by a single 7-brane of the same charge $(-3,1)$, respecting the S-rule.}}
\label{fig:so12wcs}
\end{figure}
A 5-brane web is constructed by a combination of $(p, q)$ 5-branes in type IIB string theory \cite{Aharony:1997ju, Aharony:1997bh} and it realizes a 5d theory on the brane web. As for the brane configuration, our convention is that a D5-brane extends in the $(x^0, x^1, x^2, x^3, x^4, x^5)$-directions and an NS5-brane extends in the $(x^0, x^1, x^2, x^3, x^4, x^6)$-directions in the ten-dimensional spacetime in type IIB string theory. A $(p, q)$ 5-brane extends in the $(x^0, x^1, x^2, x^3, x^4)$-directions and also in a one-dimensional space in the $(x^5, x^6)$-plane. The one-dimensional space is given by a line with slope $\frac{q}{p}$ in the $(x^5, x^6)$-plane. 7-branes in the $(x^0, x^1, x^2, x^3, x^4, x^7, x^8, x^9)$-directions may be also introduced in the configuration by ending a $(p, q)$ 7-brane on top of an external $(p, q)$ 5-brane. 7-branes are useful to see a global symmetry of the theory realized on a web and also to consider a Higgsing. Since a non-trivial structure of the brane appear in the $(x^5, x^6)$-plane, we only write the configuration in the two-dimensional plane where we choose the horizontal direction as the $x^5$-direction and the vertical direction as the $x^6$-direction.

 A 5-brane web for the $SO(12)$ gauge theory with a half-hypermultiplet in the spinor or the conjugate spinor representation has been proposed in \cite{Zafrir:2015ftn} and we depict the diagram in Figure \ref{fig:so12wcs}. The ``$Sp(0)$" part in-between the $(2, -1)$ 5-brane and the $(2, 1)$ 5-brane yields ``$Sp(0)$" instantons and they can be interpreted as a half-hypermultiplet in the spinor or the conjugate spinor representation depending on the discrete theta angle of the $Sp(0)$.

\begin{figure}
\centering
\includegraphics[width=10cm]{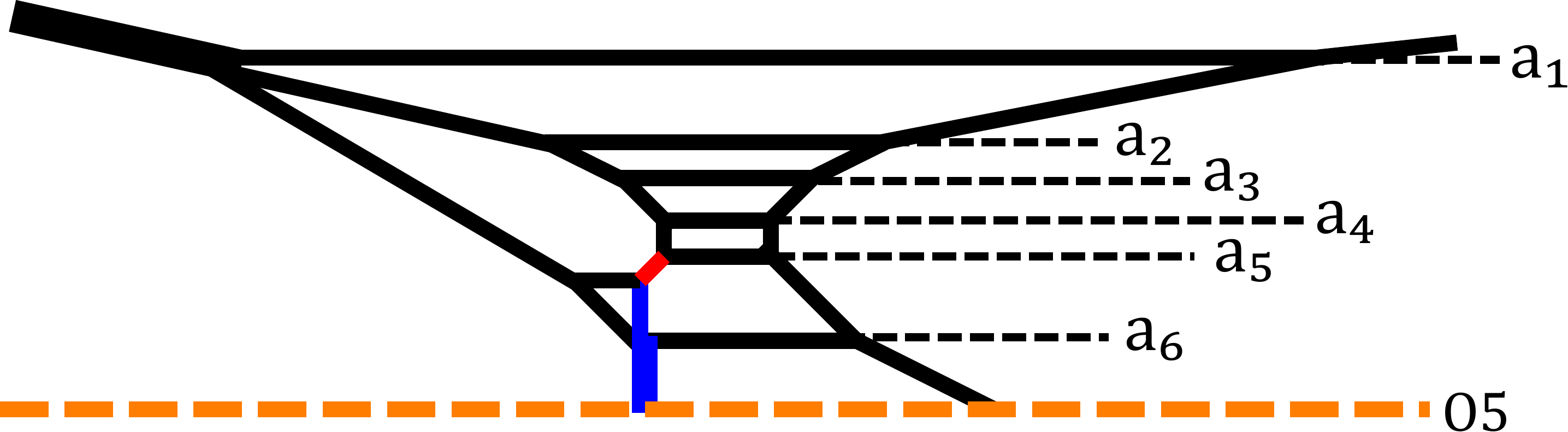} 
\caption{A 5-brane diagram realizing the $SO(12)$ gauge theory with a conjugate spinor which is obtained after performing a generalized flop transition to the diagram in Figure \ref{fig:so12wcs}.}
\label{fig:so12wcs1}
\end{figure}
In order to explicitly see if the configuration contains the spinor or the conjugate spinor representation, we may consider a diagram after a generalized flop transition for the $Sp(0)$ part, which can distinguish the discrete theta angle of the $Sp(0)$ \cite{Hayashi:2017btw}. It turns out that the generalized flop transition in the case of the conjugate spinor representation of $SO(12)$ yields the diagram depicted in Figure \ref{fig:so12wcs1}. To see that, we identify a weight of a representation for the matter in the theory from the length of an internal 5-brane in the diagram. 
We first label the height of the six color D5-branes as $a_1, a_2, a_3, a_4, a_5, a_6$ as in Figure \ref{fig:so12wcs1} and identify 
 them with the six Coulomb branch moduli of the $SO(12)$. This parameterization corresponds to being in a Weyl chamber specified by positive roots
\be
e_i \pm e_j, \qquad  (1 \leq i < j \leq 6). \label{SO12positiveroot}
\ee
With the parameterization, the length of the $(1, 1)$ 5-brane depicted as a red line in Figure \ref{fig:so12wcs1} is $\frac{1}{2}(a_1 - a_2 - a_3 - a_4 + a_5 + a_6)$. Also the length of the $(0, 1)$ 5-brane that is reflected in the O5-plane depicted as a blue line in Figure \ref{fig:so12wcs1} is $\frac{1}{2}(-a_1 + a_2 + a_3 + a_4 + a_5 + a_6)$. Hence, a string with the length $\frac{1}{2}(a_1 - a_2 - a_3 - a_4 + a_5 + a_6)$ connecting two D5-branes in the diagram yields a hypermultiplet for a weight
\be
\frac{1}{2}(e_1 - e_2 - e_3 - e_4 + e_5 + e_6), \label{SO12minimalweight1}
\ee
while a string with the length $\frac{1}{2}(-a_1 + a_2 + a_3 + a_4 + a_5 + a_6)$ connecting two D5-branes through the O5-plane in the diagram yields a hypermultiplet for a weight
\be
\frac{1}{2}(-e_1 + e_2 + e_3 + e_4 + e_5 + e_6), \label{SO12minimalweight2}
\ee
where $e_i, (i=1, \cdots, 6)$ are the orthonormal basis of $\mathbb{R}^6$. The weight of \eqref{SO12minimalweight1} and \eqref{SO12minimalweight2} are indeed weights in the conjugate spinor representation of $SO(12)$. Combining the 5-brane lines corresponding to the weight \eqref{SO12minimalweight1} and \eqref{SO12minimalweight2} with 5-brane lines for the positive roots of \eqref{SO12positiveroot} gives a half of the weights of the conjugate spinor representation which are given by
\begin{small}
\begin{align}
&\frac{1}{2}(-e_1+e_2+e_3+e_4+e_5+e_6),&&\frac{1}{2}(e_1-e_2+e_3+e_4+e_5+e_6),& &\frac{1}{2}(e_1+e_2-e_3+e_4+e_5+e_6), \nn\\
&\frac{1}{2}(e_1+e_2+e_3-e_4+e_5+e_6),& &\frac{1}{2}(e_1+e_2+e_3+e_4-e_5+e_6),&  &\frac{1}{2}(e_1+e_2+e_3+e_4+e_5-e_6),\label{SO12CS1}
\end{align}
\end{small}
and
\begin{small}
\begin{align}
&\frac{1}{2}(e_1 - e_2 - e_3 - e_4 + e_5 + e_6), &&\frac{1}{2}(e_1 - e_2 - e_3 + e_4 - e_5 + e_6),&&\frac{1}{2}(e_1 - e_2 - e_3 + e_4 + e_5 - e_6),\nn\\
&\frac{1}{2}(e_1 - e_2 + e_3 - e_4 - e_5 + e_6),&&\frac{1}{2}(e_1 - e_2 + e_3 - e_4 + e_5 - e_6),&&\frac{1}{2}(e_1 + e_2 - e_3 - e_4 - e_5 + e_6),\nn\\
&\frac{1}{2}(e_1 - e_2 + e_3 + e_4 - e_5 - e_6),& &\frac{1}{2}(e_1 + e_2 - e_3 - e_4 + e_5 - e_6),& &\frac{1}{2}(e_1 + e_2 - e_3 + e_4 - e_5 - e_6),\nn\\
&\frac{1}{2}(e_1+e_2+e_3-e_4-e_5-e_6). && && \label{SO12CS2}
\end{align}
\end{small}
Hence, the diagram in Figure \ref{fig:so12wcs1} yields hypermultiplets corresponding to the weights \eqref{SO12CS1} and \eqref{SO12CS2} or equivalently a half-hypermultiplet in the conjugate spinor representation of $SO(12)$.

\begin{figure}
\centering
\includegraphics[width=10cm]{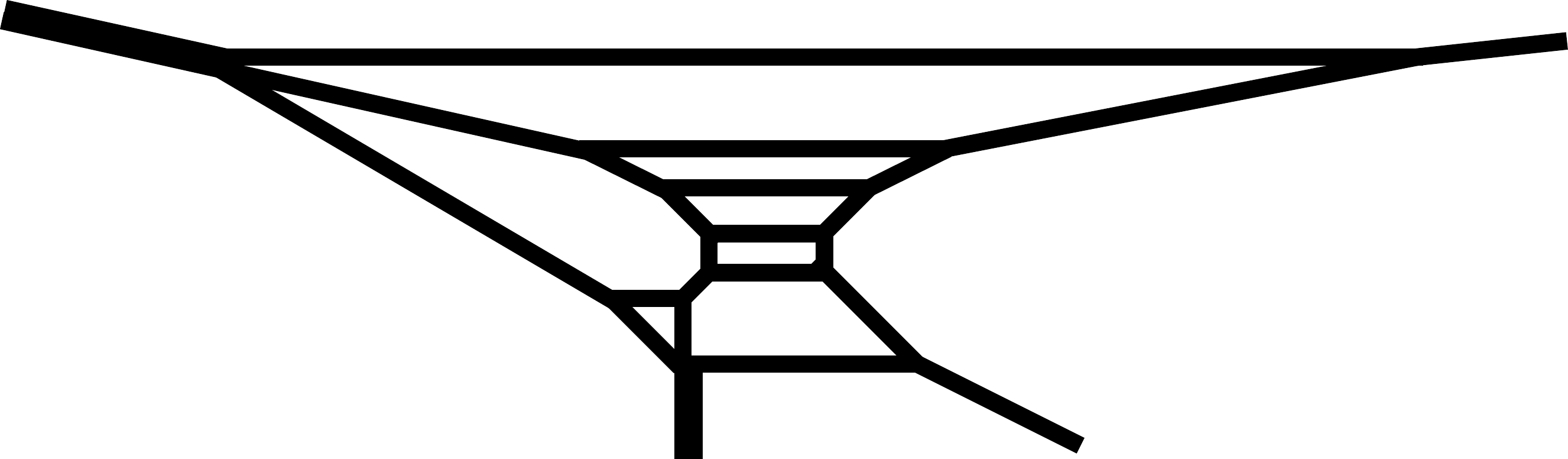} 
\caption{A 5-brane diagram realizing an $SU(6)$ gauge theory with a half-hypermultiplet in the rank-3 antisymmetric representation. It will turn out that the Chern-Simons level of this theory is $\kappa = \frac{5}{2}$ in section \ref{sec:su6wtsatension}. }
\label{fig:su6whtsa}
\end{figure}
In order to obtain a diagram for the $SU(6)$ gauge theory with a half-hypermultiplet in the rank-3 antisymmetric representation, we need to take a limit where the Coulomb branch modulus for the $U(1)$ in the decomposition $SO(12) \supset SU(6) \times U(1)$ becomes infinitely large. 
 It is in fact straightforward to identify this $U(1)$ degree of freedom from the diagram in Figure \ref{fig:so12wcs1}.  Due to the presence of the O5-plane, individual height of the six color D5-branes can be independent parameters. 
 {
 The $U(1)$ part (or ${\bf 1_0}$ of $SU(6)$) is the center of mass position of the color branes with respect to an O5-plane.
 One can adjust the bare coupling and increase the Coulomb parameter of $U(1)$ so that the orientifold gets pushed down while the $SU(6)$ part remains steady.} Therefore, the $U(1)$ part {becomes non-dynamical} 
 when we separate the brane configuration in the upper half-plane infinitely far away from the O5-plane. In this limit, the O5-plane is infinitely far from the other brane configuration and strings between them are decoupled. The resulting brane diagram without the O5-plane is depicted in Figure \ref{fig:su6whtsa} and the diagram should realize an $SU(6)$ gauge theory with a half-hypermultiplet in the rank-3 antisymmetric representation. From the diagram in Figure \ref{fig:su6whtsa}, the right part of the digram is identical to the one for the pure $SU(6)$ gauge theory. Hence the matter contribution comes only from the left part of the diagram. 

\begin{figure}[t]
\centering
\subfigure[]{\label{fig:so12wcs2}
\includegraphics[width=6.5cm]{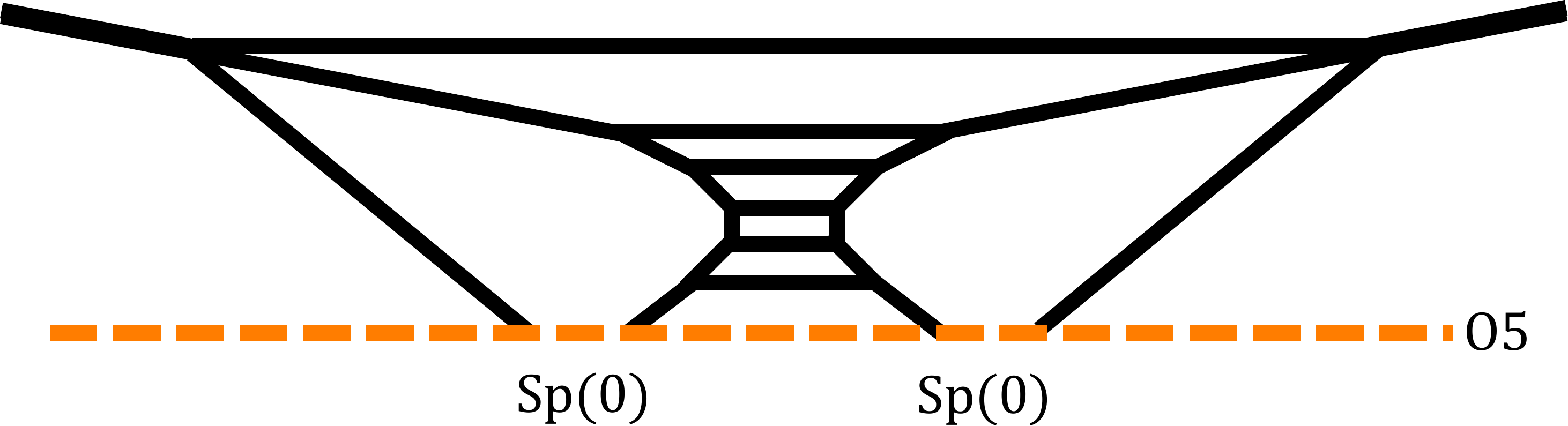}}\hspace{1cm}
\subfigure[]{\label{fig:su6wtsa}
\includegraphics[width=6.5cm]{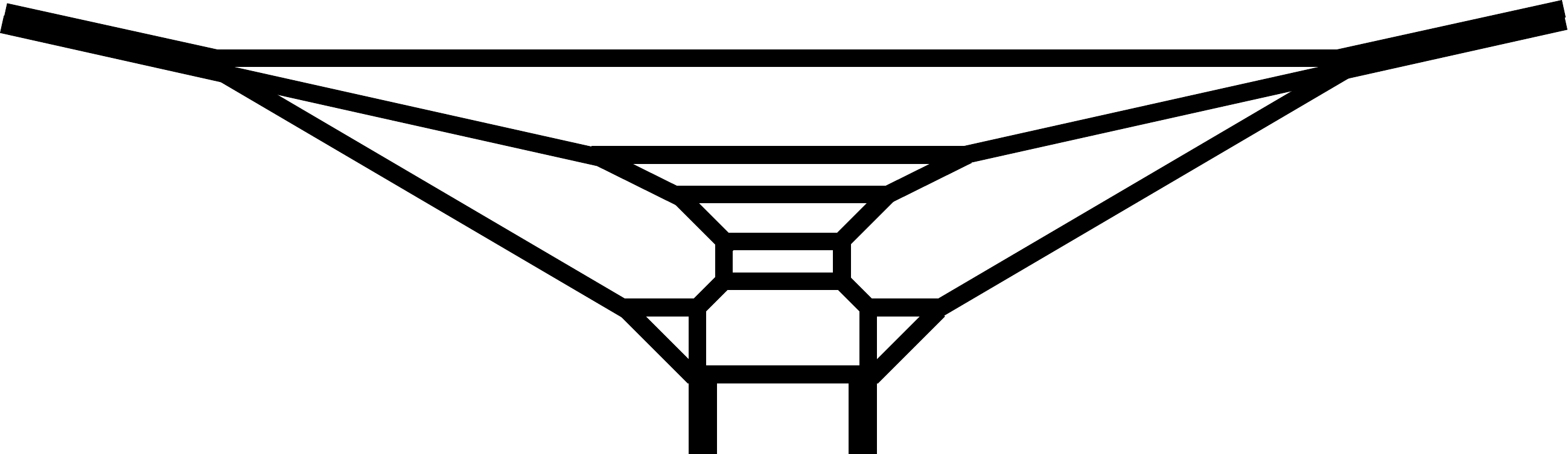}}
\caption{(a): A 5-brane diagram for the $SO(12)$ gauge theory with a massless hypermultiplet in the conjugate spinor representation. (b): A 5-brane diagram for an $SU(6)$ gauge theory with a massless hypermultiplet in the rank-3 antisymmetric representation, obtained by applying generalized flop transitions and decoupling to the diagram in Figure \ref{fig:so12wcs2}. It will turn out that the Chern-Simons level of this theory is $\kappa = 3$ in section \ref{sec:su6wtsatension}.}
\label{fig:so12su6wtsa}
\end{figure}

It is straightforward to construct a 5-brane diagram for an $SU(6)$ gauge theory with a hypermultiplet in the rank-3 antisymmetric representation. When we obtained a half-hypermultiplet in the rank-3 antisymmetric representation, we started from the diagram which realizes the $SO(12)$ gauge theory with a half-hypermultiplet in the conjugate spinor representation. Hence, we can start from the $SO(12)$ gauge theory with a hypermultiplet in the conjugate spinor representation in order to obtain a diagram for a rank-3 antisymmetric hypermultiplet. The proposed diagram in \cite{Zafrir:2015ftn} for the $SO(12)$ gauge theory with a conjugate spinor is depicted in Figure \ref{fig:so12wcs2}. The discrete theta angle for the two $Sp(0)$ parts should be chosen so that the diagram contains matter in the conjugate spinor representation of $SO(12)$. In this case, we can only realize massless hypermultiplet in the conjugate spinor representation. We then perform generalized flop transitions for the two $Sp(0)$ parts in the diagram in Figure \ref{fig:so12wcs2} and decouple the $U(1)$ degree of freedom. The procedure yields a diagram in Figure \ref{fig:su6wtsa} which should realize an $SU(6)$ gauge theory with a hypermultiplet in the rank-3 antisymmetric representation. Since we started from massless matter, the rank-3 antisymmetric hypermultiplet after the decoupling is also massless. It is indeed natural that the diagram in Figure \ref{fig:su6wtsa} gives a massless hypermultiplet in the rank-3 antisymmetric representation given that a half-hypermultiplet in the rank-3 antisymmetric representation comes from the left part of the diagram in Figure \ref{fig:su6whtsa}. The diagram consists of two copies of the left part of the diagram in Figure \ref{fig:su6whtsa} and hence it should give two half-hypermultiplets in the rank-3 antisymmetric representation, which correspond to a massless hypermultiplet in the rank-3 antisymmetric representation. 

\subsection{Monopole string tension}
\label{sec:su6wtsatension} 

In the previous subsection, we obtained 5-brane diagrams for $SU(6)$ gauge theories with rank-3 antisymmetric matter. We give further support for the claim by comparing the monopole string tension computed from the diagram in Figure \ref{fig:su6whtsa} with that calculated from the prepotential in the gauge theory. We have not yet determined the Chern-Simons (CS) level for the theories and the CS level can be also fixed from the monopole string tension computation.

\begin{figure}[t]
\centering
\subfigure[]{\label{fig:su6whtsacb}
\includegraphics[width=6.5cm]{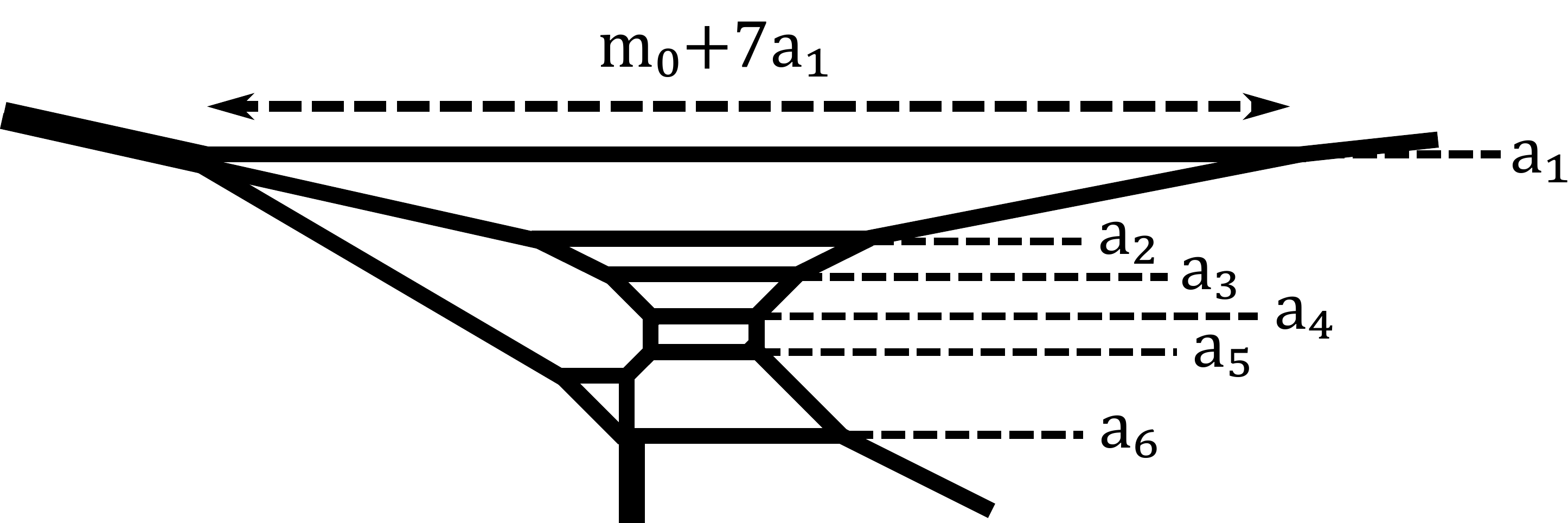}}\hspace{1cm}
\subfigure[]{\label{fig:su6whtsaarea}
\includegraphics[width=6.5cm]{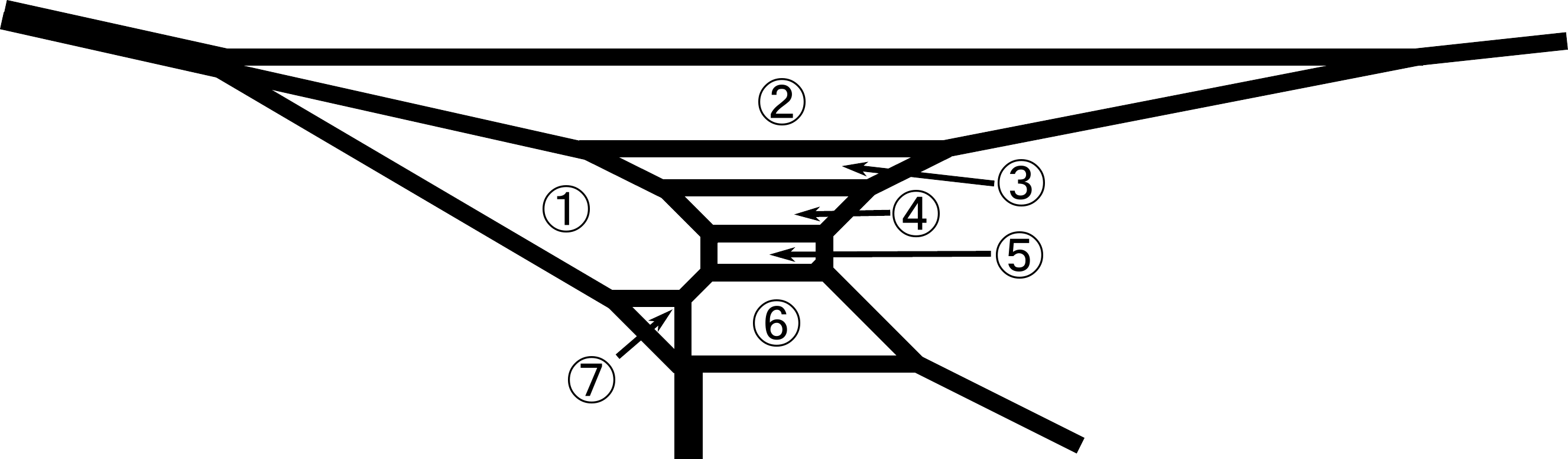}}
\caption{(a): A parameterization of Coulomb branch moduli for the diagram in Figure \ref{fig:su6whtsa}. (b): A labeling for the area of faces in the diagram in Figure \ref{fig:su6whtsa}. {As the external 5-branes are bound by 7-branes such that they satisfy the S-rule, some of the faces are in fact connected. For instance, $\textcircled{\scriptsize 1}$ and $\textcircled{\scriptsize 2}$ are connected and so are $\textcircled{\scriptsize 6}$ and $\textcircled{\scriptsize 7}$.}}
\label{fig:su6whtsa1}
\end{figure}

We first compute the monopole string tension from the diagram in Figure \ref{fig:su6whtsa}. A monopole string in a 5d theory can be realized by a D3-brane stretched on a face bounded by 5-brane segments in the corresponding 5-brane web. Hence the tension of the monopole string is given by the area of the face on which the D3-brane is stretched. In order to compute the area, we label the height of the six color D5-branes as $a_1, a_2, a_3, a_4, a_5, a_6$ as in Figure \ref{fig:su6whtsacb}. Contrary to the diagram for the $SO(12)$ gauge theory in Figure \ref{fig:so12wcs1}, the overall height is irrelevant and the parameters satisfy $\sum_{i=1}^6a_i = 0$, which can be solved by setting
\begin{align}
a_1 = \phi_1,&& a_2 = -\phi_1 + \phi_2,&& a_3=-\phi_2+\phi_3, &&a_4=-\phi_3+\phi_4,&& a_5=-\phi_4+\phi_5,&& a_6=-\phi_5 \label{CB.SU6whTSA}
\end{align}
On the other hand, the inverse of the squared classical gauge coupling $m_0$ is the length of D5-branes in the limit where all the Coulomb branch moduli are turned off. It turns out that the length of the top D5-brane is parameterized by $m_0+7a_1$ as in Figure \ref{fig:su6whtsacb}. 

We can then compute the area of faces in the diagram in Figure \ref{fig:su6whtsaarea}. {We note that the external 5-branes in Figure \ref{fig:su6whtsaarea} are bound by 7-branes in such a way that they satisfy the S-rule \cite{Benini:2009gi}, and hence though some of regions appear as distinct regions, they are in fact a single face. For instance,} 
the region $\textcircled{\scriptsize 1}$ is connected to $\textcircled{\scriptsize 2}$ and it forms a single face on which a D3-brane is stretched. Similarly the region $\textcircled{\scriptsize 7}$ is connected to $\textcircled{\scriptsize 6}$. Therefore we have in total five faces in the diagram in Figure \ref{fig:su6whtsaarea}, agreeing with the number of the Coulomb branch moduli of the $SU(6)$. The area of the five faces parameterized by $m_0, a_i, (i=1, \cdots, 6)$ is 
\begin{align}
\textcircled{\scriptsize 1} + \textcircled{\scriptsize 2} &= m_0(2\phi_1 - \phi_2) + \frac{5}{2}\phi_1^2 + 6\phi_1\phi_2 - 4\phi_2^2+\phi_2\phi_3 - \phi_3^2 + \phi_3\phi_4 - \phi_4^2 + \phi_4\phi_5 - \phi_5^2,\label{su6whtsaarea1}\\
\textcircled{\scriptsize 3} &= (m_0 - 3\phi_1 + 2\phi_2 + 2\phi_3)(-\phi_1 + 2\phi_2 - \phi_3),\\
\textcircled{\scriptsize 4} &= (m_0 - \phi_1 - \phi_2 + 2\phi_3 + \phi_4)(-\phi_2 + 2\phi_3 - \phi_4),\\
\textcircled{\scriptsize 5} &=(m_0 - \phi_1 + 2\phi_4)(-\phi_3 + 2\phi_4 - \phi_5),\\
\textcircled{\scriptsize 6} + \textcircled{\scriptsize 7} &=(m_0 - \phi_1 + \phi_4 + 2\phi_5)(-\phi_4 + 2\phi_5).\label{su6whtsaarea5}
\end{align}

We can compare the area \eqref{su6whtsaarea1}-\eqref{su6whtsaarea5} with the monopole string tension computed from the effective prepotential. In general the effective prepotential on a Coulomb branch of a 5d gauge theory with a gauge group $G$ and matter $f$ in a representation $R_f$ is given by \cite{Seiberg:1996bd, Morrison:1996xf, Intriligator:1997pq}\footnote{In \cite{Closset:2018bjz}, the authors approach the prepotential differently.}
\begin{align}
\mathcal{F}(\phi) = \frac{1}{2}m_0h_{ij}\phi_i\phi_j + \frac{\kappa}{6}d_{ijk}\phi_i\phi_j\phi_k + \frac{1}{12}\left(\sum_{r\in\text{roots}}\left|r\cdot \phi\right|^3 - \sum_f\sum_{w \in R_f}\left|w\cdot \phi - m_f\right|^3\right). \label{prepotential}
\end{align}
Here, $m_0$ is the inverse of the squared gauge coupling, $\kappa$ is the classical Chern-Simons level and $m_f$ is a mass parameter for the matter $f$. $r$ is a root of the Lie algebra $\mathfrak{g}$ associated to $G$ and $w$ is a weight of the representation $R_f$ of $\mathfrak{g}$. Furthermore, we defined $h_{ij} = \text{Tr}(T_iT_j), d_{ijk} = \frac{1}{2}\text{Tr}\left(T_i\{T_j, T_k\}\right)$ where $T_i$ are the Cartan generators of the Lie algebra $\mathfrak{g}$.

The sign for the one-loop correction terms in \eqref{prepotential} is fixed from the parameterization of the Coulomb branch moduli in the diagram in Figure \ref{fig:su6whtsacb}. Namely, the positive roots are given by $e_i - e_j, (1 \leq i < j \leq 6)$, and the positive weights of the rank-3 antisymmetric representation  are reduced from the positive weights of the conjugate spinor representation in \eqref{SO12CS1} and \eqref{SO12CS2}, $e_1 + e_i + e_j, (2 \leq i < j \leq 6)$. On this phase, the effective prepotential for the $SU(6)$ gauge theory with a half-hypermultiplet in the rank-3 antisymmetric representation ({\bf TAS}) becomes
\begin{align}
\mathcal{F}^{SU(6)_{\kappa}}_{N_{{\bf TAS}}=\frac{1}{2}} = \frac{1}{2}m_0\sum_{i=1}^6a_i^2 + \frac{\kappa}{6}\sum_{i=1}^6a_i^3 +\frac{1}{12}\left(2\sum_{1 \leq i < j \leq 6}(a_i - a_j)^3 - \sum_{2 \leq i < j \leq 6}(a_1 + a_i + a_j)^3\right), \label{pre.SU6whTSA}
\end{align}
where $\kappa$ is the CS level. We then rewrite the effective prepotential \eqref{pre.SU6whTSA} in terms of the Coulomb branch moduli $\phi_i, (i=1, \cdots, 5)$ in \eqref{CB.SU6whTSA}, and the monopole string tension is given by taking the derivative of the effective prepotential with respect to the $\phi_i$. Then the comparison with \eqref{su6whtsaarea1} which corresponds to taking the derivative with respect to $\phi_1$,
\begin{align}
\frac{\partial \mathcal{F}^{SU(6)_{\kappa}}_{N_{{\bf TAS}}=\frac{1}{2}} }{\partial \phi_1} = \textcircled{\scriptsize 1} + \textcircled{\scriptsize 2},
\end{align}
yields $\kappa = \frac{5}{2}$. Hence the diagram in Figure \ref{fig:su6whtsa} realizes the $SU(6)$ gauge theory with a half-hypermultiplet in the rank-3 antisymmetric representation and the CS level $\kappa = \frac{5}{2}$. Fixing the CS level to $\frac{5}{2}$, the other comparison between the area and the monopole string tension may be interpreted as support for our claim that the diagram in Figure \ref{fig:su6whtsa} yields the $SU(6)$ gauge theory with $N_{{\bf TAS}} = \frac{1}{2}$ and $\kappa = \frac{5}{2}$. Indeed the explicit comparison gives
\begin{align}
\frac{\partial \mathcal{F}^{SU(6)_{\frac{5}{2}}}_{N_{{\bf TAS}} =\frac{1}{2}}}{\partial \phi_2} = \textcircled{\scriptsize 3},\quad
\frac{\partial \mathcal{F}^{SU(6)_{\frac{5}{2}}}_{N_{{\bf TAS}} =\frac{1}{2}}}{\partial \phi_3} = \textcircled{\scriptsize 4},\quad
\frac{\partial \mathcal{F}^{SU(6)_{\frac{5}{2}}}_{N_{{\bf TAS}} =\frac{1}{2}}}{\partial \phi_4} = \textcircled{\scriptsize 5},\quad
\frac{\partial \mathcal{F}^{SU(6)_{\frac{5}{2}}}_{N_{{\bf TAS}} =\frac{1}{2}}}{\partial \phi_5} = \textcircled{\scriptsize 6} + \textcircled{\scriptsize 7}.
\end{align}

It is also possible to make a comparison between the area and the monopole string tension for the diagram in Figure \ref{fig:su6wtsa}. We checked the agreement and the CS level of the $SU(6)$ gauge theory realized by the diagram in Figure \ref{fig:su6wtsa} is $\kappa = 3$.

\subsection{Nekrasov partition function}

As we have seen in the previous subsection, the computation of the monopole string tension confirms that the diagrams in Figure \ref{fig:su6whtsa} and in Figure \ref{fig:su6wtsa} realize the $SU(6)$ gauge theory with $N_{{\bf TAS}} = \frac{1}{2}$ and $\kappa = \frac{5}{2}$ and the $SU(6)$ gauge theory with $N_{{\bf TAS}} = 1$ and $\kappa = 3$ respectively. We can now use power of 5-brane web diagrams to compute various physical quantities from the 5-brane web diagrams. One important application is to compute the Nekrasov partition function or the topological string partition function from the 5-brane webs using the topological vertex \cite{Aganagic:2003db, Iqbal:2007ii}. Although the topological vertex was originally formulated to compute the all genus topological string partition function for toric Calabi-Yau threefolds, we can also apply the topological vertex to non-toric diagrams obtained from a Higgsing of toric diagrams \cite{Hayashi:2013qwa, Hayashi:2014wfa, Kim:2015jba, Hayashi:2015xla} and also to diagrams with an O5-plane \cite{Kim:2017jqn}. 

\begin{figure}[t]
\centering
\includegraphics[width=8cm]{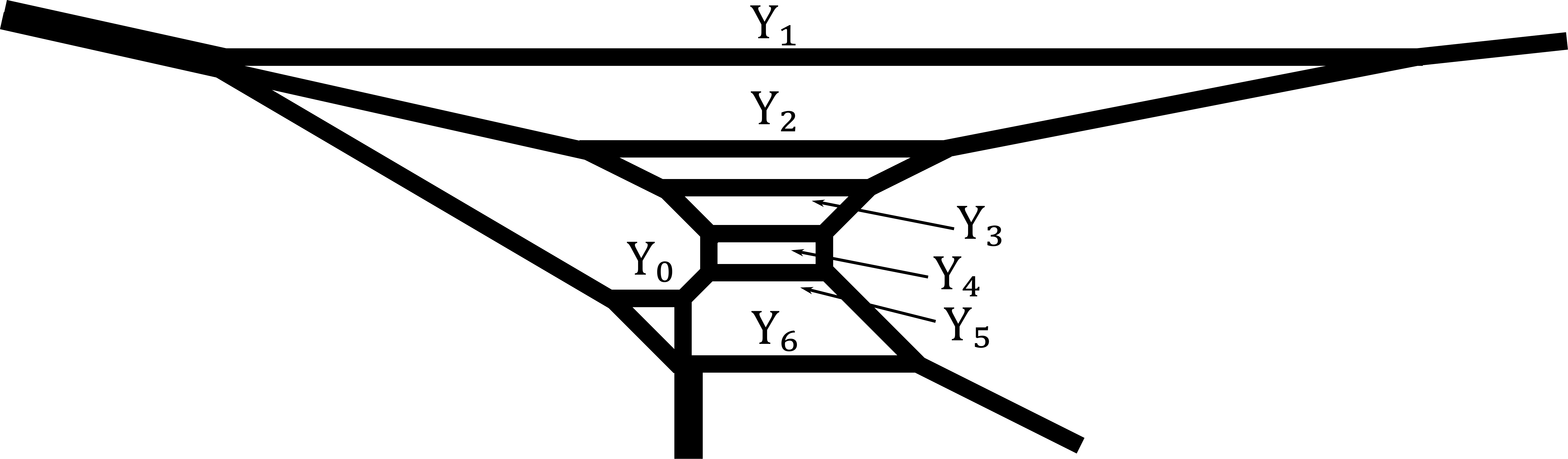}
\caption{A labeling of Young diagrams assigned to the horizontal lines in Figure \ref{fig:su6whtsa}.}
\label{fig:su6whtsay}
\end{figure}

By using the techniques, it is straightforward to apply the topological vertex for the 5-brane diagrams in Figure \ref{fig:su6whtsa} and in Figure \ref{fig:su6wtsa}. We here illustrate the computation by using the diagram in Figure \ref{fig:su6whtsa} and we calculate the Nekrasov partition function for the $SU(6)$ gauge theory with $N_{\bf TAS} = \frac{1}{2}$ and $\kappa = \frac{5}{2}$. 
To this end, we first assign Young diagrams $Y_0, Y_1, \cdots, Y_6$ to the horizontal lines as in Figure \ref{fig:su6whtsay}.
The topological vertex based on  the diagram in Figure \ref{fig:su6whtsay} yields 
\begin{align}
Z_{\text{Nek}} 
=& \sum_{\vec{Y}}q^{\sum_{i=1}^6|Y_i|} (-A_1^6)^{|Y_1|}(-A_2^6)^{|Y_2|}(-A_1A_2^2A_3^4)^{|Y_3|}\nn\\
&\times (-A_1A_2^2A_3^2A_4^2)^{|Y_4| + |Y_5|}(-A_1^2A_2^2A_3^2A_4^2A_5^2)^{|Y_6|}f_{Y_1}(g)^6f_{Y_2}(g)^5f_{Y_3}(g)^3f_{Y_4}(g)\nn\\
&\times f_{Y_5}(g)^{-1}f_{Y_6}(g)^{-1}Z_{\text{left}}(\vec{Y})Z_{\text{right}}(\vec{Y}), \label{Znek1}
\end{align}
where $\vec{Y}=(Y_1, Y_2, Y_3, Y_4, Y_5, Y_6)$.
$Z_{\text{left}}(\vec{Y})$ and $Z_{\text{right}}(\vec{Y})$ are contributions of the left part and the right part of the web in Figure \ref{fig:su6whtsay} respectively when we cut the diagram at the horizontal lines with the Young diagrams $Y_i, (i=1, \cdots, 6)$ and they are given by
\begin{align}
Z_{\text{left}}(\vec{Y})=& 
\sum_{Y_0} ( - A_1{}^{-1} A_6{}^{-2})^{|Y_0|} 
g^{\frac{||Y_0^t||^2+||Y_0||^2}{2}} \tilde{Z}_{Y_0}^2 f_{Y_0}^2(g)
\prod_{i=1}^6 g^{\frac{||Y_i||^2}{2}} \tilde{Z}_{Y_i} 
\cr 
& 
\times 
R_{Y_1 Y_6^t}^{-1} (A_1 A_6{}^{-1}) \prod_{2 \le i <  j \le 5} R_{Y_i Y_j^t}^{-1} (A_i A_j{}^{-1})
\cr 
& 
\times  
R_{Y_0 Y_6^t}^{-1} (A_1{}^{-1} A_6{}^{-2})R_{Y_1Y_0^t }^{-1} (A_1^2A_6) \prod_{i=2}^5 R_{Y_0^t Y_i} (A_1 A_i  A_6), 
\\
Z_{\text{right}}(\vec{Y})=&
\prod_{i=1}^6  g^{\frac{||Y_i^t||^2}{2}} \tilde{Z}_{Y^t_i} \times
\prod_{1 \le i <  j \le 6} R_{Y_i Y_j^t}^{-1} (A_i A_j{}^{-1})
\end{align}
with 
\begin{align}
A_6 
= \prod_{i=1}^5 A_i{}^{-1}, 
\quad
\tilde{Z}_{\lambda} 
= \prod_{(i,j) \in \lambda} \frac{1}{1 - g^{\lambda_i + \lambda^t_j - i - j +1} } , 
\quad
R_{\lambda \mu } (Q)= \prod_{i.j=1}^{\infty} (1 - Q g^{i+j-\lambda_j - \mu_i -1}).
\end{align}
$f_{Y}(g)$ is the framing factor defined by
\begin{align}
f_Y(g) = (-1)^{|Y|}g^{\frac{1}{2}(g^{||Y^t||^2 - ||Y||^2})},
\end{align}
and the Coulomb branch parameters $A_i, (i=1, \cdots, 6)$, the instanton fugacity $q$ and the unrefined $\Omega$-deformation parameter $g$ are defined by 
\begin{align}
A_i = e^{-a_i}, \qquad q = e^{-m_0}, \qquad g=e^{-\epsilon}.
\end{align}
We argue that the topological string partition function \eqref{Znek1} is the Nekrasov partition function for the the 5d $SU(6)$ gauge theory with $N_{\bf TAS} = \frac{1}{2}$ and $\kappa = \frac{5}{2}$.
The partition function can be written as a sum of the instanton partition functions 
\begin{align}
Z_{\text{Nek}} = Z_{\text{pert}}\left(1 + \sum_{k=1}^{\infty}q^kZ_k\right) ,
\end{align}
where $Z_{\text{pert}}$ represents the perturbative part of the partition function given by the order $q^0$ in \eqref{Znek1}, while $Z_k$ stands for the $k$-instanton partition function. 

Let us first look at the perturbative part. 
This is obtained from the contribution of $Y_1=Y_2=\cdots = Y_6 = \emptyset$ in  \eqref{Znek1} and is given as
\begin{align}
Z_{\rm pert} 
=& \,
Z_{\text{left}}( (\emptyset,\emptyset,\emptyset,\emptyset,\emptyset,\emptyset ) )
Z_{\text{right}}( (\emptyset,\emptyset,\emptyset,\emptyset,\emptyset,\emptyset ) )
\cr
=&\, 
\text{PE} \Biggl[ 
\frac{g}{(1-g)^2} 
\Bigl( A_1 A_6{}^{-1} 
+ \sum_{2 \le i <  j \le 5} A_i A_j{}^{-1}
+ \sum_{1 \le i <  j \le 6} A_i A_j{}^{-1}
\cr
& \qquad \qquad \qquad \qquad 
+ A_1{}^{-1} A_6{}^{-2}+A_1{}^{2} A_6{}^{} 
- \sum_{i=2}^5 A_1 A_i  A_6
\Bigr)
\Biggr]
\cr 
& 
\times \bigg(\sum_{Y_0} ( - A_1{}^{-1} A_6{}^{-2})^{|Y_0|} 
g^{\frac{||Y_0^t||^2+||Y_0||^2}{2}} \tilde{Z}_{Y_0}^2 f_{Y_0}^2(g)
\cr 
&\qquad \times N_{Y_0^t \emptyset }^{-1} (A_1{}^{-1} A_6{}^{-2})
 N^{-1}_{Y_0 \emptyset } (A_1^2  A_6)\prod_{i=2}^5 N_{Y_0 \emptyset } (A_1 A_i  A_6)\bigg) ,
\end{align}
where we used the identity
\begin{align}
R_{\lambda \mu} (Q)
&= \text{PE} \left[ - \frac{g}{(1-g)^2} Q \right]
\times N_{\lambda^t \mu} (Q)
\end{align}
with PE representing the Plethystic exponential
 and 
\begin{align}
N_{\lambda \mu} (Q) 
= \prod_{(i,j) \in \lambda} \left( 1 - Q g^{\lambda_i + \mu_j^t -i-j+1} \right)
\prod_{(i,j) \in \mu} \left( 1 - Q g^{-\lambda^t_j - \mu_i + i + j - 1} \right).
\end{align}
Note that in order to obtain the exact expression for the perturbative part we still need to sum over the Young diagram $Y_0$. We can still evaluate the summation in terms of an expansion by $A_1$. Namely when we sum over the Young diagram until $|Y_0| \leq k$, the expression is exact until the order $A_1^k$. The summation of the Young diagram $Y_0$ until $|Y_0| = 7$ yields the expression  
\begin{align}
Z_{\rm pert} 
&= \text{PE}\left[\frac{g}{(1-g)^2} 
\left( 2\sum_{1 \le i < j \le 6} A_i A_j{}^{-1} 
- \sum_{1 = i < j < k \le 6} A_i A_j A_k
+ \mathcal{O} (A_1{}^8)\right)\right].\label{Zpert1}
\end{align} 
We observed that the the series expansion by $A_1$ gives an expression which stops at the order $A_1$ inside the Plethystic exponential as far as we checked. Hence, we claim that $\mathcal{O} (A_1{}^8)$ term is actually exactly zero. Indeed, the partition function \eqref{Zpert1} is exactly equal to the perturbative part of the partition function of the $SU(6)$ gauge theory with a half-hypermultiplet in the rank-3 antisymmetric representation. We can also see that the charge of the BPS states counted by the perturbative partition function agrees with the charge of the positive weights used in the prepotential computation for \eqref{pre.SU6whTSA}.

Next, we compute the $1$-instanton part. The $1$-instanton part can be read off from the coefficient of the $q^1$ order part in \eqref{Znek1} divided by the perturbative part given in \eqref{Zpert1}.
Hence the order $q^1$ contribution is given by combinations where $|Y_i| = 1$ for one of the $Y_i, (i=1, \cdots, 6)$ and the others are trivial. Furthermore, we still need to sum over $Y_0$ and evaluate the summation in terms of a series expansion by $A_1$\footnote{The expression \eqref{Znek1} contains factors with $A_1$ in the denominator. We perform a series expansion by $A_1$ only for the numerator of \eqref{Znek1}. }. $A_1$ is a good expansion parameter since the explicit summation of Young diagrams in \eqref{Znek1} involves only positive powers of $A_1$. 
For example, the contribution from $|Y_1| = 1 , |Y_j| = 0 \,\, (j=2,3,4,5,6)$ to the 1-instanton part is given by 
\begin{align}
 - \frac{g}{(1-g)^2} \frac{A_1^5}{\prod_{i= 2}^6 (A_i - A_1)^2} 
 \Bigl[
1 - \sum_{i = 2}^6  A_i{}^{-1} A_1 + \sum_{i = 2}^6  A_i A_1^2 - A_1^3
+ \mathcal{O} (A_1{}^8)
\Bigr]. \label{Z1inst1}
\end{align} 
We again observed that the stop of the series expansion by $A_1$ in the numerator of \eqref{Z1inst1} and we claim that the $\mathcal{O} (A_1{}^8)$ term is actually exactly zero. 
Similarly we can also compute the other combinations of the Young diagrams which contribute to the $1$-instanton part. 
Summing up all the contributions from the other combinations of the Young diagrams which contribute to the $1$-instanton part, we obtain
\begin{align}
Z_{1} 
&= - \sum_{\ell=1}^6
\frac{g}{(1-g)^2} \frac{A_\ell^5}{\prod_{i \neq \ell} (A_i - A_\ell)^2} 
\Bigl[
1 - \sum_{i \neq \ell}  A_i{}^{-1} A_\ell + \sum_{i \neq \ell}  A_i A_\ell^2 - A_\ell^3
\Bigr]
\nn\\
&= - \sum_{\ell=1}^6
\frac{e^{-\frac{5}{2}a_\ell}}{(2 \sinh \frac{\epsilon}{2} )^2\prod_{i \neq \ell} (2 \sinh \frac{a_i-a_\ell}{2})^2} 
\Bigl[ (2\sinh \frac{3 a_\ell}{2}) -
 \sum_{i \neq \ell}  (2\sinh \frac{2a_i +a_\ell}{2}) \Bigr].
\label{Zinst1}
\end{align}
This is the explicit expression for the $1$-instanton part of the partition function for the $SU(6)$ gauge theory with $N_{\bf TAS} = \frac{1}{2}$ and $\kappa = \frac{5}{2}$.

The two-instanton contribution can be written in the following form:
\begin{align}\label{eq:2inst}
& Z_2=  Z_{\{ 1,1 \}} + Z_{\{ 2 \}} + Z_{\{ 1\} , \{ 1 \}}
+ \mathcal{O}(A_1^{11})
\end{align}
with 
\begin{align}
& Z_{\{ 1,1 \}} 
= 
\sum_{\ell=1}^6 A_\ell{}^5 g{}^{\frac{5}{2}}
Z_{\{ 1,1 \}_\ell}^{\text{\bf Vec}} Z_{\{ 1,1 \}_\ell}^{\text{\bf TAS}},
\qquad
Z_{\{ 2 \}} 
= Z_{\{ 1,1 \}} (g \to g^{-1}),
\cr
& Z_{\{ 1\} , \{ 1 \}}
=  \sum_{1 \le \ell < m \le 6}
A_\ell {}^{\frac{5}{2}} A_m{}^{\frac{5}{2}} 
Z_{\{ 1\}_\ell ,\{1 \}_m}^{\text{\bf Vec}} Z_{\{ 1\}_\ell ,\{1 \}_m}^{\text{\bf TAS}}.
\end{align}
Here, the lower indices $\{ 1,1 \}_\ell$ indicates the contribution from the Young diagrams $Y_{\ell}=\{1,1\}$ and $Y_i = \emptyset$ $(i\neq \ell)$, while $\{ 1\}_\ell ,\{1 \}_m$ indicates the contribution from $Y_{\ell}=\{1\}$, $Y_{m}=\{1\}$ ($m \neq \ell$), and $Y_i = \emptyset$ $(i\neq \ell, m)$. 
The contributions from the vector multiplets are given as
\begin{align}
Z_{\{ 1,1 \}_\ell}^{\text{\bf Vec}}
&= 
\frac{g^8}{(1-g)^2(1-g^2)^2} 
\frac{A_\ell{}^{8}  }{ \prod_{i \neq \ell} (A_i - A_\ell)^2  (A_i -  g A_\ell)^2}
\cr
& =
\frac{1}{
\left( 2 \sinh \frac{\epsilon}{2} \right){}^2 \left( 2 \sinh \epsilon \right){}^2
\prod_{i \neq \ell} \left( 2 \sinh \frac{a_i - a_\ell}{2} \right){}^2
\left( 2 \sinh \frac{a_i - a_\ell - \epsilon}{2} \right){}^2
} 
\end{align}
and
\begin{align}
Z_{\{ 1\} , \{1 \}_\ell}^{\text{\bf Vec}}
&= \frac{g^2}{(1-g)^4} 
\frac{A_\ell {}^{4} A_m{}^{4} }{ 
(A_m-g A_\ell)^2(A_m-g^{-1} A_\ell)^2
\prod_{i \neq \ell,m} (A_i - A_\ell)^2  (A_i -  A_m)^2}
\cr
&= 
\frac{1}{ \left( 2 \sinh \frac{\epsilon}{2} \right)^4
\left( 2 \sinh \frac{a_\ell - a_m + \epsilon}{2} \right){}^2
\left( 2 \sinh \frac{a_\ell - a_m - \epsilon}{2} \right){}^2}
\cr
& \qquad \times 
\frac{1}{\prod_{i \neq \ell,m} 
\left( \sinh \frac{a_i-a_\ell}{2} \right){}^2
\left( \sinh \frac{a_i-a_m}{2} \right){}^2 }
\end{align}
while the contributions from the hypermultiplet in rank-3 antisymmetric tensor representation are
\begin{align}
Z_{\{ 1,1 \}_\ell}^{\text{\bf TAS}}
= & 
g^{-\frac{3}{2}} A_\ell{}^{-3}
\Biggl[
(g^3 A_\ell^3 - 1) 
\left( 
-1 + A_\ell \chi_\ell^{\overline{\tiny\yng(1)}} - A_\ell^2 \chi_\ell^{\tiny\yng(1)}  + A_\ell^3
\right) 
\cr
&   + g^2 A_\ell{}^2 
\left(
\chi_\ell^{\overline{\tiny\yng(1,1)}} - A_\ell \chi_\ell^{\tiny\yng(1)} \chi_\ell^{\overline{\tiny\yng(1)}}
+  A_\ell^2 
\left( \chi_\ell^{\overline{\tiny\yng(1)}} + (\chi_\ell^{\tiny\yng(1)})^2 - \chi_\ell^{{\tiny\yng(1,1)}} \right) 
-  A_\ell^3 \chi_\ell^{\tiny\yng(1)}
\right)
\cr
&   + g A_\ell 
\left( 
-\chi_\ell^{\overline{\tiny\yng(1)}} + A_\ell (\chi_\ell^{\tiny\yng(1)} + (\chi_\ell^{\overline{\tiny\yng(1)}})^2 - \chi_\ell^{\overline{\tiny\yng(1,1)}}) - A_\ell{}^2 \chi_\ell^{\tiny\yng(1)} \chi_\ell^{\overline{\tiny\yng(1)}}  + A_\ell{}^3 \chi_\ell^{{\tiny\yng(1,1)}} )
\right)
\Biggr]
\cr
= & 
\left( 2 \sinh \frac{3(a_\ell+\epsilon)}{2} \right)
\left( 2 \sinh \frac{3a_\ell}{2} - \sum_{i \neq \ell} 2 \sinh \frac{a_\ell + 2a_i}{2} \right)
\cr
& 
\qquad - \left( \sum_{i\neq \ell} 2 \sinh\frac{a_\ell+2a_i}{2} \right)
\left( 2\sinh \frac{3a_\ell+\epsilon}{2} - \sum_{i \neq \ell} 2 \sinh \frac{a_\ell+2a_i+\epsilon}{2} \right)
\cr
&
\qquad - \left( 2 \sinh \frac{\epsilon}{2} \right)
\left( \sum_{1\le i < j \le 6 \atop i , j \neq \ell} 2 \sinh (a_i+a_j+a_\ell) \right)
\end{align}
and
\begin{align}
Z_{\{ 1\}_\ell ,\{1 \}_m}^{\text{\bf TAS}}
= & 
A_\ell{}^{-\frac{3}{2}} A_m{}^{-\frac{3}{2}}
\Biggl[
(g + g^{-1}) 
\left( 
A_\ell {}^3 A_m{}^3 
+ A_\ell {}^2 A_m{}^2 
\left( \chi_{\ell,m}^{\tiny\yng(1,1)} - \chi_{\ell,m}^{\overline{\tiny\yng(1)}} \right) 
- A_\ell A_m \chi_{\ell,m}^{\tiny\yng(1)}  
+ 1
\right)
\cr
& 
\quad 
- (A_\ell {}^3 + A_m{}^3 + A_\ell {}^4 A_m{}^2 + A_\ell {}^2 A_m{}^4 
- A_\ell{}^2 A_m - A_\ell A_m{}^2
+ A_\ell {} A_m{}^{-1} + A_\ell {}^{-1} A_m)
\cr
& 
\quad
+ A_\ell A_m ( A_\ell{}^2 + A_m{}^2) \chi_{\ell,m}^{\overline{\tiny\yng(1)}}  
- A_\ell A_m (A_\ell + A_m) \chi_{\ell,m}^{\tiny\yng(1)} \chi_{\ell,m}^{\overline{\tiny\yng(1)}}  
+ (A_\ell {}^2 + A_m{}^2) \chi_{\ell,m}^{\tiny\yng(1)}
\cr
&
\quad
+  A_\ell A_m (\chi_{\ell,m}^{\overline{\tiny\yng(1)}} )^2  
+ A_\ell{}^2 A_m{}^2 
 \left( (\chi_{\ell,m}^{\tiny\yng(1)})^2 - 2 \chi_{\ell,m}^{\tiny\yng(1,1)} \right)
 \Biggr]
\cr
= &
\left( \sum_{i \neq \ell,m}  2 \sinh \frac{2 a_i + a_\ell}{2} \right)
\left( \sum_{i \neq \ell,m}  2 \sinh \frac{2 a_i + a_m}{2} \right)
\cr
&
+ 
\left( 2 \sinh \frac{a_\ell - a_m + \epsilon}{2} \right)
\left( 2 \sinh \frac{a_\ell - a_m - \epsilon}{2} \right)
\left( \sum_{i \neq \ell,m} 2 \cosh \frac{a_\ell+a_m - 2a_i}{2} \right)
\cr
& 
+ 
\left( 2 \sinh \frac{\epsilon}{2} \right){}^2
\Biggl( 
2 \cosh \frac{3a_\ell+3a_m}{2}
+ \sum_{1 \le i < j \le 6 \atop {i \neq \ell, m \atop j \neq \ell,m} } 
\cosh \frac{a_\ell + a_m + 2a_i + 2a_j}{2} \Biggr)
\cr
& 
- 
\left( 2 \sinh \frac{a_m - a_\ell}{2} \right){}^2
\left( 2 \cosh \frac{2 a_\ell + a_m}{2} \right)
\left( 2 \cosh \frac{a_\ell + 2a_m}{2} \right),
\end{align}
Here, we have introduced the following $U(5)$ characters
\begin{align}
& \chi_\ell^{\tiny\yng(1)
} = \sum_{i \neq \ell} A_i, 
\quad
\chi_\ell^{\overline{\tiny\yng(1)}
} = \sum_{i \neq \ell} A_i{}^{-1}, 
\quad
\chi_\ell^{
{\tiny\yng(1,1)}
} = \sum_{1 \le i < j \le 6 \atop i \neq \ell, j \neq \ell} A_i A_j,
\quad
\chi_\ell^{
\overline{\tiny\yng(1,1)}
} = \sum_{1 \le i < j \le 6 \atop i \neq \ell, j \neq \ell} A_i{}^{-1} A_j{}^{-1},
\end{align}
and the following $U(4)$ characters
\begin{align}
& \chi_{\ell,m}^{
\tiny\yng(1)
} = \sum_{i \neq \ell,m} A_i, 
\quad
\chi_{\ell,m}^{
\overline{\tiny\yng(1)}
} = \sum_{i \neq \ell,m} A_i{}^{-1}, 
\quad
\chi_{\ell,m}^{
\tiny\yng(1,1)
} = \sum_{1 \le i < j \le 6 \atop {i \neq \ell,m \atop j \neq \ell,m}} A_i A_j.
\end{align}
Analogous to the case of the perturbative and the 1-instanton contribution, 
we claim that the $\mathcal{O}(A_1^{11})$ term in $\eqref{eq:2inst}$ is exactly zero.

\subsection{$SU(6)$ gauge theories with $N_{\text{{\bf TAS}}} = \frac{3}{2}$ and $2$}\label{sec:su6+2TAS}

\begin{figure}
\centering
\subfigure[]{\label{fig:su6su3}
\includegraphics[width=6.5cm]{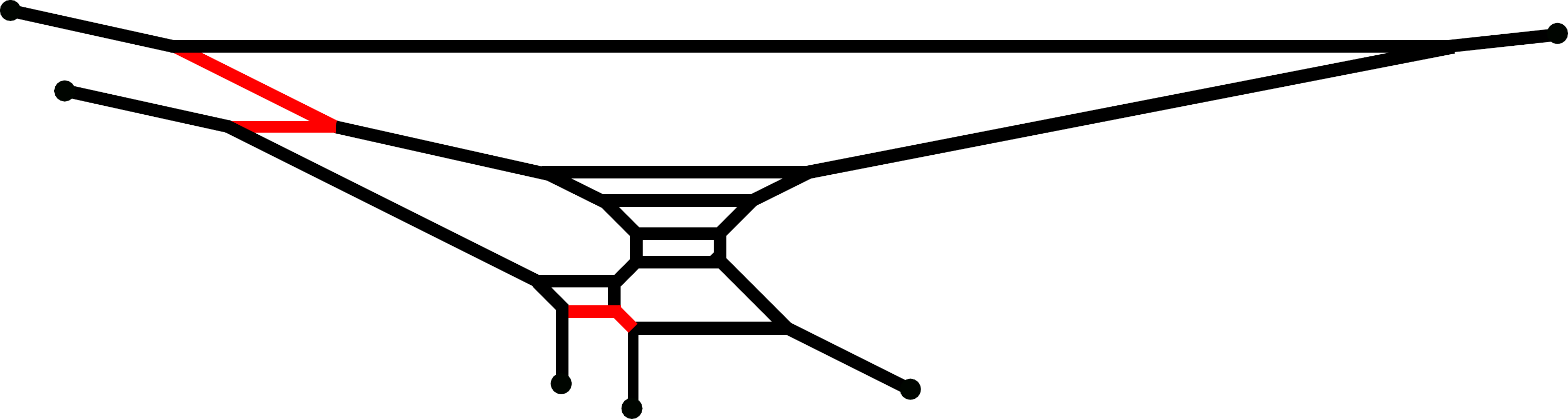}}\hspace{1cm}
\subfigure[]{\label{fig:su6su3v2}
\includegraphics[width=6.5cm]{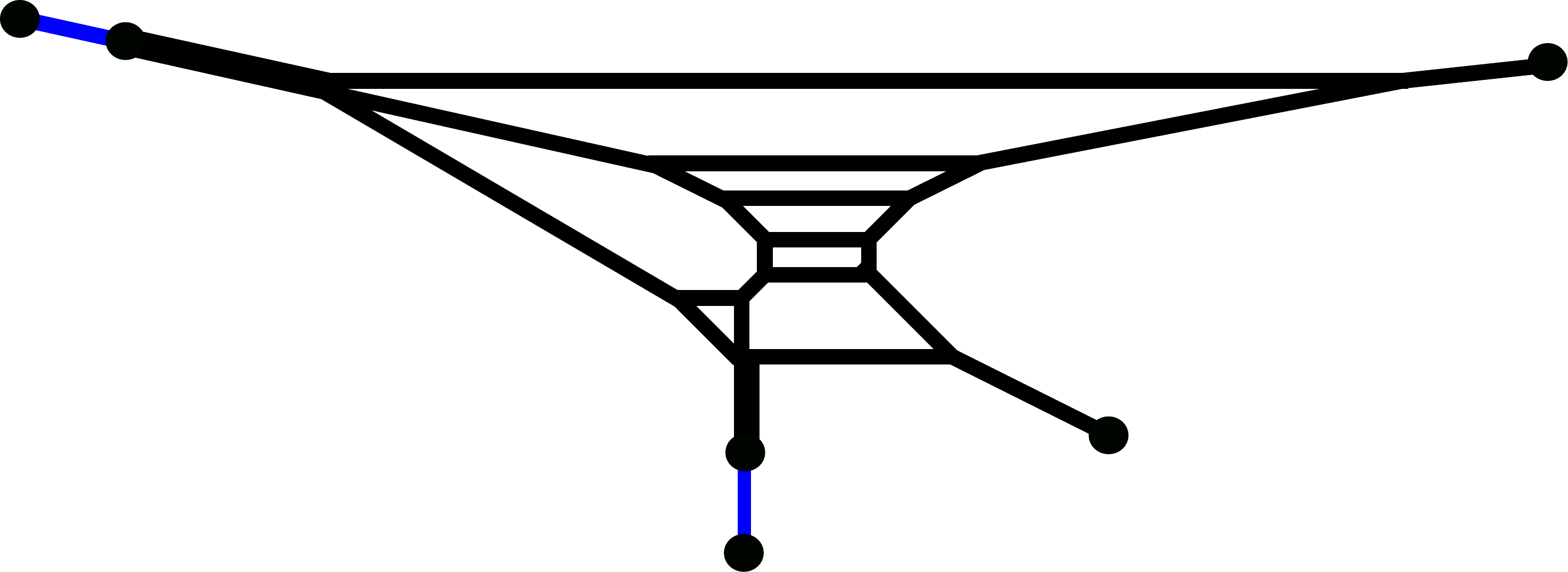}}
\caption{(a): A 5-brane diagram realizing the $SU(6)_{\frac{5}{2}}-SU(3)_0$ quiver theory. (b): The diagram obtained after shrinking red lines in Figure \ref{fig:su6su3}.}
\label{fig:su6su3quiver}
\end{figure}
\begin{figure}
\centering
\subfigure[]{\label{fig:su6su3su3}
\includegraphics[width=6.5cm]{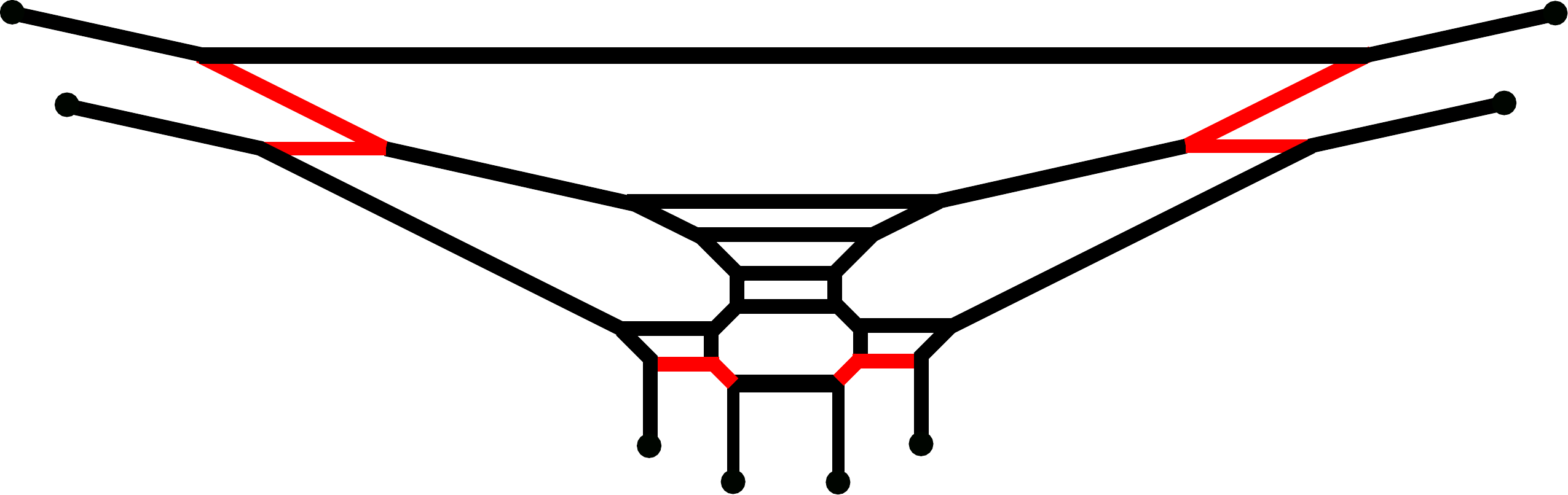}}\hspace{1cm}
\subfigure[]{\label{fig:su6su3su3v2}
\includegraphics[width=6.5cm]{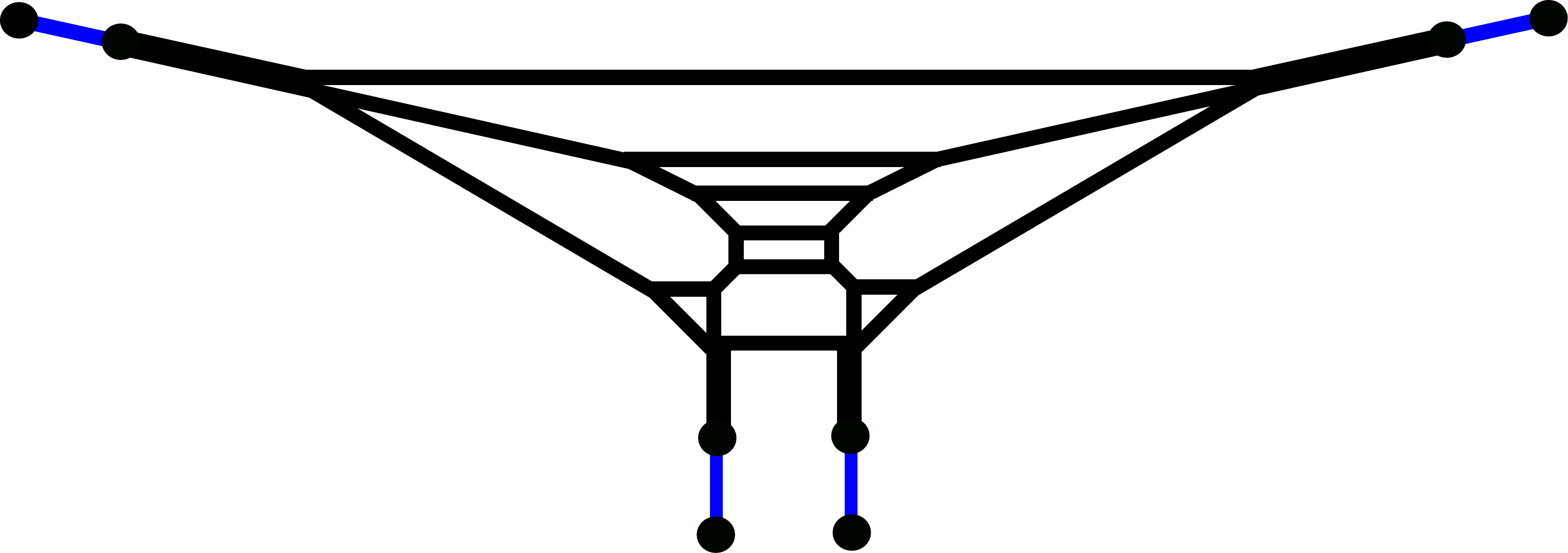}}
\caption{(a): A 5-brane diagram realizing the $SU(3)_0-SU(6)_{3}-SU(3)_0$ quiver theory. (b): The diagram obtained after shrinking red lines in Figure \ref{fig:su6su3su3}.}
\label{fig:su6su3su3quiver}
\end{figure}

We have constructed 5-brane webs for an $SU(6)$ gauge theory with one or two half-hypermultiplets in the rank-3 antisymmetric representation. It is natural to ask if we can add more half-hypermultiplets in the rank-3 antisymmetric representation. For that it is useful to take a different view for the diagrams in Figure \ref{fig:su6whtsa} and Figure \ref{fig:su6wtsa}. In fact, the diagram for the $SU(6)_{\frac{5}{2}}$\footnote{The subscript of the $SU$ gauge group represents the Chern-Simons level.} gauge theory with $N_{\bf TAS} = \frac{1}{2}$ in Figure \ref{fig:su6whtsa} may be also obtained from a Higgsing of a diagram for the $SU(6)_{\frac{5}{2}} -SU(3)_0$ quiver theory in Figure \ref{fig:su6su3}. In the diagram we introduced 7-branes ending on the external 5-branes to see the Higgsing explicitly. The Higgsing procedure can be done as follows. The $SU(6)_{\frac{5}{2}} - SU(3)_0$ quiver theory has an $SU(2)\times SU(2)$ flavor symmetry, which can be seen from parallel external 5-branes in the diagram in Figure \ref{fig:su6su3}, and we will give vevs to hypermultiplets associated to the flavor symmetry. For that we first set the length of 5-branes depicted as red lines in Figure \ref{fig:su6su3} to be zero. The resulting diagram is given by the one in Figure \ref{fig:su6su3v2}. Then giving the vevs corresponds to decoupling pieces of 5-branes in between 7-branes which are drawn as blue lines in Figure \ref{fig:su6su3v2}.
{Then moving the blue lines in Figure 
\ref{fig:su6su3v2}  
to the transverse  $(x^7, x^8, x^9)$ direction, which corresponds to Higgsing procedure\footnote{Unlike the usual Higgsing discussed e.g. in \cite{Hanany:1996ie}, we move $(2,-1)$ and $(0,1)$ 5-branes instead of D5 branes. Such type of non-perturbative Higgsing is not realized by giving vev to the hypermultiplets appearing in the Lagrangian. Instead, we expect that it corresponds to giving a vev to certain instanton operator.}, reduces to the diagram in Figure \ref{fig:su6whtsa} 
which yields a 5-brane realization of the $SU(6)$ gauge theory with half hyper in the rank-3 antisymmetric representation, as we discussed in the previous subsections.}
 After this procedure, coincident 5-branes end on the same 7-brane and then the diagram is equivalent to the one in Figure \ref{fig:su6whtsa}. The Higgsing from the $SU(3)_0-SU(6)_3-SU(3)_0$ quiver theory to $SU(6)_3$ gauge theory with $N_{\bf TAS} = 1$ may be obtained in  similar way. We first shrink the length of the red lines in a diagram for the $SU(3)_0 - SU(6)_3 - SU(3)_0$ quiver theory in Figure \ref{fig:su6su3su3}, which gives rise to the diagram in Figure \ref{fig:su6su3su3v2}. 
%
%
Then decoupling the blue lines in Figure
 \ref{fig:su6su3su3v2} 
reduces to the diagram for the $SU(6)_3$ gauge theory with $N_{\bf TAS} = 1$ in Figure \ref{fig:su6wtsa}. 

To summarize, the Higgsing of the $SU(6)_{\frac{5}{2}} - SU(3)_0$ quiver theory yields the $SU(6)_{\frac{5}{2}}$ gauge theory with $N_{\bf TAS} = \frac{1}{2}$ and also the Higgsing of the $SU(3)_0 - SU(6)_{3} - SU(3)_0$ quiver theory gives the $SU(6)_{3}$ gauge theory with $N_{\bf TAS} = 1$. Namely, a Higgsing of one $SU(3)_0$ coupled to an $SU(6)$ gauge node introduces a half-hypermultiplet in the rank-3 antisymmetric representation and it does not change the Chern-Simons level of the $SU(6)$ gauge theory,
\begin{align}
[SU(6)_{\kappa}] - SU(3)_0\quad \xrightarrow{\text{Higgsing}} \quad [SU(6)_{\kappa}] - [1/2\,{\bf TAS}].
\label{HiggstoTAS}
\end{align}
Hence we can add rank-3 antisymmetric matter by coupling more $SU(3)_0$ gauge nodes and then Higgsing them. For the original theory to be UV complete, one can consider two more types of quiver theories which are given by $SU(6)_{\frac{1}{2}} - \left[SU(3)_0\right]^3$\footnote{In this case, we need half-integer Chern-Simons level for the $SU(6)$ since nine fundamental hypermultiplets are effectively coupled to the $SU(6)$ gauge node.} or $SU(6)_0 - \left[SU(3)_0\right]^4$. The former one is an $D_4$ quiver theory and the latter one is an affine $D_4$ quiver theory. The Higgsing of the $D_4$ quiver theory will yields the $SU(6)_{\frac{1}{2}}$ gauge theory with $N_{\bf TAS} = \frac{3}{2}$ and the Higgsing of the affine $D_4$ quiver theory will give the $SU(6)_0$ gauge theory with $N_{\bf TAS} = 2$ which is supposed to have a 6d UV completion \cite{Jefferson:2017ahm}. The latter Higgsing realizes a renormalization group flow from an affine $D_4$ Dynkin quiver theory which has a 6d UV completion to another 5d theory which also has a 6d UV completion.

\begin{figure}
\centering
\subfigure[]{\label{fig:su6su3cube}
\includegraphics[width=7.5cm]{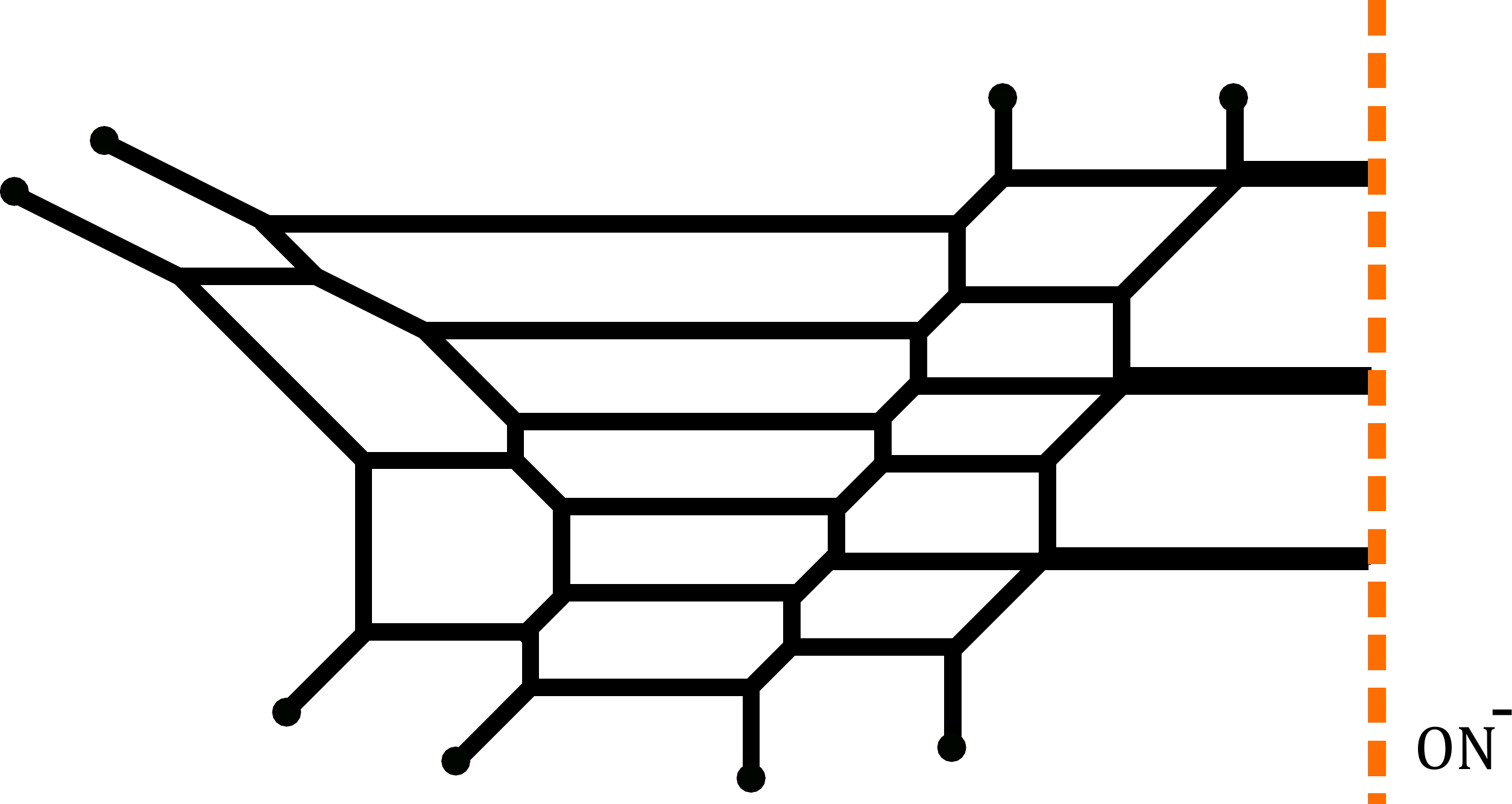}}\hspace{1cm}
\subfigure[]{\label{fig:su6w3htsa}
\includegraphics[width=5.5cm]{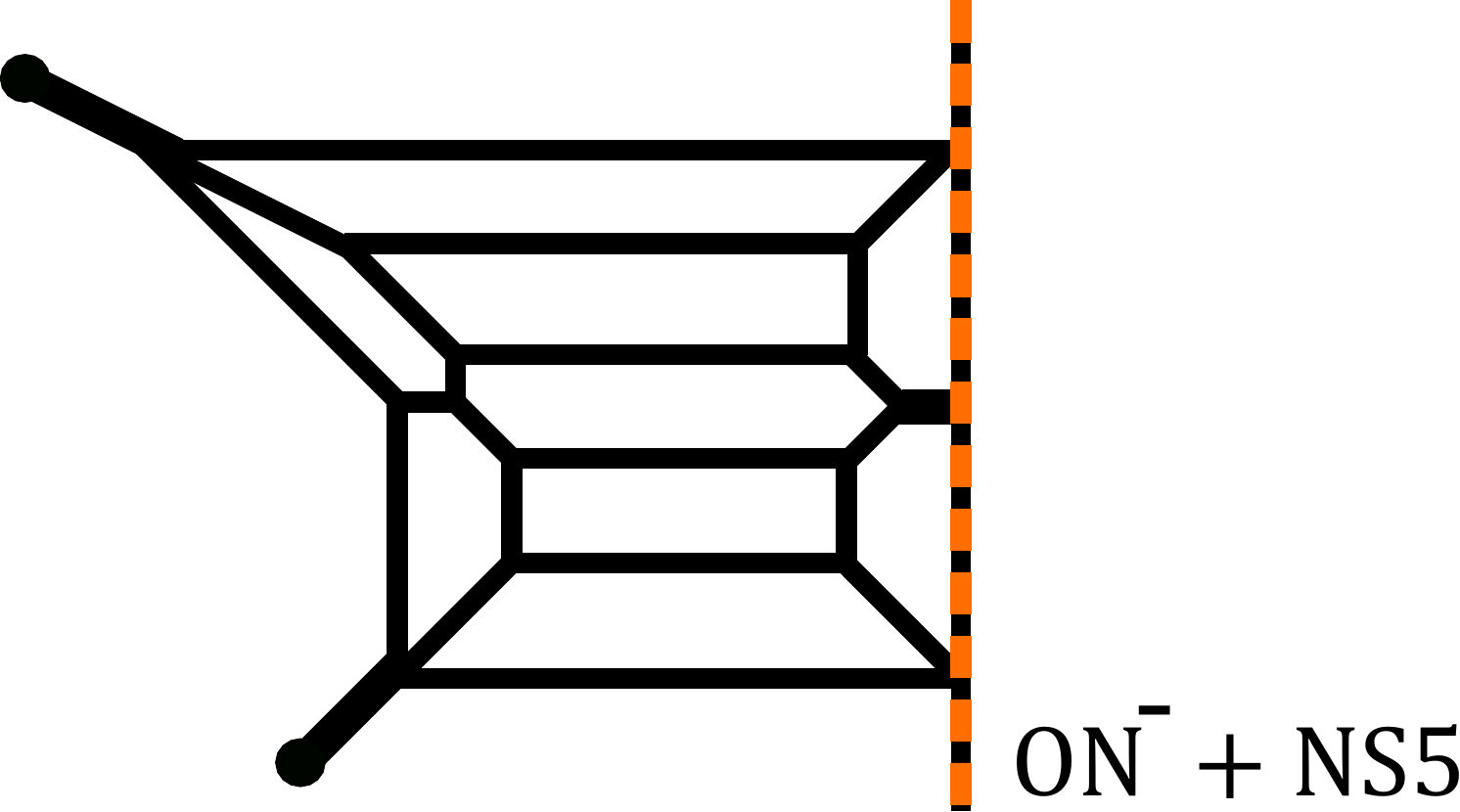}}
\caption{(a): A 5-brane diagram for the $SU(6)_{\frac{1}{2}} - \left[SU(3)_0\right]^3$ quiver theory. (b): A 5-brane web for the $SU(6)_{\frac{1}{2}}$ gauge theory with $N_{\bf TAS} = \frac{3}{2}$ which is obtained from a Higgsing from the diagram in Figure 
\ref{fig:su6su3cube}. }
\label{fig:D4totsa}
\end{figure}
We here make use of the Higgsing procedure to construct 5-branes webs for an $SU(6)$ gauge theory with $N_{\bf TAS} = \frac{3}{2}, 2$. We first start from the construction of a web for the $SU(6)$ gauge theory with $N_{\bf TSA} = \frac{3}{2}$ and $\kappa = \frac{1}{2}$ which will be obtained from a Higgsing of the $SU(6)_{\frac{1}{2}} - \left[SU(3)_0\right]^3$ theory. A 5-brane diagram for the $SU(6)_{\frac{1}{2}} - \left[SU(3)_0\right]^3$ may be realized by introducing an ON-plane \cite{Kutasov:1995te, Sen:1996na, Kapustin:1998fa, Hanany:1999sj} and it is depicted in Figure \ref{fig:su6su3cube}. One of the $SU(3)_0$ gauge nodes in Figure \ref{fig:su6su3cube} is given by the left part of the diagram in Figure \ref{fig:su6su3cube} and two of the $SU(3)_0$ gauge nodes are realized by the right part of the diagram in Figure \ref{fig:su6su3cube} using an ON-plane. We can also see an $SU(2) \times SU(2)$ flavor symmetry from the left part and also an $SO(4) \times SO(4) \cong SU(2)^4$ flavor symmetry from the right part. Then we apply the same Higgsing procedure in Figure \ref{fig:su6su3quiver} to the diagram in Figure \ref{fig:su6su3cube}. The Higgsing associated to the flavor symmetry yields the diagram in Figure \ref{fig:su6w3htsa}. We claim that the diagram in Figure \ref{fig:su6w3htsa} gives rise to the $SU(6)$ gauge theory with three half-hypermultiplets in the rank-3 antisymmetric representation and $\kappa = \frac{1}{2}$. 

\begin{figure}
\centering
\subfigure[]{\label{fig:affineD4}
\includegraphics[width=8cm]{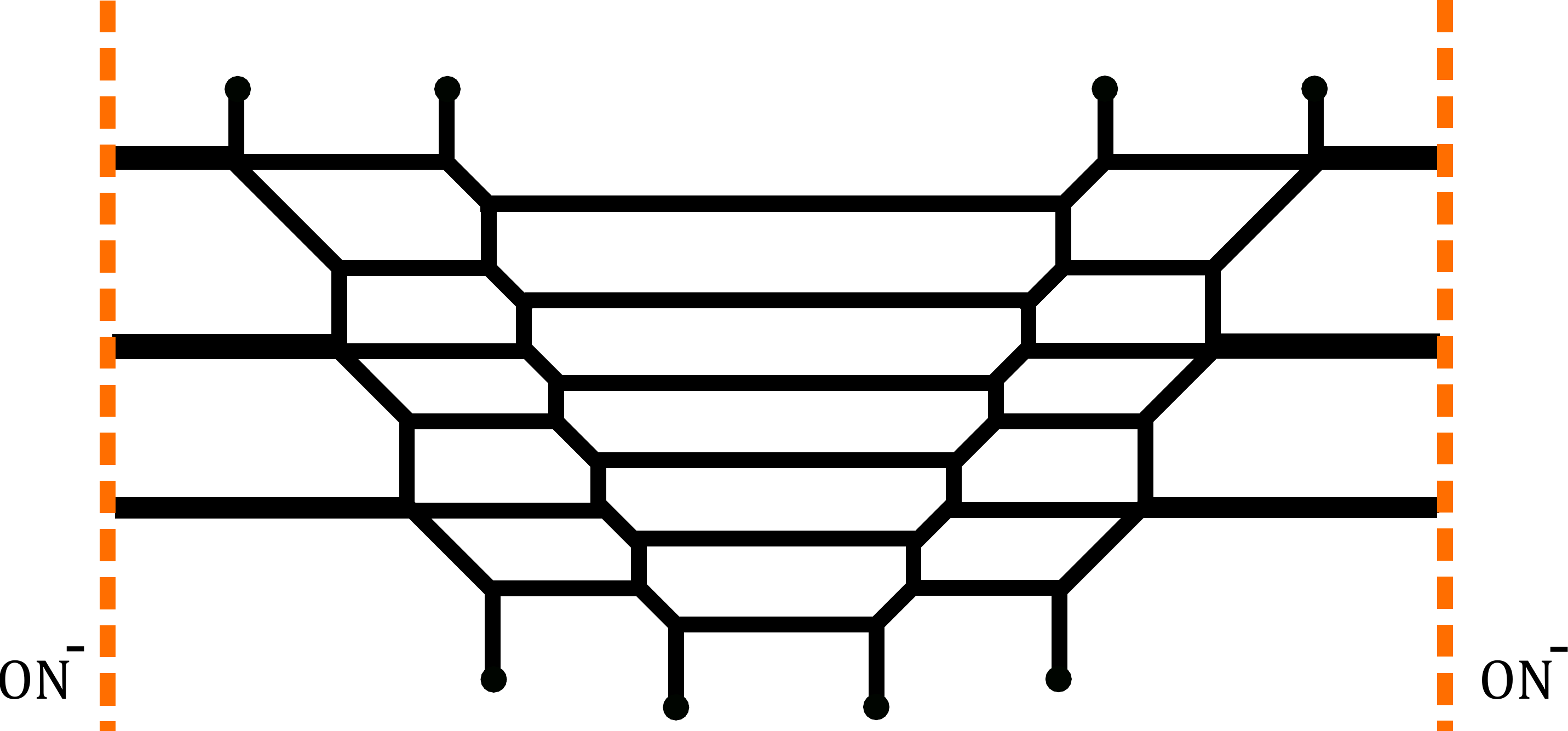}}\hspace{1cm}
\subfigure[]{\label{fig:su6w4htsa}
\includegraphics[width=5cm]{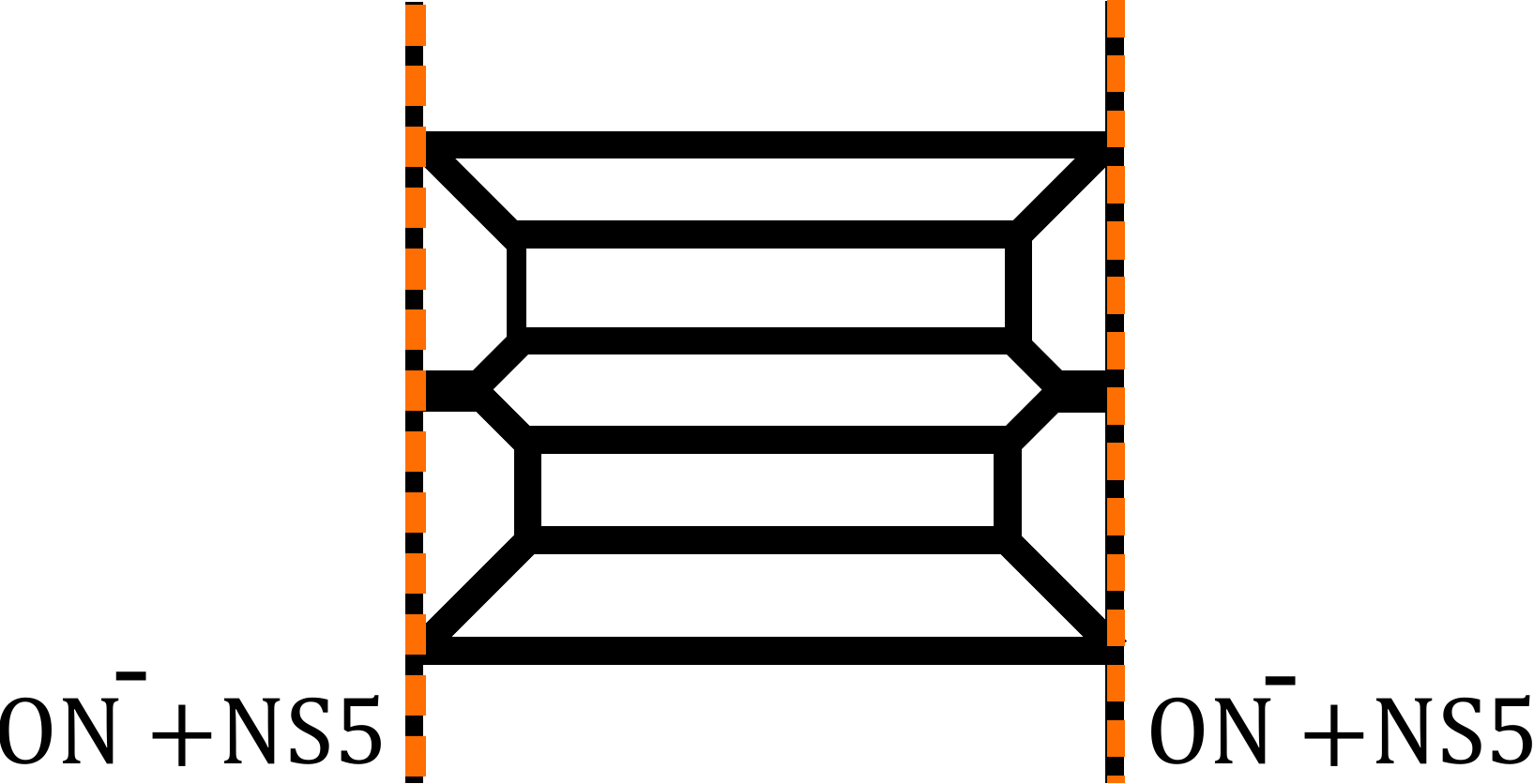}}
\caption{(a): A 5-brane diagram for the $SU(6)_{0} - \left[SU(3)_0\right]^4$ affine $D_4$ quiver theory. (b): A 5-brane web for the $SU(6)_0$ gauge theory with $N_{\bf TAS} = 2$ which is obtained from a Higgsing from the diagram in Figure \ref{fig:affineD4}. }
\label{fig:affinD4totsa}
\end{figure}
In order to obtain a diagram for the $SU(6)_0$ gauge theory with $N_{\bf TAS} = 2$, we start from a diagram for the $SU(6)_0-\left[SU(3)_0\right]^4$ affine $D_4$ quiver theory in Figure \ref{fig:affineD4}. The two ON-planes in Figure \ref{fig:affineD4} realizes the four $SU(3)_0$ gauge nodes coupled to the middle $SU(6)$ gauge theory, Applying the Higgsing done in Figure \ref{fig:su6su3quiver} to the both sides of the diagram in Figure \ref{fig:affineD4} gives rise to the diagram in Figure \ref{fig:su6w4htsa}. We argue that the diagram in Figure \ref{fig:su6w4htsa} realizes the $SU(6)$ gauge theory with two massless hypermultiplets in the rank-3 antisymmetric representation with zero Chern-Simons level. After performing S-duality to the diagram in Figure \ref{fig:su6w4htsa}, which is equivalent to rotating the diagram by $90$ degrees, the diagram contains two O5-planes on the upper side and the lower side. The two O5-planes implies a periodic direction in the vertical direction, suggesting a 6d UV completion.

\begin{figure}[t]
\centering
\subfigure[]{\label{fig:su6w3htsacb}
\includegraphics[width=6.5cm]{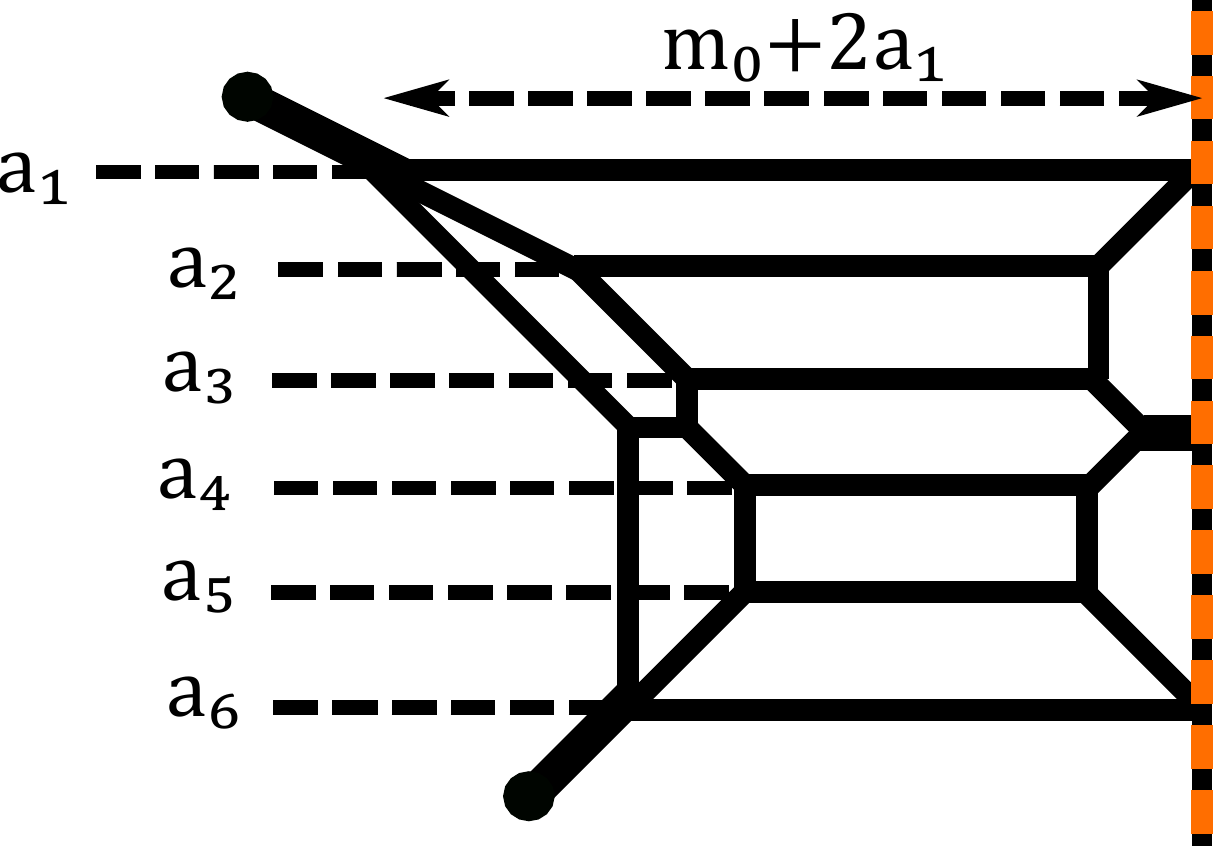}}\hspace{1cm}
\subfigure[]{\label{fig:su6w3htsaarea}
\includegraphics[width=6.5cm]{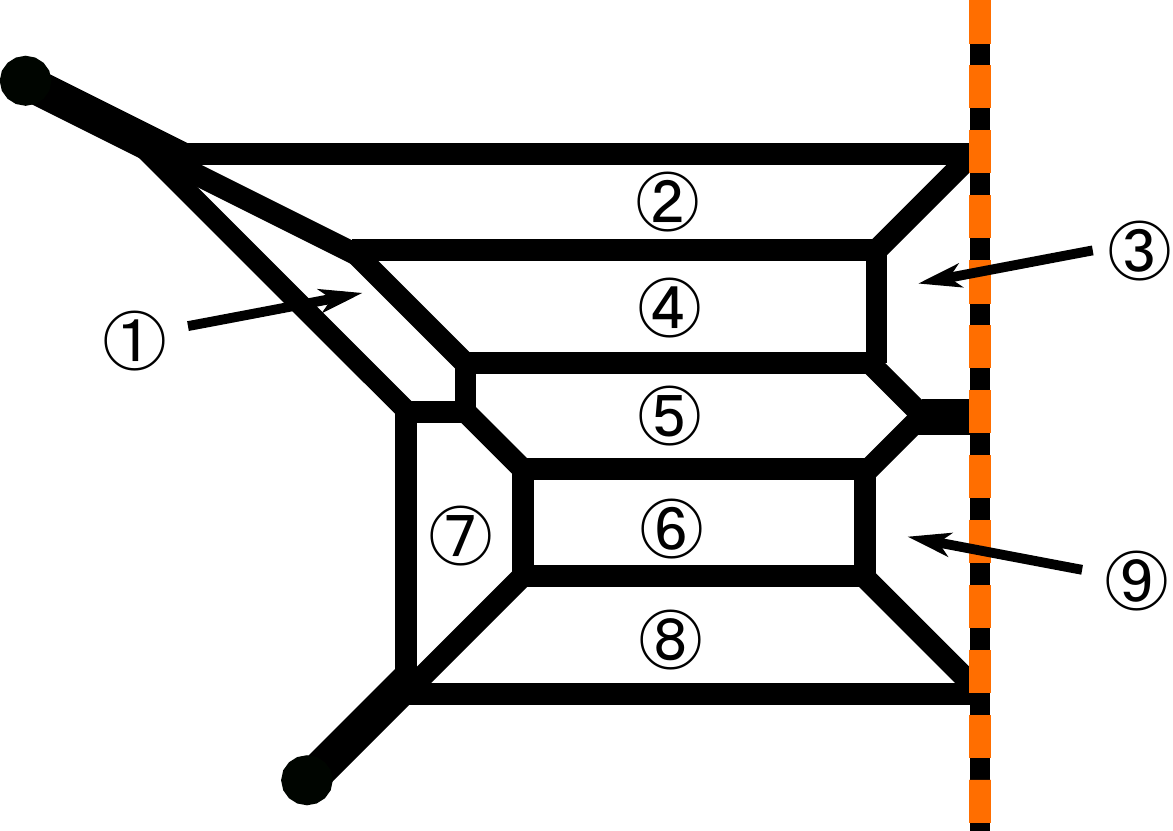}}
\caption{(a): A parameterization of Coulomb branch moduli for the diagram in Figure \ref{fig:su6w3htsa}. (b): A labeling for the area of faces in the diagram in Figure \ref{fig:su6w3htsa}.}
\label{fig:su6w3htsa1}
\end{figure}
We can confirm the claim by comparing the area with the monopole string tension as done in section \ref{sec:su6wtsatension}. In order to compute the area of the faces in the diagram in Figure \ref{fig:su6w3htsa}, we label the height of the six color D5-branes as $a_1, a_2, a_3, a_4, a_5, a_6$ with $\sum_{i=1}^6a_i=0$, which are the Coulomb brach moduli of the $SU(6)_{\frac{1}{2}}$ gauge theory with $N_{\bf TAS} = \frac{3}{2}$. The inverse of the squared gauge coupling $m_0$ is determined by the non-trivial length of the D5-brane after turning off the Coulomb branch moduli. Then the length of the top color D5-brane in Figure \ref{fig:su6w3htsa} is $m_0 + 2a_1$. The parameterization is summarized in Figure \ref{fig:su6w3htsacb}. With this parameterization we can compute the area of the faces of the diagram in Figure \ref{fig:su6w3htsa}. A labeling of the faces is given in Figure \ref{fig:su6w3htsaarea}. As in the case of the diagram in Figure \ref{fig:su6whtsaarea}, some of them are connected to each other. In fact, $\textcircled{\scriptsize 1}, \textcircled{\scriptsize 2}$ and  $\textcircled{\scriptsize 3}$ are a single face and the area of the region $\textcircled{\scriptsize 3}$ should be doubled due to the presence of the ON$^-$-plane \cite{Hayashi:2018bkd, Hayashi:2018lyv}. Simliarly, the region$\textcircled{\scriptsize 7},  \textcircled{\scriptsize 8}$ and $\textcircled{\scriptsize 9}$ are connected to each other and the area of the region $\textcircled{\scriptsize 9}$ needs to be doubled. Then we have in total five faces and the corresponding area is given by
\begin{align}
\textcircled{\scriptsize 1} + \textcircled{\scriptsize 2} + 2 \textcircled{\scriptsize 3}&= m_0(2\phi_1 - \phi_2) + \frac{5}{2}\phi_1^2 + 7\phi_1\phi_2 - \frac{9}{2}\phi_2^2+3\phi_2\phi_3 - \frac{3}{2}(\phi_3 - \phi_5)^2 - 3\phi_1(\phi_3 + \phi_5),\label{su6w3htsaarea1}\\
\textcircled{\scriptsize 4} &= \frac{1}{2}(m_0 - 7\phi_1 + 4\phi_2 + 2\phi_3)(-\phi_1 + 2\phi_2 - \phi_3),\\
\textcircled{\scriptsize 5} &= m_0(\phi_2 - 2\phi_3 + \phi_4)\nn\\
& +\frac{1}{2}(-3\phi_1^2 -\phi_2^2 - 2\phi_2\phi_3 + 5\phi_3^2 - 2\phi_4^2 -6\phi_3\phi_5 + 6\phi_4\phi_5 -3\phi_5^2 + 6\phi_1(\phi_2 - \phi_3 + \phi_5)),\\
\textcircled{\scriptsize 6} &=(m_0 + 2\phi_4 - 3\phi_5)(-\phi_3 + 2\phi_4 - \phi_5),\\
\textcircled{\scriptsize 7} + \textcircled{\scriptsize 8} + 2 \textcircled{\scriptsize 9}  &=m_0(-\phi_4 + 2\phi_5)\nn\\
&+ \frac{1}{2}\left(-3\phi_1^2 - 3\phi_3^2 - 8\phi_4^2 + 6\phi_1(\phi_3-\phi_5) + 6\phi_3(\phi_4 - \phi_5) + 12\phi_4\phi_5 + 5\phi_5^2\right).\label{su6w3htsaarea5}
\end{align}

We then compare the area \eqref{su6w3htsaarea1}-\eqref{su6w3htsaarea5} with the monopole string tension calculated, using the effective prepotential \eqref{prepotential}. The parameterization in Figure \ref{fig:su6w3htsacb} fixes the phase of the $SU(3)_{\frac{1}{2}}$ gauge theory with $N_{\bf TAS} = \frac{3}{2}$. Positive roots are $e_i - e_j, (1 \leq i < j \leq 6)$ and positive weights are $e_1 + e_i + e_j, (2 \leq i < j \leq 5)$ and $e_1 + e_i + e_6, (i=2, 3)$, $-e_1 -e_i - e_6, (i=4, 5)$. Then the effective prepotential \eqref{prepotential} is given by
\begin{align}
\mathcal{F}^{SU(6)_\frac{1}{2}}_{N_{{\bf TAS}}=\frac{3}{2}} = &\;\frac{1}{2}m_0\sum_{i=1}^6a_i^2 + \frac{1}{12}\sum_{i=1}^5a_i^3 + \frac{1}{6}\sum_{1 \leq i < j \leq 6}(a_i - a_j)^3\nn\\
& - \frac{3}{12}\left(\sum_{2 \leq i < j  \leq 5}(a_1 + a_i + a_j)^3 - \sum_{i=2, 3}\left\{(a_1 + a_i + a_6)^3 - (a_1 + a_{i+2} + a_6)^3\right\}\right), \label{pre.SU6w3hTSA}
\end{align}
After rewriting \eqref{pre.SU6w3hTSA} in terms of the Coulomb branch moduli $\phi_i, (i=1, \cdots, 5)$ in \eqref{CB.SU6whTSA}, taking the derivative of \eqref{pre.SU6w3hTSA} with respect to the $\phi_i$ gives the monopole string tension. Indeed we found that 
\begin{align}
&\frac{\partial \mathcal{F}^{SU(6)_{\frac{1}{2}}}_{N_{{\bf TAS}} =\frac{3}{2}}}{\partial \phi_1} = \textcircled{\scriptsize 1} + \textcircled{\scriptsize 2} + 2 \textcircled{\scriptsize 3}, \qquad 
\frac{\partial \mathcal{F}^{SU(6)_{\frac{1}{2}}}_{N_{{\bf TAS}} =\frac{3}{2}}}{\partial \phi_2} = \textcircled{\scriptsize 4},\quad
\frac{\partial \mathcal{F}^{SU(6)_{\frac{1}{2}}}_{N_{{\bf TAS}} =\frac{3}{2}}}{\partial \phi_3} = \textcircled{\scriptsize 5},\nn\\
&\frac{\partial \mathcal{F}^{SU(6)_{\frac{1}{2}}}_{N_{{\bf TAS}} =\frac{3}{2}}}{\partial \phi_4} = \textcircled{\scriptsize 6},\quad
\frac{\partial \mathcal{F}^{SU(6)_{\frac{1}{2}}}_{N_{{\bf TAS}} =\frac{3}{2}}}{\partial \phi_5} = \textcircled{\scriptsize 7} + \textcircled{\scriptsize 8} + 2 \textcircled{\scriptsize 9},
\end{align}
which supports the claim that the diagram in Figure \ref{fig:su6w3htsa} yields the $SU(6)$ gauge theory with three half-hypermultiplets in the rank-3 antisymmetric representation and the Chern-Simons level $\kappa = \frac{1}{2}$. 

Extending the comparison to the case of the $SU(6)_0$ gauge theory with $N_{\bf TAS} = 2$ realized in the diagram in Figure \ref{fig:su6w4htsa} is straightforward. We checked that the area of the faces in the diagram in Figure \ref{fig:su6w4htsa} reproduces the monopole string tension calculated from the effective prepotential of the $SU(6)$ gauge theory with two massless hypermultiplets in the rank-3 antisymmetric representation and zero Chern-Simons level.

\bigskip

\section{Marginal $SU(6)$ gauge theory with rank-3 
antisymmetric matter}\label{sec:SU6}
In this section, we provide more 5-brane diagrams for $SU(6)$ gauge theories with rank-3 antisymmetric matter by including hypermultiplets in other representations. 
In particular we present 5-brane web diagrams for $SU(6)$ marginal theories 
with half-hypermultiplets in the rank-3 antisymmetric representation and other matter which have the UV completion as a 6d theory. Possible $SU(6)$ marginal theories with rank-3 antisymmetric hypermultiplets and other hypermultiplets are classified in \cite{Jefferson:2017ahm} based on the Coulomb branch analysis. Though we do not find all the marginal $SU(6)$ marginal theories with rank-3 antisymmetric matter listed in \cite{Jefferson:2017ahm}, 5-brane webs for the marginal theories that we found precisely agree with the matter content 
and also show the periodic structure which supports that the theories can be understood as 6d theory on a circle with or without a twist \cite{Kim:2015jba,Hayashi:2015fsa,Hayashi:2015zka,Hayashi:2015vhy}. 

Moreover, one can put the 7-branes appearing in our 5-brane webs into the 5-brane loops, and then from which one can read off a global symmetry of the theory. Such characterization of a global symmetry is only possible for symmetry group of ADE type \cite{Gaberdiel:1997ud, Gaberdiel:1998mv, DeWolfe:1999hj}. For instance, given a 7-brane configuration where one allocates 7-branes into a 5-brane loop, the corresponding (non-abelian part of) global symmetry is read off from the Kodaira classification
\begin{align}
	A_m: \textsf{\textbf{A}}^{m+1},\qquad D_{m\ge 4}: \textsf{\textbf{A}}^{m}\textsf{\textbf{BC}},\qquad E_{m\ge 6}: \textsf{\textbf{A}}^{m-1}\textsf{\textbf{BCC}},
\end{align}
where the following shorthand notation is used to denote the 7-brane charges
\begin{align}
	\textsf{\textbf{ A}} = (1,0),\qquad \textsf{\textbf{B}}=(1,-1), \qquad \textsf{\textbf{C}}=(1,1).
\end{align} 
For other types of global symmetry, one may infer it from possible maximal subgroups of ADE type via various Hanany-Witten transitions on a given 5-brane web. We remark that in the way, we perform 7-brane monodromy analysis for those 5-brane webs which do not have orientifolds to find global symmetries for the marginal theories, and we see that the obtained global symmetries are consistent with those given in \cite{Jefferson:2017ahm}. 

In Table \ref{Tab:su6marginal}, we summarize marginal 5-brane web diagrams that we obtained. We note that as decoupling of hypermultiplets from the marginal theories, one can also perform decoupling of hypermultiplets on 5-brane webs, as discussed in \cite{Hayashi:2018lyv}, which  would give rise to various 5-brane webs for other genuine 5d SCFTs. 

\begin{table}
\begin{center}
\begin{tabular}{ |c|c|c|c|c|c| } 
 \hline
 $N_{\bf TAS}$ & $N_{\bf Sym}$ & $N_{\bf AS}$& $N_{\bf F}$ &
 CS&5-brane web \\ 
 \hline \hline
  2 & . & .	& . & 0 & Figure \ref{fig:affinD4totsa}
\\ \hline
3/2 & . & . & 5 & 0 & ?\\ \hline
3/2 & . & . & 3 & 2 & ?\\ \hline
3/2 & . & . & . & 9/2 & ?\\ \hline
  1 & . & 1 & 4 & 0 & Figure \ref{fig:SU61AS4F}\\ \hline
  1 & . & 1 & 3 & 3/2 & ?\\ \hline
  1 & . & 1 & . & 4 & ? \\ \hline
  1 & . & . & 10 & 0 & Figure \ref{fig:SU6+1TAS+10F}\\ \hline
  1 & . & . & 9 & 3/2 & Figure \ref{fig:SU6+1TAS+9F}\\ \hline
1/2 & 1 & . & 1 & 0 & Figure \ref{fig:SU6+1over2TAS+1Sym+1F-1}\\ \hline
1/2 & 1 & . & . & 3/2 & Figure \ref{fig:SU6+1over2TAS+1Sym-1}\\ \hline
1/2 & . & 2 & 2 & 3/2 & ?\\ \hline
1/2 & . & 2 & 2 & 1/2 & ?\\ \hline
1/2 & . & 2 & . & 7/2 & ?\\ \hline
1/2 & . & 1 & 9 & 0 & Figure \ref{fig:SU6+1over2TAS+1AS+9F}\\ \hline
1/2 & . & 1 & 8 & 3/2 & Figure \ref{fig:SU6+1over2TAS+1AS+8F}\\ \hline
1/2 & . & . & 13 & 0 & Figure \ref{fig:SU6+1over2TAS+13F}\\ \hline
1/2 & . & . & 9 & 3 & Figure \ref{fig:SU6+1over2TAS+9F}\\ \hline
\end{tabular}
\caption{Table for $SU(6)$ marginal theories with rank-3 antisymmetric matter and other matter. $N_{\rm\bf TAS}$ denotes the number of  hypermultiplet in the rank-3 antisymmetric representation, $N_{\rm \bf Sym}$ the number of hypermultiplet in the symmetric representation,  $N_{\rm \bf AS}$ the number of hypermultiplet in the antisymmetric representation, $N_{\rm\bf  F}$ the number of hypermultiplet in the fundamental representation, and CS the Chern-Simons level.}
\label{Tab:su6marginal}
\end{center}
\end{table}

\subsection{5-brane web for $SU(6)_0+2\,\mathbf{TAS}$}
The maximum number of the hypermultiplet in the rank-3 antisymmetric hypermultiplet is two, which is itself marginal. The 5-brane web for $SU(6)_0$ theory with two rank-3 antisymmetric hypermultiplets was already discussed in section \ref{sec:su6+2TAS}, and the corresponding web diagram is given in Figure \ref{fig:affinD4totsa}.

\subsection{5-brane webs for $SU(6)+1\,\mathbf{TAS}$ with various hypermultiplets}
\label{sec:SU6w1TAS}
\begin{figure}
\centering
\includegraphics[width=13cm]{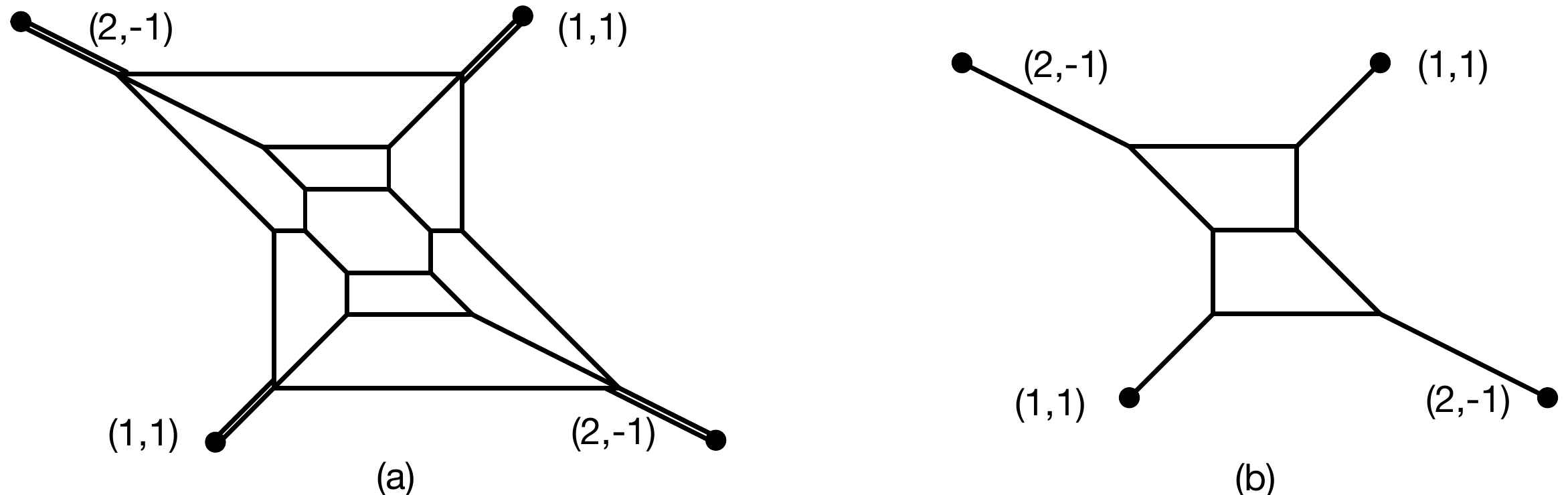} 
\caption{(a) A 5-brane configuration for  $SU(6)_{0}+1{\bf TAS}$ and asymptotic charges of external 5-branes. (b) A 5-brane configuration for the pure $SU(3)_{0}$ gauge theory. Asymptotic charges of external 5-branes/7-branes for both 5-brane web diagrams are the same.}
\label{fig:SU3}
\end{figure}

Following section \ref{sec:SU6TSA}, it is straightforward to get a 5-brane web diagram for an $SU(6)$ gauge theory with one rank-3 antisymmetric hypermultiplet. For instance, in Figure \ref{fig:su6whtsa1}, we presented a 5-brane web for  the $SU(6)$ gauge theory with one rank-3 antisymmetric hypermultiplet, which has the Chern-Simons level $\kappa=3$. It is then possible to express a 5-brane web for the $SU(6)$ theory with one rank-3 antisymmetric hypermultiplet which is of the Chern-Simons level $\kappa=0$ by suitably choosing the asymptotic $(p, q)$ charges for the external 5-branes, as depicted in Figure \ref{fig:SU3}(a).
Notice that asymptotic 5-brane charges for this $SU(6)_0$ gauge theory with one rank-3 antisymmetric hypermultiplet is the same as those for the pure $SU(3)_0$ gauge theory. See Figure \ref{fig:SU3}. With this observation in mind, one can easily construct 5-brane configurations for some of the marginal $SU(6)$ gauge theories with one rank-3 antisymmetric hypermultiplet by referring to the 5-brane construction for the corresponding marginal $SU(3)$ gauge theories with proper numbers of flavors.
\begin{figure}
\centering
\includegraphics[width=13cm]{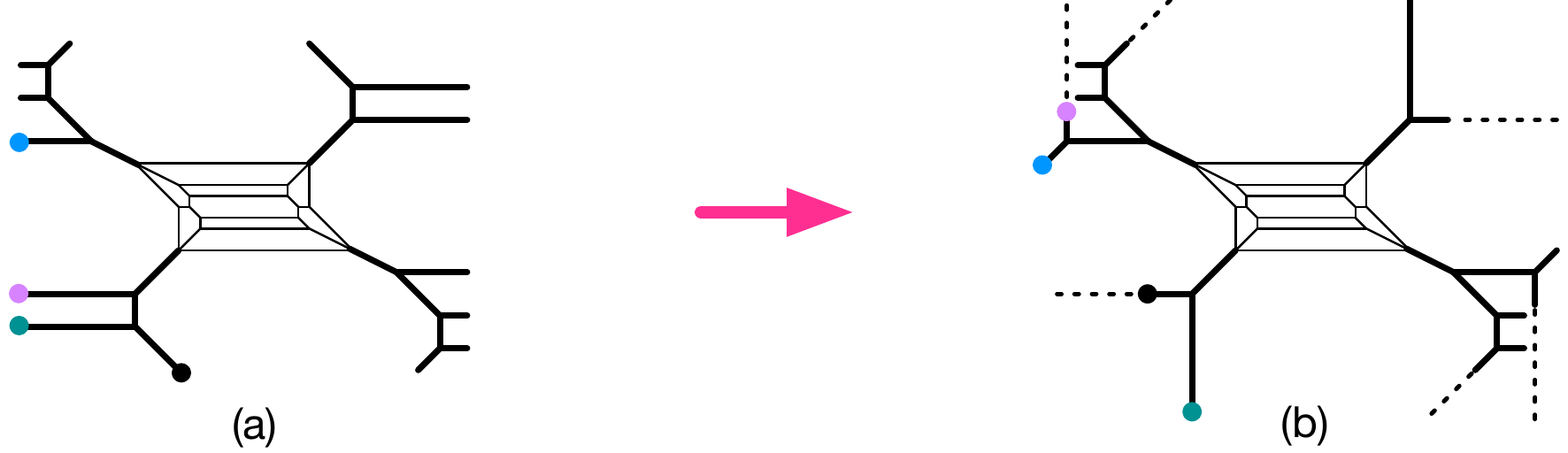} 
\caption{(a) A 5-brane configuration for $SU(6)_{0}+1{\bf TAS}+10\bF$. (b)  $SU(6)_{0}+1{\bf TAS}+10\bF$ Tao diagram. Here the thick line means two coincident 5-branes and the dotted lines are the monodromy cuts. 7-brane motions from (a) to (b) are explained in \cite{Hayashi:2016abm}.}
\label{fig:SU6+1TAS+10F}
\end{figure}
\paragraph{\underline{$SU(6)_0+1{\bf TAS}+10\bF$}.}
A 5-brane web for $SU(3)_0+10\bF$ theory is constructed in \cite{Hayashi:2015fsa}, and then it follows that a 5-brane web for the $SU(6)_0+1{\bf TAS}+10\bF$ theory can be constructed by adding 10 flavors in the same way as done for 5-brane web for the $SU(3)_0+10\bF$ theory. The resulting 5-brane web is depicted in Figure \ref{fig:SU6+1TAS+10F}. We note that as shown in Figure \ref{fig:SU6+1TAS+10F}(b), 
the 5-brane web of the $SU(6)_0+1{\bf TAS}+10\bF$ theory has an infinite repeated periodic structure, named Tao web diagrams \cite{Kim:2015jba}, which is expected as the 5-brane web for the $SU(3)_0+10\bF$ theory also has the periodic structure \cite{Hayashi:2015fsa}. The period 
of a Tao web diagram is expressed in terms of the coupling and mass parameters, which corresponds to the inverse 
of the compactification radius. Hence such Tao web diagrams imply that these 5d theories are realized as a 5-brane for a Kaluza-Klein (KK) theory where a 6d theory is compactified on a circle associated with the period on the Tao web diagram. We will discuss the 6d origin of the $SU(6)_0+1{\bf TAS}+10\bF$ theory later in section \ref{sec:SU6w1TAS10F}.

\begin{figure}
\includegraphics[width=15cm]{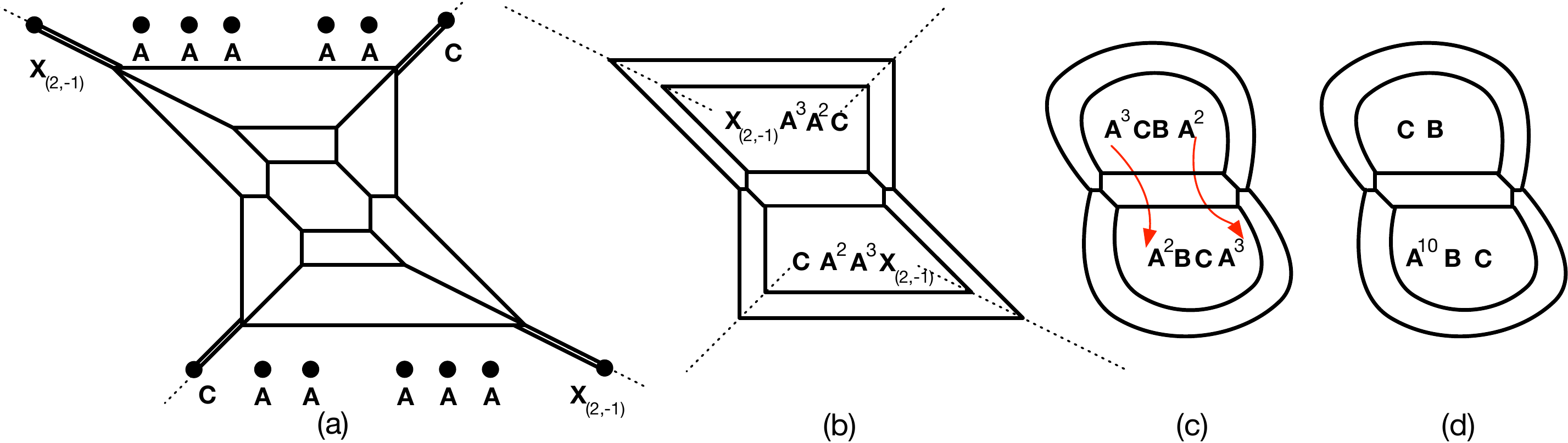} 
\caption{(a) 7-brane configurations for 5-brane web in Figure \ref{fig:SU6+1TAS+10F}, where $\textsf{\textbf{A}} = (1,0)$ 7-brane, $\textsf{\textbf{C}}=(1,1)$ 7-brane, and $\textsf{\textbf{X}}_{(2,-1)}=(2,-1)$ 7-brane. (b) 7-brane configuration where all the 7-branes are put in 5-brane loops. (c) Rearrangement of 7-branes. (d) 7-brane configurations showing an $\textsf{\textbf{A}}^{10}\textsf{\textbf{BC}}=SO(20)$ symmetry, after allocating five $\textsf{\textbf{A}}$'s to the lower 5-brane loops.}
\label{fig:SU610Fsymm}
\end{figure}

The enhanced global symmetry of the $SU(6)_0+1{\bf TAS}+10\bF$ theory can be read off from 7-brane monodromy analysis as shown in Figure \ref{fig:SU610Fsymm}. Starting from the 5-brane web for the $SU(6)_0+1{\bf TAS}+10\bF$ theory given in Figure \ref{fig:SU6+1TAS+10F}, one puts flavor D7-branes $\textsf{\textbf{A}}$'s together as in Figure \ref{fig:SU610Fsymm}(a). As D7-branes can cross D5-branes, one can put all the D7-branes inside 5-brane loops, which gives 7-brane configuration given in Figure \ref{fig:SU610Fsymm}(b). Using 7-brane monodromy analysis (counterclockwise) like
\begin{align}
\textsf{\textbf{A}}\textsf{\textbf{X}}_{(p,q)}=	\textsf{\textbf{X}}_{(p+q,q)}\textsf{\textbf{A}} \qquad \Longrightarrow \qquad
\textsf{\textbf{A}}^3\textsf{\textbf{X}}_{(2,-1)}=	\textsf{\textbf{C}}\textsf{\textbf{A}}^3,\qquad
	\textsf{\textbf{CA}}^2= 
	\textsf{\textbf{ A}}^2\textsf{\textbf{B}},
	\label{eq:7Bmonodromy}
\end{align} 
one can relocate the 7-branes to obtain the configuration in Figure \ref{fig:SU610Fsymm}(c), which leads to the 7-brane configuration yielding an  $SO(20)$ symmetry as $D_{10}= \textsf{\textbf{A}}^{10}\textsf{\textbf{BC}}$ as shown in Figure \ref{fig:SU610Fsymm}(d). This agrees with the propsed global symmetry in \cite{Jefferson:2017ahm}. It is in fact the same global symmetry structure as that for the $SU(3)_0+10\bF$ theory \cite{Hayashi:2015fsa}, which is expected as their asymptotic 7-brane configurations are identical. 
%

\begin{figure}
\centering
\includegraphics[width=14cm]{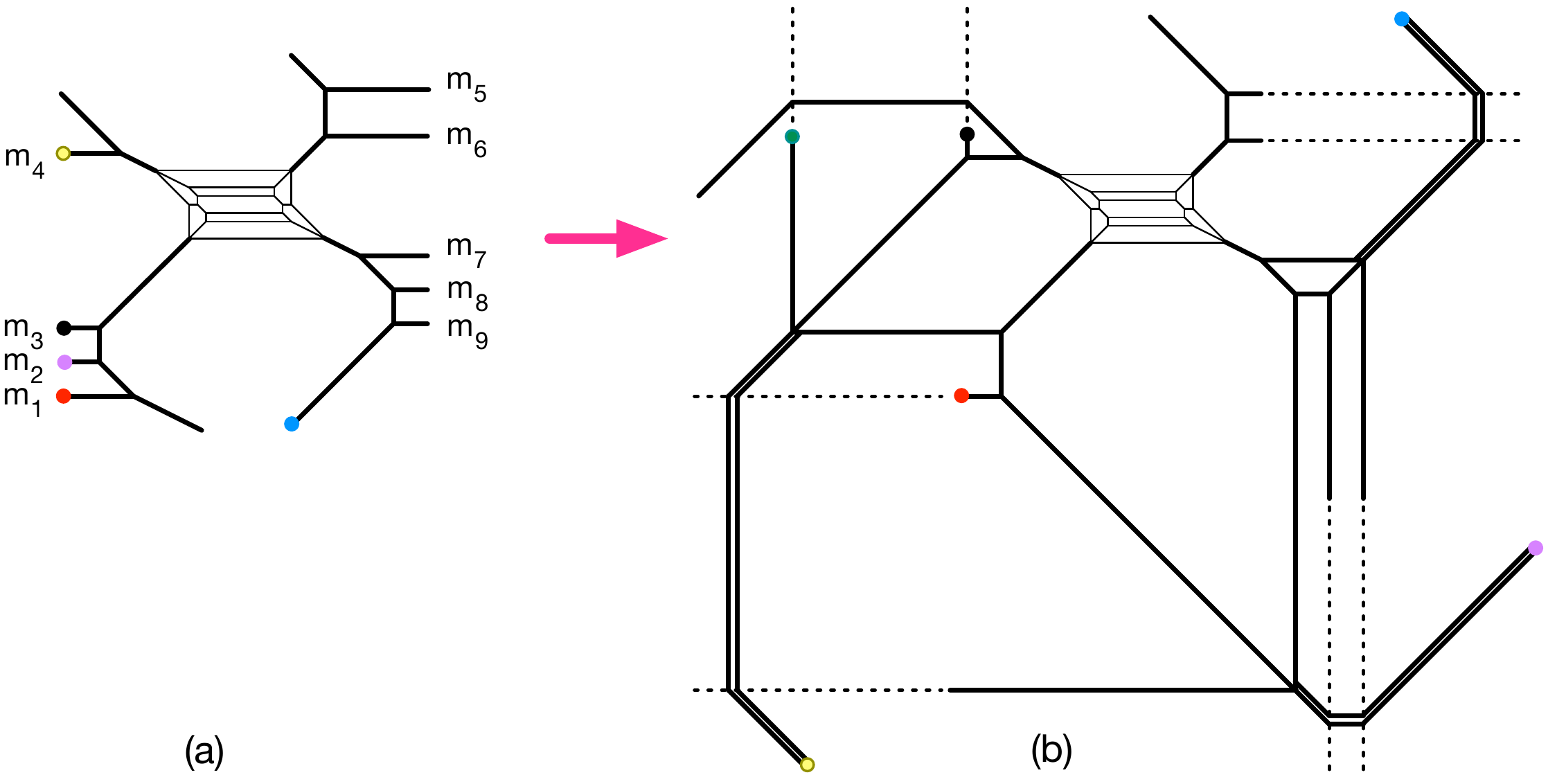} 
\caption{(a) $SU(6)_{
\frac32}+1{\bf TAS}+9\bF$ configuration. (b) $SU(6)_{\frac32}+1{\bf TAS}+9\bF$ Tao diagram. Here we denoted that the thick line means two coincident 5-branes and the dotted lines are the monodromy cuts.  7-brane motions from (a) to (b) are explained in \cite{Hayashi:2018lyv}.}
\label{fig:SU6+1TAS+9F}
\end{figure}
\paragraph{\underline{$SU(6)_\frac32+1{\bf TAS}+9\bF$}.} Another example of such sort is 
the $SU(6)_\frac32$ marginal theory with one rank-3 antisymmetric hypermultiplet and nine flavors. Its asymptotic configuration is  the same as that of the $SU(3)_\frac32+9\bF$ theory, whose 5-brane web is constructed in  \cite{Zafrir:2015rga, Hayashi:2018lyv}. Hence, in the same way, one can construct a 5-brane configuration for  the $SU(6)_\frac32+1{\bf TAS}+9\bF$ theory by introducing 9 D7-branes such that the Chern-Simons level is $\kappa=3/2$. For instance see Figure \ref{fig:SU6+1TAS+9F}. As expected, it is also also a Tao diagram. 

\begin{figure}
\includegraphics[width=13cm]{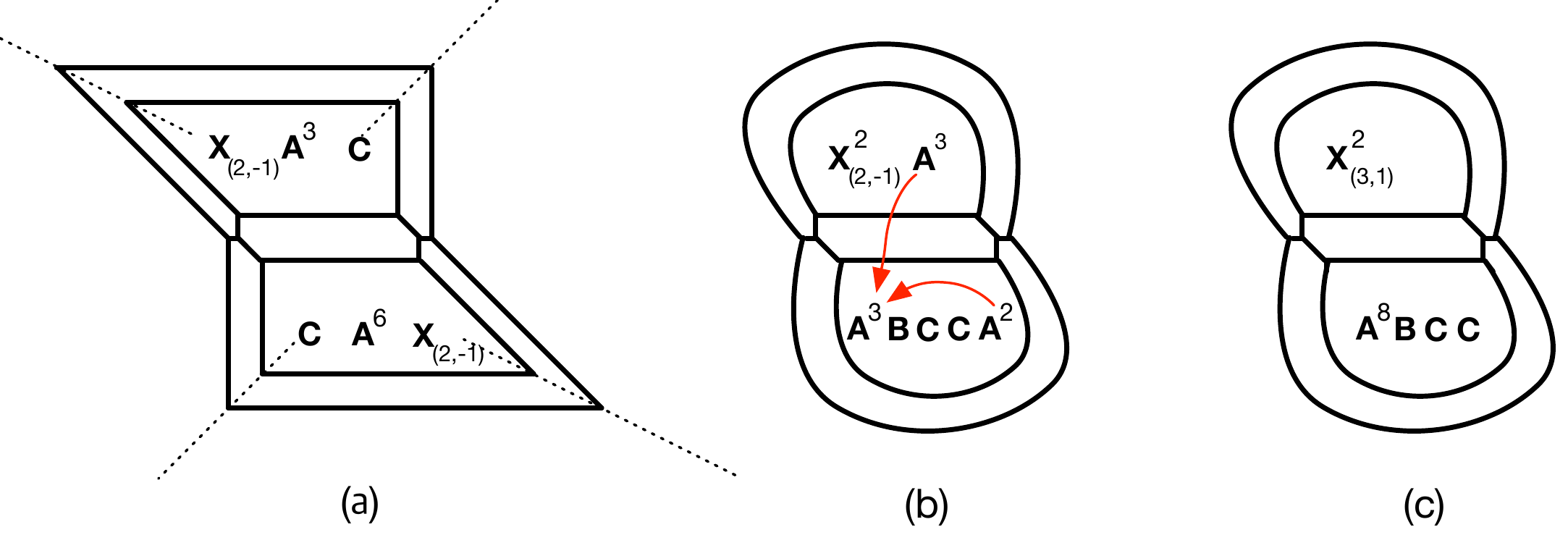} 
\caption{(a) 7-brane configurations for 5-brane web in Figure \ref{fig:SU6+1TAS+9F} where all the 7-branes are put in the 5-brane loops. (b) Rearrangement of 7-branes. (c) 7-brane configurations showing the non-abelian part of the global symmetry being an 
$E_8\times SU(2)$ symmetry.}
\label{fig:SU6+9Fsymm}
\end{figure}

One can perform a similar 7-brane monodromy analysis to read off the global symmetry. As in Figure \ref{fig:SU6+9Fsymm}, one can rearrange the 7-branes with \eqref{eq:7Bmonodromy} and also 
\begin{align}
\textsf{\textbf{A}}\textsf{\textbf{C}}=	\textsf{\textbf{C}}\textsf{\textbf{X}}_{(2,1)}, \qquad
\textsf{\textbf{B}}\textsf{\textbf{C}}\textsf{\textbf{A}}^n =
\textsf{\textbf{A}}^n\textsf{\textbf{B}}\textsf{\textbf{C}}
	\label{eq:7Bmonodromy2}
\end{align}
to find the global symmetry of the $SU(6)_\frac32+1{\bf TAS}+9\bF$ theory. Given a 7-brane configuration for the $SU(6)_\frac32+1{\bf TAS}+9\bF$ theory, for instance, Figure \ref{fig:SU6+9Fsymm}(a), one can use \eqref{eq:7Bmonodromy} to obtain
\begin{align}
\textsf{\textbf{C}}\textsf{\textbf{A}}^3\textsf{\textbf{X}}_{(2,-1)}
=\textsf{\textbf{A}}^3\textsf{\textbf{X}}_{(2,-1)}^2, 
\end{align} 
from which one rearranges the 7-brane in the upper 5-brane loop chamber in Figure \ref{fig:SU6+9Fsymm}(b). The rearrangement of the 7-brane in the lower chamber in Figure \ref{fig:SU6+9Fsymm}(b) is obtained from
\begin{align}
\textsf{\textbf{C}}\textsf{\textbf{A}}^2\;\textsf{\textbf{A}}\;\textsf{\textbf{A}}^3\textsf{\textbf{X}}_{(2,-1)}
=\textsf{\textbf{A}}^2\textsf{\textbf{B}}\;\textsf{\textbf{A}}\textsf{\textbf{C}}\;\textsf{\textbf{A}}^3
=\textsf{\textbf{A}}^2\textsf{\textbf{B}}\textsf{\textbf{C}}\; \textsf{\textbf{X}}_{(2,1)}\textsf{\textbf{A}}\;\textsf{\textbf{A}}^2
=\textsf{\textbf{A}}^2\textsf{\textbf{B}}\textsf{\textbf{C}}\textsf{\textbf{A}}\textsf{\textbf{C}}\textsf{\textbf{A}}^2
=\textsf{\textbf{A}}^3\textsf{\textbf{B}}\textsf{\textbf{C}}\textsf{\textbf{C}}\textsf{\textbf{A}}^2,
\end{align}
where the first and third equalities are due to \eqref{eq:7Bmonodromy} and the second and fourth equalities come from \eqref{eq:7Bmonodromy2}.
One then finally relocates three $\textsf{\textbf{A}}$'s in the upper chamber to the lower chamber as well as brings two $\textsf{\textbf{A}}
$'s in front as depicted in Figure \ref{fig:SU6+9Fsymm}(b). The resulting configuration is given in Figure \ref{fig:SU6+9Fsymm}(c),
\begin{align}
(~	\textsf{\textbf{X}}_{(3,1)}^2 ~|~\emptyset~|~\textsf{\textbf{A}}^8\textsf{\textbf{B}}\textsf{\textbf{C}}\textsf{\textbf{C}}~), 
\end{align}
from which we find that the non-Abelian part of the global symmetry\footnote{It was discussed in \cite{DeWolfe:1998pr} that the 7-brane configuration $\textsf{\textbf{A}}^8\textsf{\textbf{B}}\textsf{\textbf{C}}\textsf{\textbf{C}}$ is equivalent to that of $\textsf{\textbf{A}}^7\textsf{\textbf{B}}\textsf{\textbf{C}}\textsf{\textbf{B}}\textsf{\textbf{C}}$, where the corresponding global symmetry is $E_8$.}
 is $E_8\times SU(2)$ .

We note that as the 7-brane analysis is insensitive for an abelian symmetry, here $U(1)$ is added by hand to match with the number of the mass parameters in the 5d theory, assuming that the rank-3 antisymmetric hypermultiplet is massive in general.  
We also note that this global symmetry is slightly different from the expected global symmetry reported in \cite{Jefferson:2017ahm}, which is $E_8^{(1)}\times A_2^{(1)}$. As our 5-brane configuration is the same as that of the $SU(3)_\frac32+9\bF$ theory, and also all the 7-branes can be put in two different 5-brane loops, it is expected to show the same global symmetry as that of the $SU(3)_\frac32+9\bF$ theory. Our 5-brane construction for the $SU(6)_\frac32+1{\bf TAS}+9\bF$ theory is, in fact, the theory of massless rank-3 antisymmetric hypermultiplet. It may be that the 5-brane configuration for the massless rank-3 antisymmetric matter does not capture further enhancement from $SU(2)\times U(1)$ to $SU(3)$, since there are not enough 7-branes\footnote{There is a similar case for the 5-brane configuration for the 6d E-string on a circle, which yields a 5-brane web for the 5d $Sp(2)$ gauge theory with 8 flavors and one antisymmetric hypermultiplet. The expected global symmetry from 5d perspective is 
$E_8\times SU(2)$. Here, one has both 5-brane webs for massless \cite{Kim:2015jba} and massive \cite{Bergman:2015dpa,Hayashi:2015zka,Hayashi:2018lyv} antisymmetric hypermultiplet. For the massless case, 7-brane analysis does not capture the $SU(2)$ part, while the massive case see the full enhanced global symmetry, 
$E_8\times SU(2)$.}.

\paragraph{\underline{$SU(6)_0+1{\bf TAS}+1{\bf AS}+4\bF$}.}
\begin{figure}
\centering
\includegraphics[width=15cm]{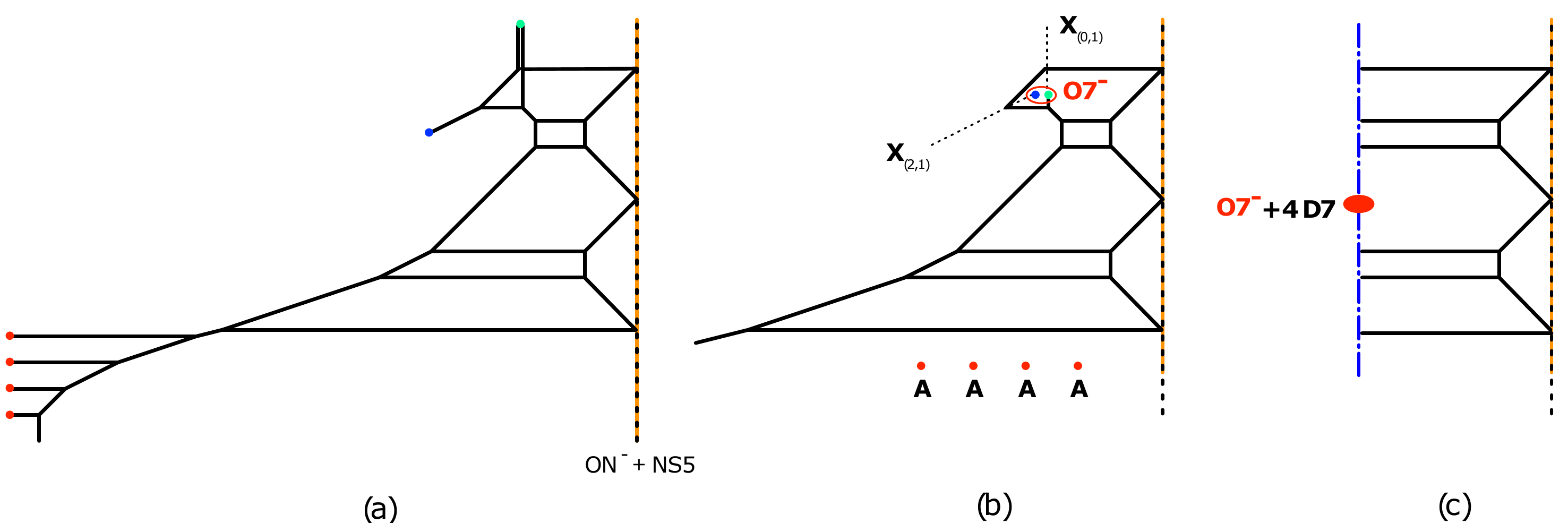} 
\caption{(a) A 5-brane configuration for the $SU(6)_0+1{\bf TAS}+1{\bf AS}+4\bF$ theory. (b) Its 7-brane configuration where 7-branes with charges $(2,-1)$ and $(0,1)$ can be converted into an O7$^-$-plane. (c) A 5-brane configuration for  the $SU(6)_0+1{\bf TAS}+1{\bf AS}+4\bF$ theory with two orientifold planes.}
\label{fig:SU61AS4F}
\end{figure}
One can also introduce a hypermultiplet in the rank-2 antisymmetric representation in addition to a hypermultiplet in the rank-3 antisymmetric representation as depicted in Figure \ref{fig:SU61AS4F}(a), which gives the $SU(6)_0 + 1{\bf TAS} + 1{\bf AS} + 4{\bf F}$. 
As shown in Figure \ref{fig:SU61AS4F}(b), 
the presence of the rank-2 antisymmetric matter 
can be understood since the diagram contains a configuration with an O7$^-$-plane attached to an NS5-brane \cite{Bergman:2015dpa}. By putting an O7$^-$-plane and four D7 branes together, one constructs a 5-brane web for the $SU(6)_0$ theory with a hypermultiplet in the antisymmetric representation and four flavors Figure \ref{fig:SU61AS4F}(c), which has two orientifolds. Since the combination of an O7$^-$-plane with four D7-branes is S-dual invariant, its S-dual diagram shows clearly that it is of a periodic structure in the vertical direction, supporting the consistency that the $SU(6)_0+1{\bf TAS}+1{\bf AS}+4\bF$ is marginal.




\subsection{5-brane webs for $SU(6)+1/2\,\mathbf{TAS}$ with various hypermultiplets}
In section \ref{sec:SU6TSA}, 5-brane webs for $SU(6)$ theories with a half-hypermultiplet in the rank-3 antisymmetric representation is discussed. For instance, Figure \ref{fig:su6whtsa} is a 5-brane web for the $SU(6)+1/2\,\mathbf{TAS}$ theory with the Chern-Simons level $\kappa=5/2$. One can readily change the Chern-Simons level by adjusting the charges of the external 5-branes just as done for the $SU(6)+1\,\mathbf{TAS}$ theory. Below we construct 5-brane webs for marginal $SU(6)+1/2\,\mathbf{TAS}$ theories with various hypermultiplets. 

\begin{figure}
\centering
\includegraphics[width=15.5cm]{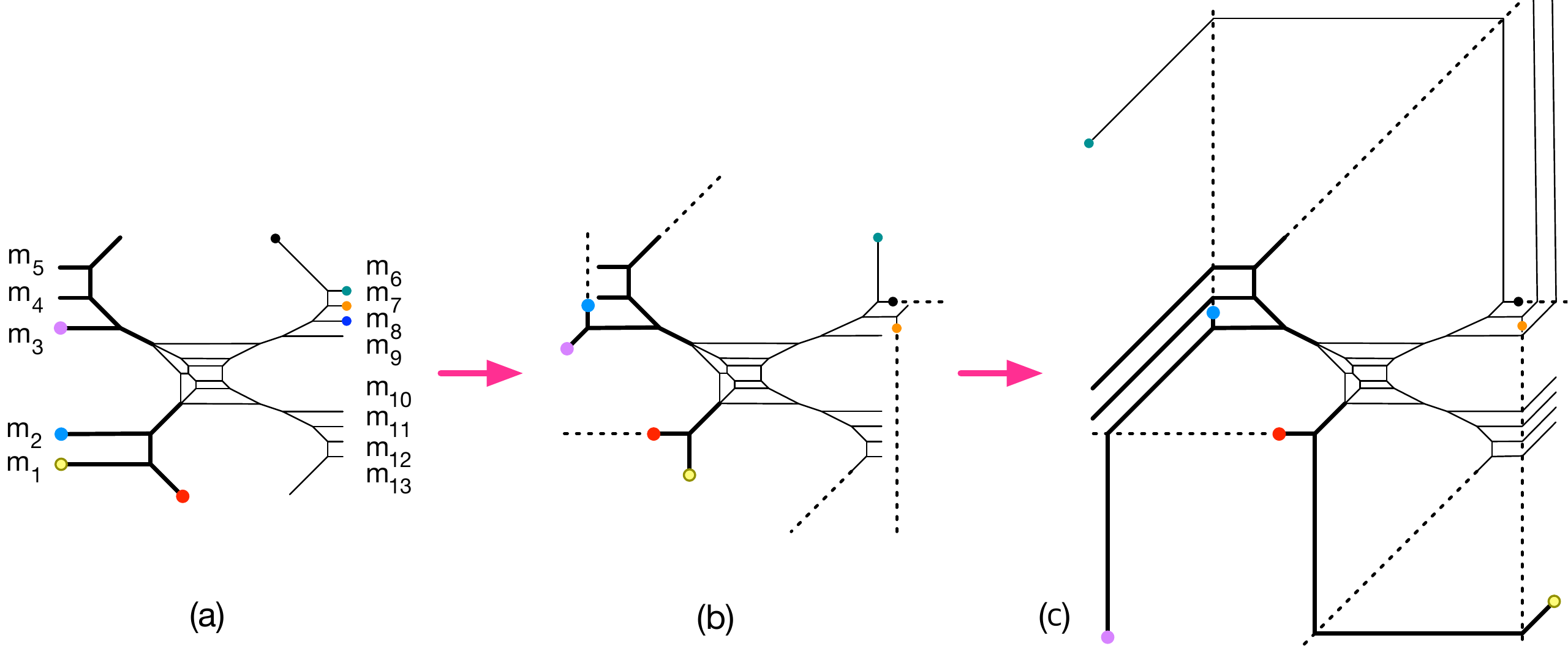}
\caption{$SU(6)_{0}+\frac12{\bf TAS}+13\bF$ Tao diagram}
\label{fig:SU6+1over2TAS+13F}
\end{figure}

\paragraph{\underline{$SU(6)_0+\frac12{\bf TAS}+13\bF$}.} 
Given a 5-brane web for the $SU(6)_{\frac52}+1/2{\bf TAS}$ theory in Figure \ref{fig:su6whtsa}, one can modify the charge of the external 5-branes and then add 13 D7 branes in a way that it leads to the Chern-Simons level $\kappa=0$. An example for 5-brane web for the $SU(6)_{0}+1/2{\bf TAS}+ 13{\bf F}$ theory is given in  
Figure \ref{fig:SU6+1over2TAS+13F}(a). As it is a marginal theory, we expect it is of a certain periodic structure. In a similar way done in the marginal $SU(6)+1{\bf TAS}$ theories with only flavors in section \ref{sec:SU6w1TAS}, we can move 7-branes and allocate the cuts of 7-branes to show a periodic structure as shown in Figure \ref{fig:SU6+1over2TAS+13F}(b). By pulling out 7-branes across the cuts arranged in \ref{fig:SU6+1over2TAS+13F}(b). One sees that it is a Tao diagram showing periodic web configuration as depicted in Figure \ref{fig:SU6+1over2TAS+13F}(c). 

\begin{figure}
\centering
\includegraphics[width=11.5cm]{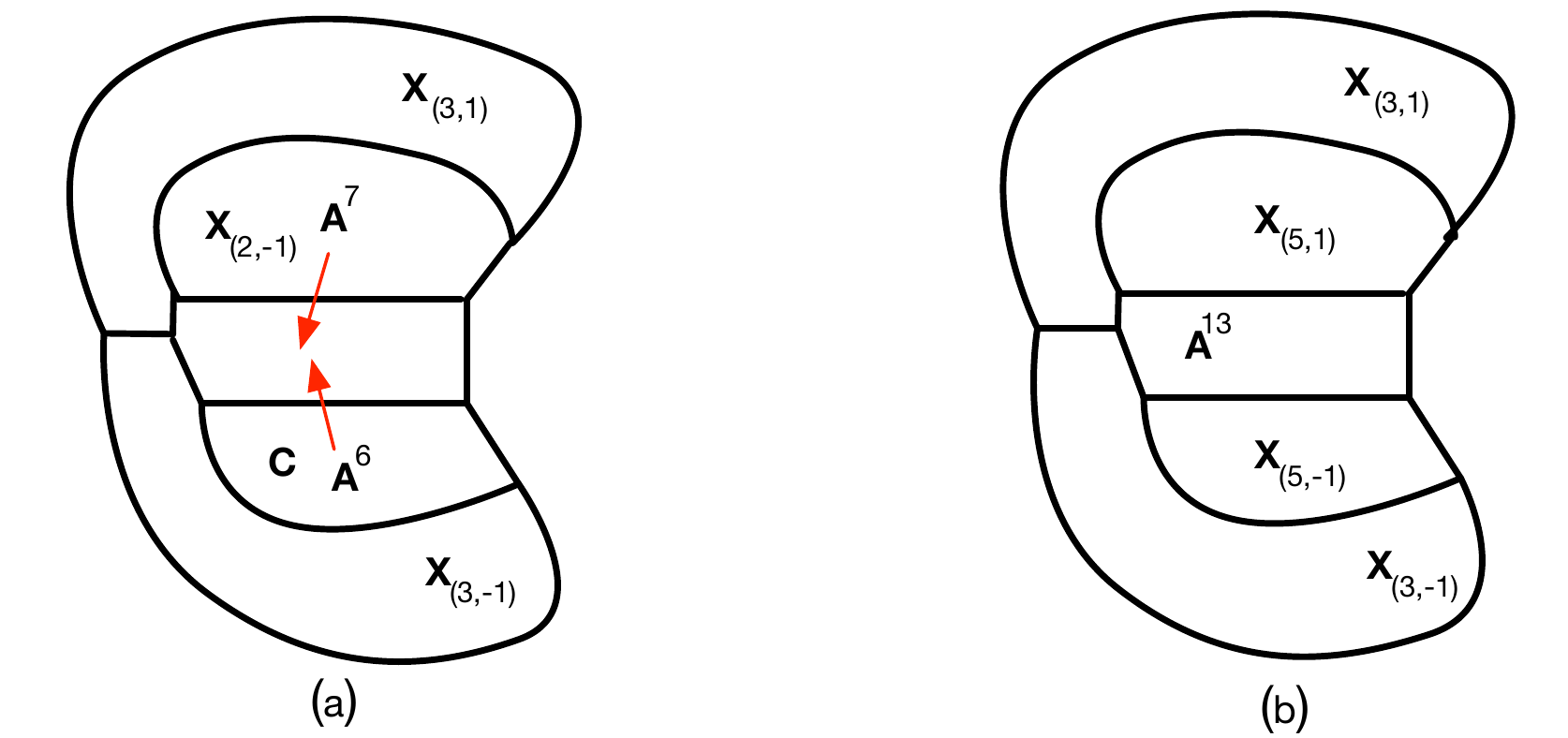}
\caption{(a) A 7-brane configuration of a web diagram for the $SU(6)_{0}+\frac12{\bf TAS}+13\bF$ theory. (b) A 7-brane configuration showing $SU(13)$ as the non-abelian part of global symmetry. }
\label{fig:SU6+1o2TAS+13FGlobal}
\end{figure}

The global symmetry in this case can be readily read off from the 7-brane configuration associated with the 5-brane web in Figure \ref{fig:SU6+1over2TAS+13F}. In Figure \ref{fig:SU6+1o2TAS+13FGlobal}, one can allocate all the D7-branes in the innermost 5-brane loop but other 7-branes are confined to all other 5-brane loops. This leads to an $SU(13)=\textsf{\textbf{A}}^{13}$ symmetry, which is the same as the non-abelian part of the perturbative global symmetry of the theory. As it is the non-abelian part of the flavor symmetry, 
the expected global symmetry would be then  $SU(13)\times U(1)\times U(1)$ since the total number of the mass parameters of the theory is $14$, agreeing with 
the global symmetry obtained in \cite{Jefferson:2017ahm}.

\begin{figure}
\centering
\includegraphics[width=15cm]{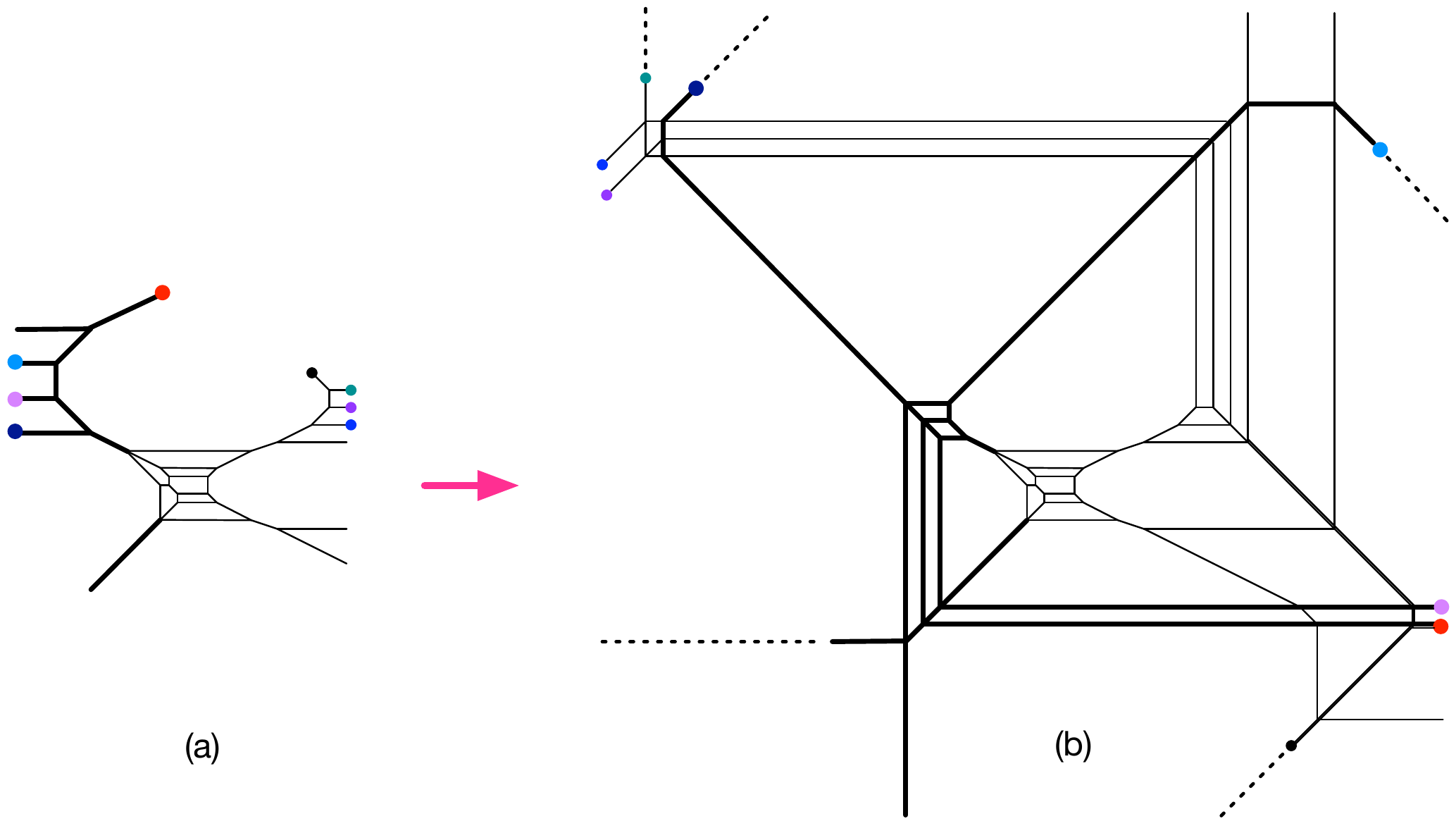}
\caption{$SU(6)_{3}+\frac12{\bf TAS}+9\bF$ Tao diagram}
\label{fig:SU6+1over2TAS+9F}
\end{figure}

\paragraph{\underline{$SU(6)_3+\frac12{\bf TAS}+9\bF$}} 
One can also construct the $SU(6)$ theory with a half-hypermultiplet in the rank-3 antisymmetric representation and 9 flavors which has the Chern-Simons level $\kappa=-3$, by introducing 9 D7-branes in such a way that it has the Chern-Simons level $\kappa=-3$. See Figure \ref{fig:SU6+1over2TAS+9F}. It can be shown that it is also a Tao diagram as depicted in Figure \ref{fig:SU6+1over2TAS+9F}(b). 

Following the 7-brane analysis for the $SU(6)_0+\frac12{\bf TAS}+13\bF$ theory in Figure \ref{fig:SU6+1o2TAS+13FGlobal}, one easily sees that the 7-brane configuration for the $SU(6)_3+\frac12{\bf TAS}+9\bF$ theory is readily manipulated to yields the non-abelian part of global symmetry is $SU(9)=\textsf{\textbf{A}}^{9}$ symmetry, which is the non-abelian part of the perturbative flavor symmetry of the theory. The expected global symmetry would be then $SU(9)\times U(1)\times U(1)$.

\begin{figure}
\centering
\includegraphics[width=15cm]{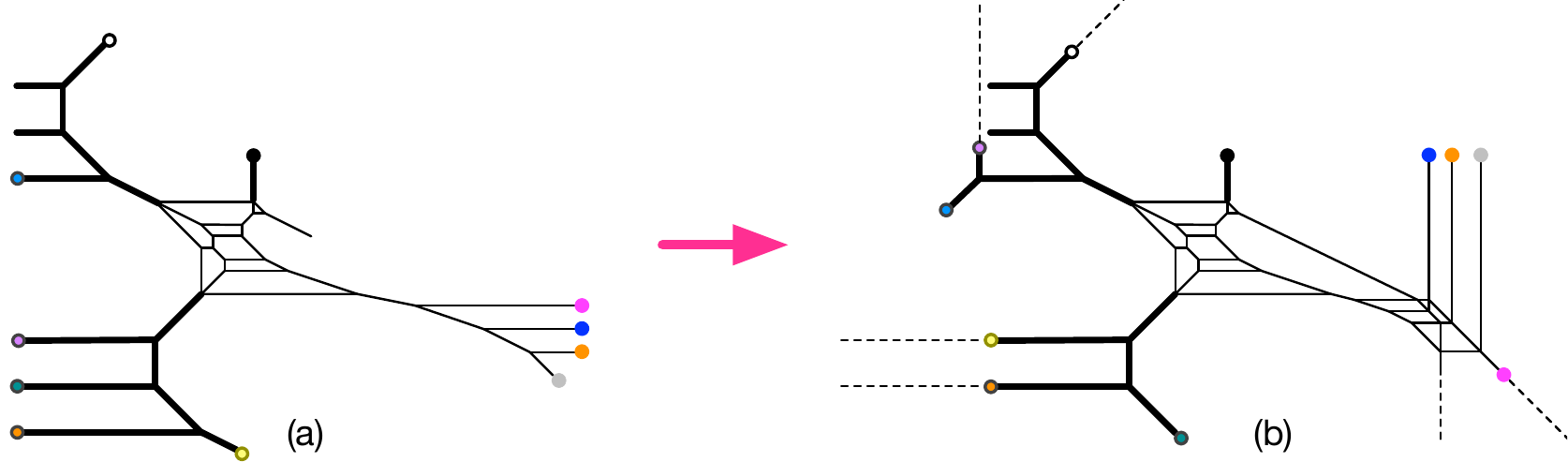}
\caption{(a) A 5-brane configuration for the $SU(6)_{0}+\frac12{\bf TAS}+1\AS+9\bF$ theory. (b) Its Tao diagram. As there are many 5-branes attached to the 7-branes which make 5-brane web complicated, here we give a sketch on how the 7-brane charges are altered. The number of 5-branes attached to a 7-brane can be read off by tracking the number of Hanany-Witten transitions taking place in Figure (b).}
\label{fig:SU6+1over2TAS+1AS+9F}
\end{figure}

\begin{figure}
\centering
\includegraphics[width=15cm]{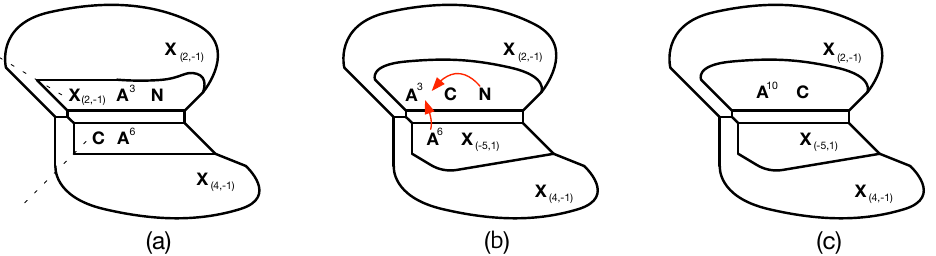}
\caption{A 7-brane configuration of a web diagram for the $SU(6)_{0}+\frac12{\bf TAS}+1{\bf AS}+9\bF$ theory. (a) A 7-brane configuration for the 5-brane web given in Figure \ref{fig:SU6+1over2TAS+1AS+9F}(a).  (b) 
In the upper chamber, using \eqref{eq:7Bmonodromy}, $\textsf{\textbf{N}}\;\textsf{\textbf{A}}^3\textsf{\textbf{X}}_{(-2,1)} 
=\textsf{\textbf{N}}\;\textsf{\textbf{C}} \textsf{\textbf{A}}^3$
(c) The resulting 7-brane configuration after moving 7-branes along the red arrows in figure (b) with 
$\textsf{\textbf{N}} \textsf{\textbf{C}}=\textsf{\textbf{C}}\textsf{\textbf{A}}$ 
. The non-abelian part of global symmetry  $SU(10)$ is obtained from the 7-brane configuration $\textsf{\textbf{A}}^{10}$ in a chamber of 5-brane loops.
}
\label{fig:7braneSU6+1o2TA+1AS+9F}
\end{figure}

\paragraph{\underline{$SU(6)_0+\frac12{\bf TAS}+1{\bf AS}+9\bF$}} 
As done in \cite{Bergman:2015dpa} and also in section \ref{sec:SU6w1TAS}, one can introduce a rank-2 antisymmetric hypermultiplet by introducing a configuration of an NS5-brane ending on an O7$^-$-plane.   
A 5-brane web for the $SU(6)_0$ theory with a rank-3 antisymmetric half-hypermultiplet and one antisymmetric hypermultiplet and 9 flavors can be constructed as in Figure \ref{fig:SU6+1over2TAS+1AS+9F}(a). It can be also shown that it is a Tao diagram as depicted in Figure \ref{fig:SU6+1over2TAS+1AS+9F}(b), implying that the theory has a 6d UV completion. As drawn in Figure \ref{fig:7braneSU6+1o2TA+1AS+9F}, the non-Abelian part of the global symmetry is $SU(10)$. 


\begin{figure}
\centering
\includegraphics[width=15cm]{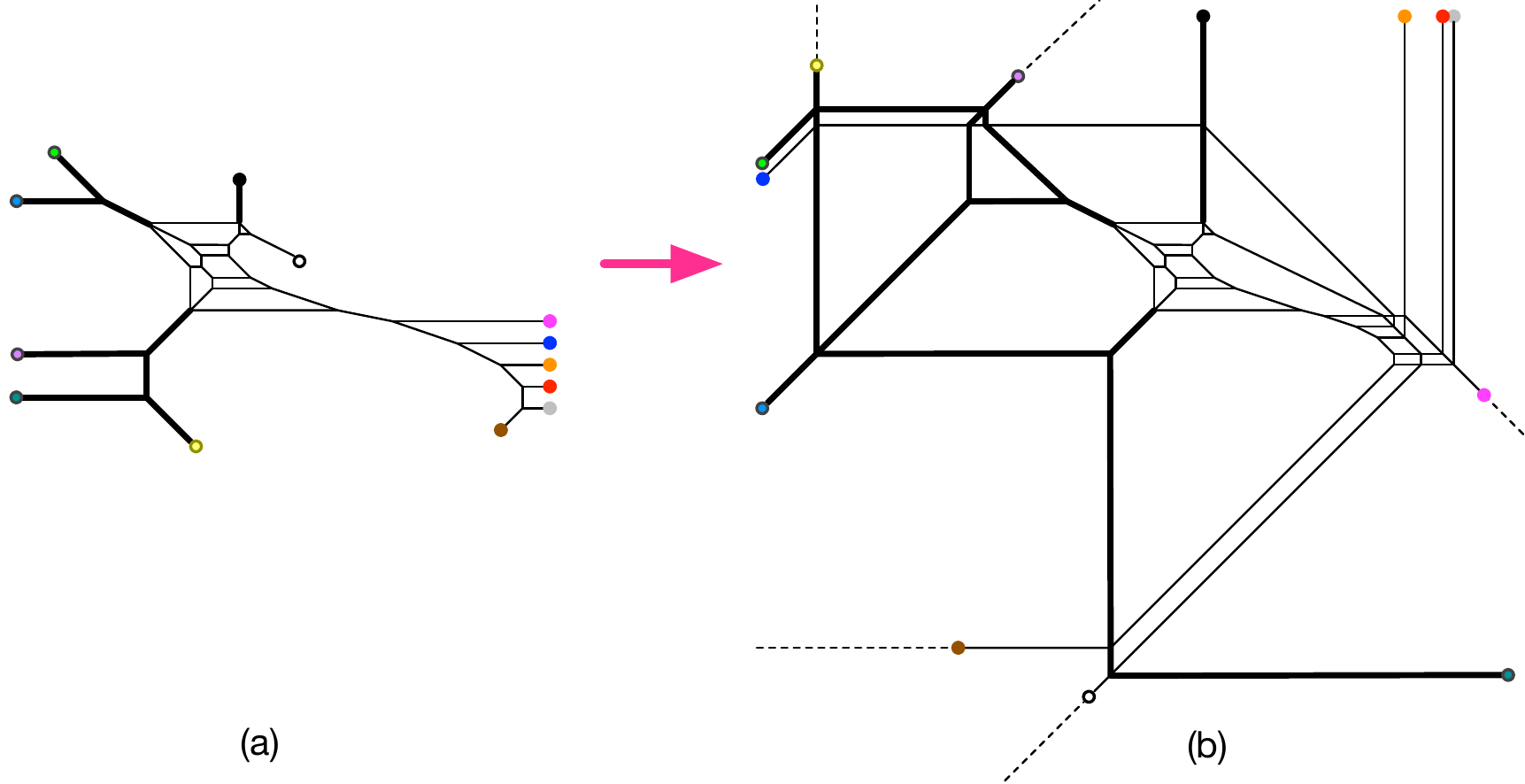} \label{fig:SU6+1over2TAS+8F}
\caption{A $SU(6)_{
\frac32}+\frac12{\bf TAS}+1\AS+8\bF$ Tao diagram. The number of 5-branes attached to a 7-brane can be read off by tracking the number of Hanany-Witten transitions taking place in Figure (b).}
\label{fig:SU6+1over2TAS+1AS+8F}
\end{figure}

\paragraph{\underline{$SU(6)_\frac32+\frac12{\bf TAS}+1{\bf AS}+8\bF$}} 
A 5-brane web for the $SU(6)_{-\frac32}+\frac12{\bf TAS}+1{\bf AS}+8\bF$ theory is depicted in
Figure \ref{fig:SU6+1over2TAS+1AS+8F}. One can show that the corresponding 5-brane web diagram is a Tao diagram, though it requires delicate arrangements of 7-branes as shown in Figure \ref{fig:SU6+1over2TAS+1AS+8F}(b). The non-abelian part of the global symmetry that we can see from the corresponding 7-brane configuration seems to be $SU(8)$, which means that this 7-brane configuration may not show any enhancement other than the perturbative symmetry of the 8 flavors.  
Hence, the rank of the global symmetry from the 7-brane configuration is smaller than that of the global symmetry $SO(16)\times SU(2)\times U(1)$ proposed in \cite{Jefferson:2017ahm}.

\begin{figure}
\centering
\includegraphics[width=14cm]{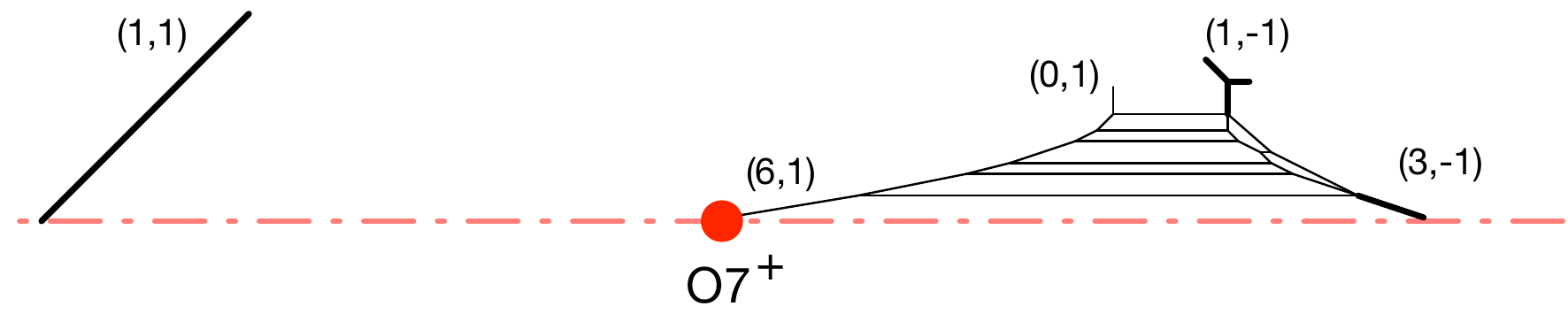} 
\caption{A 5-brane configuration for the $SU(6)_{0}+\frac12{\bf TAS}+1{\bf Sym}+1\bF$ theory.}
\label{fig:SU6+1over2TAS+1Sym+1F}
\end{figure}

\begin{figure}
\centering
\includegraphics[width=14cm]{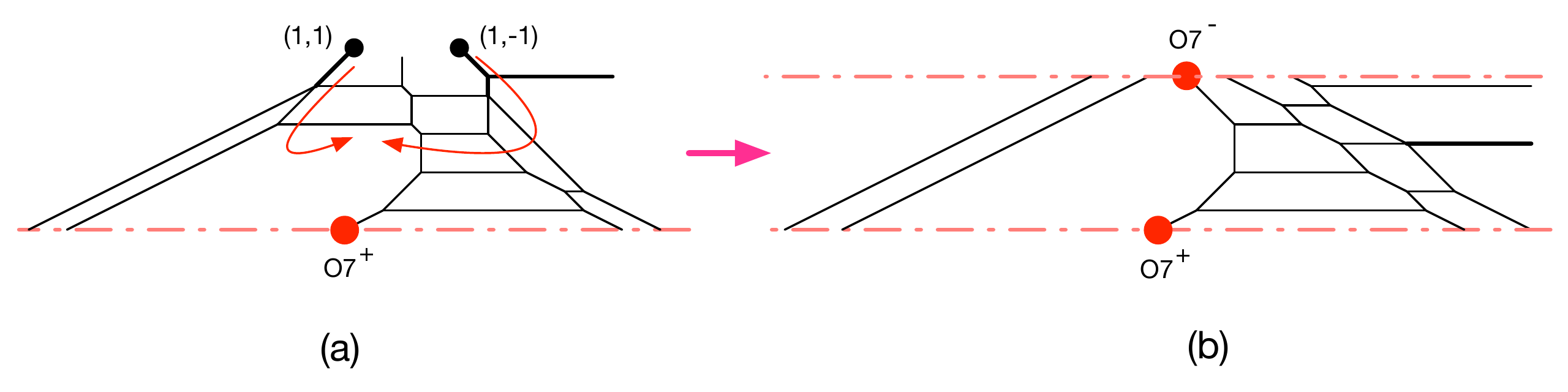} 
\caption{(a): A web diagram for $SU(6)_{0}+\frac12{\bf TAS}+1{\bf Sym}+1\bF$ obtained by moving two color branes below O7$^+$-plane in Figure \ref{fig:SU6+1over2TAS+1Sym+1F}. (b): An $SU(6)_{0}+\frac12{\bf TAS}+1{\bf Sym}+1\bF$ web diagram with both O7$^-$- and O7$^+$-planes.}
\label{fig:SU6+1over2TAS+1Sym+1F-1}
\end{figure}

\paragraph{\underline{$SU(6)_0+\frac12{\bf TAS}+1{\bf Sym}+1\bF$}} 
In 5-brane web, a hypermultiplet in the symmetric representation is represented with an NS5-brane ending on an O7$^+$-plane \cite{Bergman:2015dpa}. Examples of 5-brane webs for marginal theories with symmetric matter, $SU(3)_0+1{\bf Sym}+1\bF$ and $SU(3)_{-\frac32}+1{\bf Sym}$, are discussed in \cite{Hayashi:2018lyv}. They are, in fact, instructive examples for constructing the $SU(6)$ theories with a rank-3 antisymmetric half-hypermultiplet and a symmetric hypermultiplet as the asymptotic 7-brane configurations for both the $SU(3)$ theories and the $SU(6)$ theories are the same. We first consider a 5-brane configuration for the $SU(6)_0+\frac12{\bf TAS}+1{\bf Sym}+1\bF$ theory. Using the fact that a decoupling of a symmetric hypermultiplet for an $SU(N)$ theory gives rise to the change of the Chern-Simons level $\kappa$ by $\kappa-\frac12(N+4)$, (for $N=6$, $\kappa \to \kappa-5$), one has a 5-brane configuration for the $SU(6)_0+\frac12{\bf TAS}+1{\bf Sym}+1\bF$ theory as follows: one prepares a 5-brane web diagram for $SU(6)_{-5}+\frac12{\bf TAS}+1\bF$ and then attaches one external single 5-brane  to an O7$^+$-plane, as shown in Figure \ref{fig:SU6+1over2TAS+1Sym+1F}. We know that its 7-brane charges are the same as those appear in a 5-brane web for the $SU(3)_0+1{\bf Sym}+1\bF$ theory. (See Figure 49 in \cite{Hayashi:2018lyv}.) We note that unlike the 5-brane web for the $SU(3)_0+1{\bf Sym}+1\bF$ theory, two 5-branes are attached to some 7-branes. It is however still possible to make this 5-brane configuration to have a periodic structure, by moving 7-branes inside the 5-brane loops and also by manipulating a pair of 7-branes to be converted into an O7$^-$-plane, as shown in Figure \ref{fig:SU6+1over2TAS+1Sym+1F-1}. It is therefore a 5-brane web with an O7$^-$-plane and an O7$^-$-plane.


\begin{figure}
\centering
\includegraphics[width=15cm]{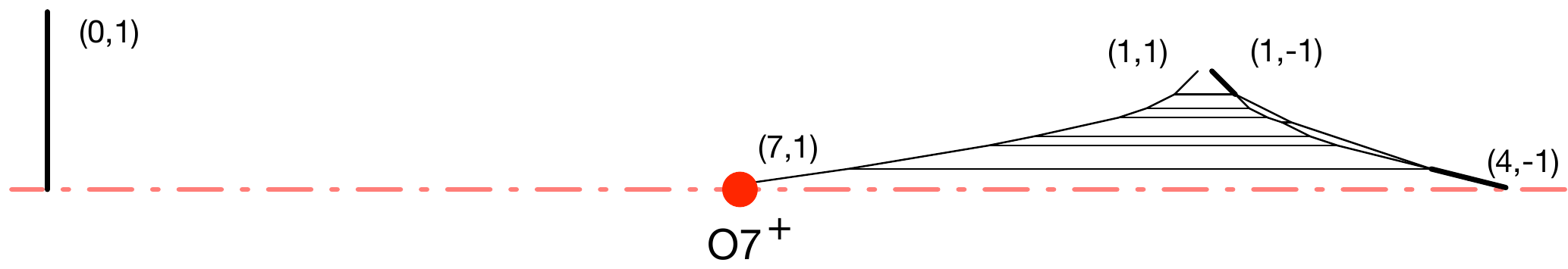}
\caption{A 5-brane configuration for the $SU(6)_{-\frac32}+\frac12{\bf TAS}+1{\bf Sym}$ theory.}
\label{fig:SU6+1over2TAS+1Sym}
\end{figure}

\begin{figure}
\centering
\includegraphics[width=15cm]{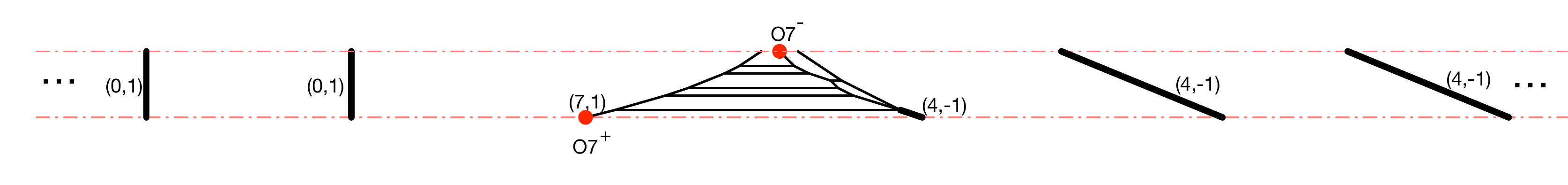}
\caption{An $SU(6)_{-\frac32}+\frac12{\bf TAS}+1{\bf Sym}$ diagram with both O7$^-$- and O7$^+$-planes.}
\label{fig:SU6+1over2TAS+1Sym-1}
\end{figure}

\paragraph{\underline{$SU(6)_\frac32+\frac12{\bf TAS}+1{\bf Sym}$}} 
Next, we consider the construction of a 5-brane for the $SU(6)_\frac32+\frac12{\bf TAS}+1{\bf Sym}$ theory. Its 7-brane charges are also the same as those for the $SU(3)_\frac32+1{\bf Sym}$ theory. In a similar way, one can have a 5-brane configuration as shown in Figure \ref{fig:SU6+1over2TAS+1Sym}, and also can make the configuration has both O7$^-$- and O7$^+$-planes as depicted in Figure \ref{fig:SU6+1over2TAS+1Sym-1}. Hence the resulting 5-brane web is also of a periodic structure. 


\bigskip
\section{$Sp(3)$ gauge theories with rank-3 
antisymmetric matter}\label{sec:Sp3}
It is also possible to introduce matter in the rank-3 antisymmetric representation to $Sp(3)$ gauge theories. In order to introduce rank-3 antisymmetric matter to an $Sp(3)$ gauge theory, we may use a Higgsing of an $SU(6)$ gauge theory by giving a vev to a hypermultiplet in the antisymmetric representation of the $SU(6)$. 
Note that the decomposition of the rank-3 antisymmetric representation of $SU(6)$ under $Sp(3)$ is given by
\begin{align}
SU(6) ~~ &\supset ~~Sp(3)\cr
{\bf 20}~~ &= ~~ {\bf 14}' + {\bf 6},
\end{align}
where ${\bf 14}'$ is the rank-3 antisymmetric representation of the $Sp(3)$\footnote{There are two $14$-dimensional representations of $Sp(3)$. One is the rank-2 antisymmetric representation whose Dynkin label is $[0, 1, 0]$ and the other is the rank-3 antisymmetric representation whose Dynkin label is $[0, 0, 1]$. We used ${\bf 14}'$ for the rank-3 antisymmetric representation. } and ${\bf 6}$ is the fundamental representation of $Sp(3)$. Hence, the Higgsing of an $SU(6)$ gauge theory with a hypermultiplet in the rank-3 antisymmetric representation yields a hypermultiplet in the rank-3 antisymmetric representation and also a hypermultiplet in the fundamental representation of an $Sp(3)$ gauge theory. 

Then, all the marginal $Sp(3)$ gauge theories with rank-3 antisymmetric matter listed in \cite{Jefferson:2017ahm} may be given by the following Higgsing\footnote{The Chern-Simons level of an $SU(6)$ gauge theory does not affect the IR $Sp(3)$ gauge theory. This can bee seen for example from the effective prepotential computation. The Higgsing of $SU(6)$ to $Sp(3)$ using a vev for an antisymmetric hypermultiplet requires the tuning $a_6 = -a_1$, $a_5 = -a_2$ and $a_4 = -a_3$. Therefore the contribution to the effective prepotential from the Chern-Simons term becomes zero after the tuning. }
\begin{align}
&SU(6)_0+ 1{\bf TAS} + 1{\bf AS}+4{\bf F}&&\to &&Sp(3)+1{\bf TAS}+5{\bf F},  \label{HiggstoSp3v1}\\
&SU(6)_{\kappa=\frac{1}{2}, \frac{3}{2}}+\frac12{\bf TAS}+2{\bf AS}+2{\bf F} &&\to &&Sp(3)+\frac12{\bf TAS}+1{\bf AS} +\frac52{\bf F}, \label{HiggstoSp3v2}\ \\
&SU(6)_0+\frac12{\bf TAS}+1{\bf AS}+9{\bf F}&&\to &&Sp(3)+\frac12{\bf TAS}+\frac{19}2{\bf F}. \label{HiggstoSp3v3}
\end{align}
The result is summarized in Table \ref{tab:Sp3marginal}.
\begin{small}
\begin{table}[t]
\begin{center}
\begin{tabular}{ |c|c|c|c|c|}  
 \hline
 $N_{\bf TAS}$ &  $N_{\bf AS}$& $N_{\bf F}$ &Web
 \\ 
 \hline \hline
  1 & . & 5 & Figure \ref{fig:sp3w1tas0}\\ \hline
1/2 & 1 &5/2& ?\\ \hline
1/2 & . &19/2& Figure 
\ref{fig:sp3w1htas0}\\ \hline
\end{tabular}
\caption{$Sp(3)$ marginal theories with rank-3 antisymmetric half-hypermultiplets and matter in other representations. }
\label{tab:Sp3marginal}
\end{center}
\end{table}
\end{small}

As we already constructed  5-brane diagrams for the $SU(6)_0$ gauge theory with $N_{\bf TAS}=1, \; N_{\bf AS}=1,\; N_{\bf F}=4$ and the $SU(6)_0$ gauge theory with $N_{\bf TAS}=\frac{1}{2}, \; N_{\bf AS}=1,\; N_{\bf F}=9$, we consider the Higgsings \eqref{HiggstoSp3v1} and \eqref{HiggstoSp3v3}.  


\begin{figure}[t]
\centering
\subfigure[]{\label{fig:sp3higgs1}
\includegraphics[width=5cm]{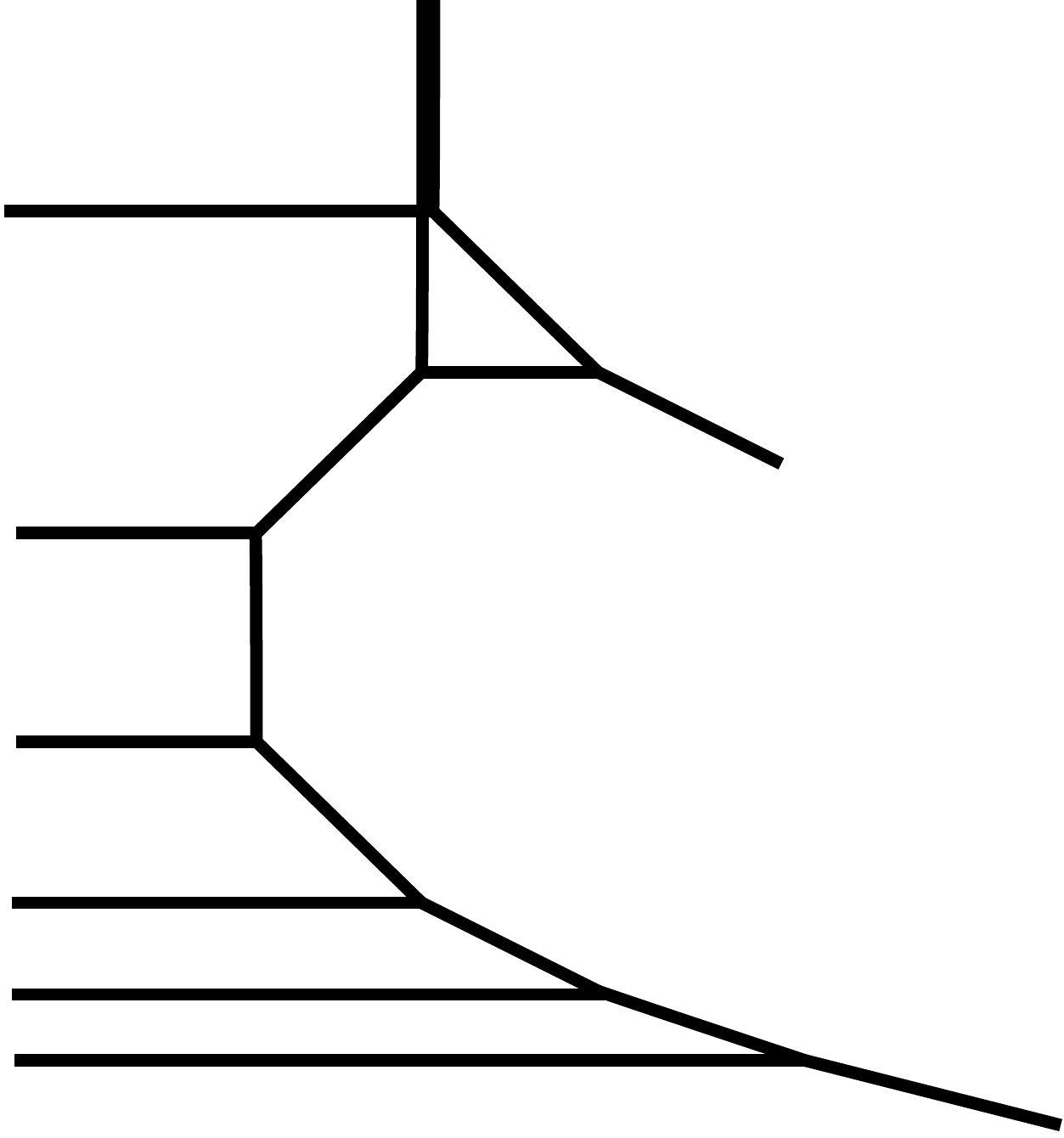}}\hspace{1cm}
\subfigure[]{\label{fig:sp3higgs2}
\includegraphics[width=5cm]{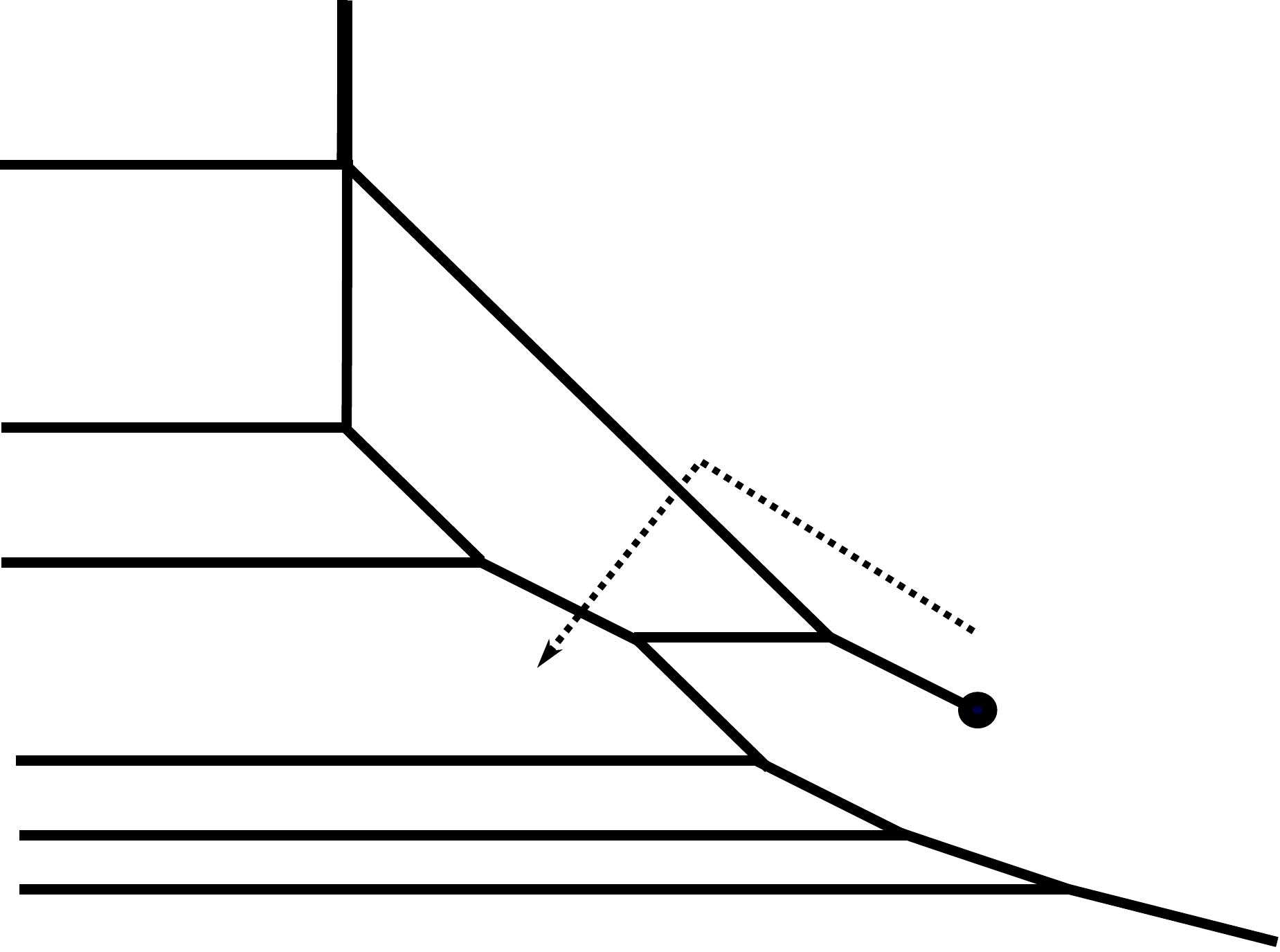}}\hspace{1cm}
\subfigure[]{\label{fig:sp3higgs3}
\includegraphics[width=5cm]{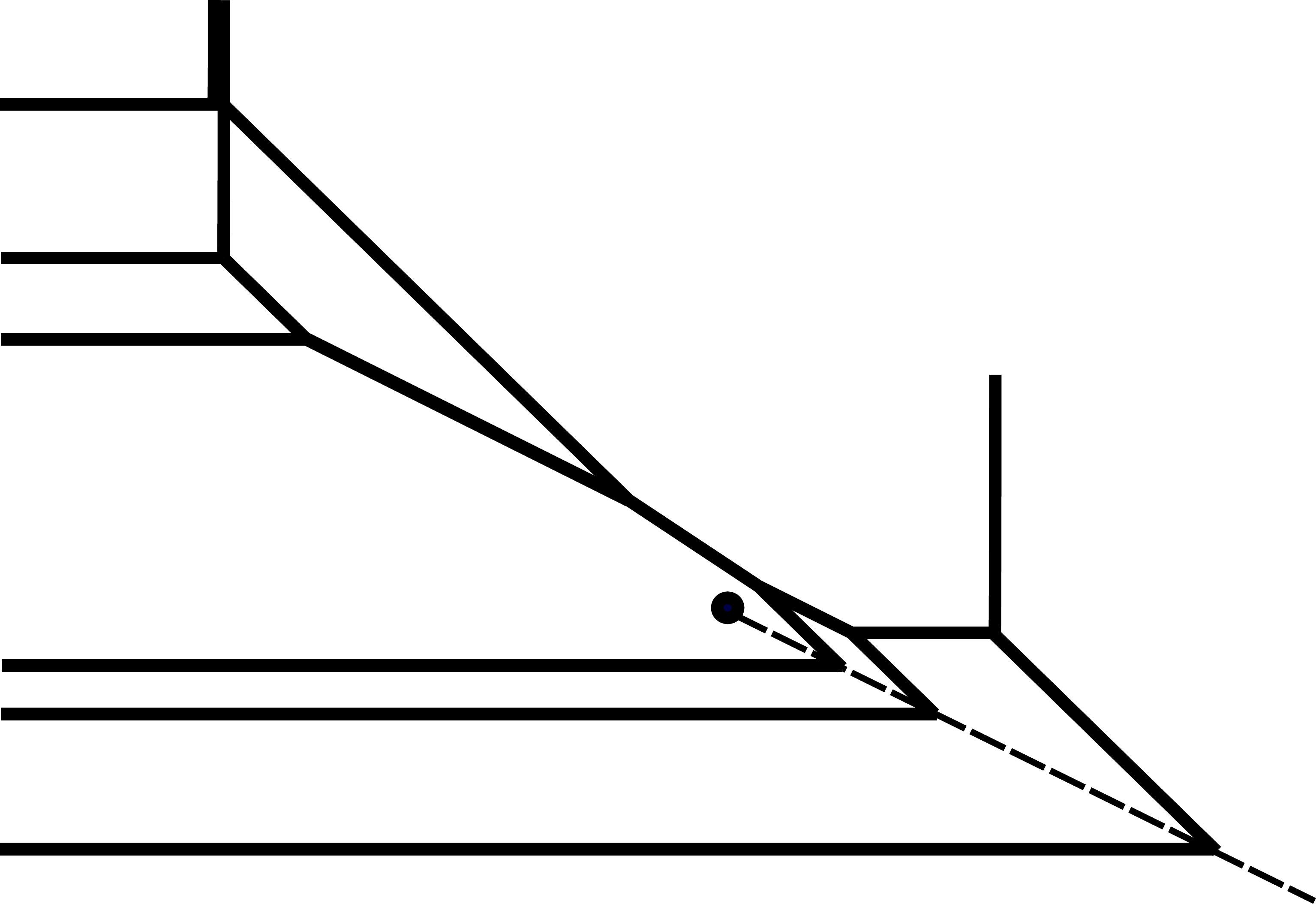}}\hspace{1cm}
\subfigure[]{\label{fig:sp3higgs4}
\includegraphics[width=5cm]{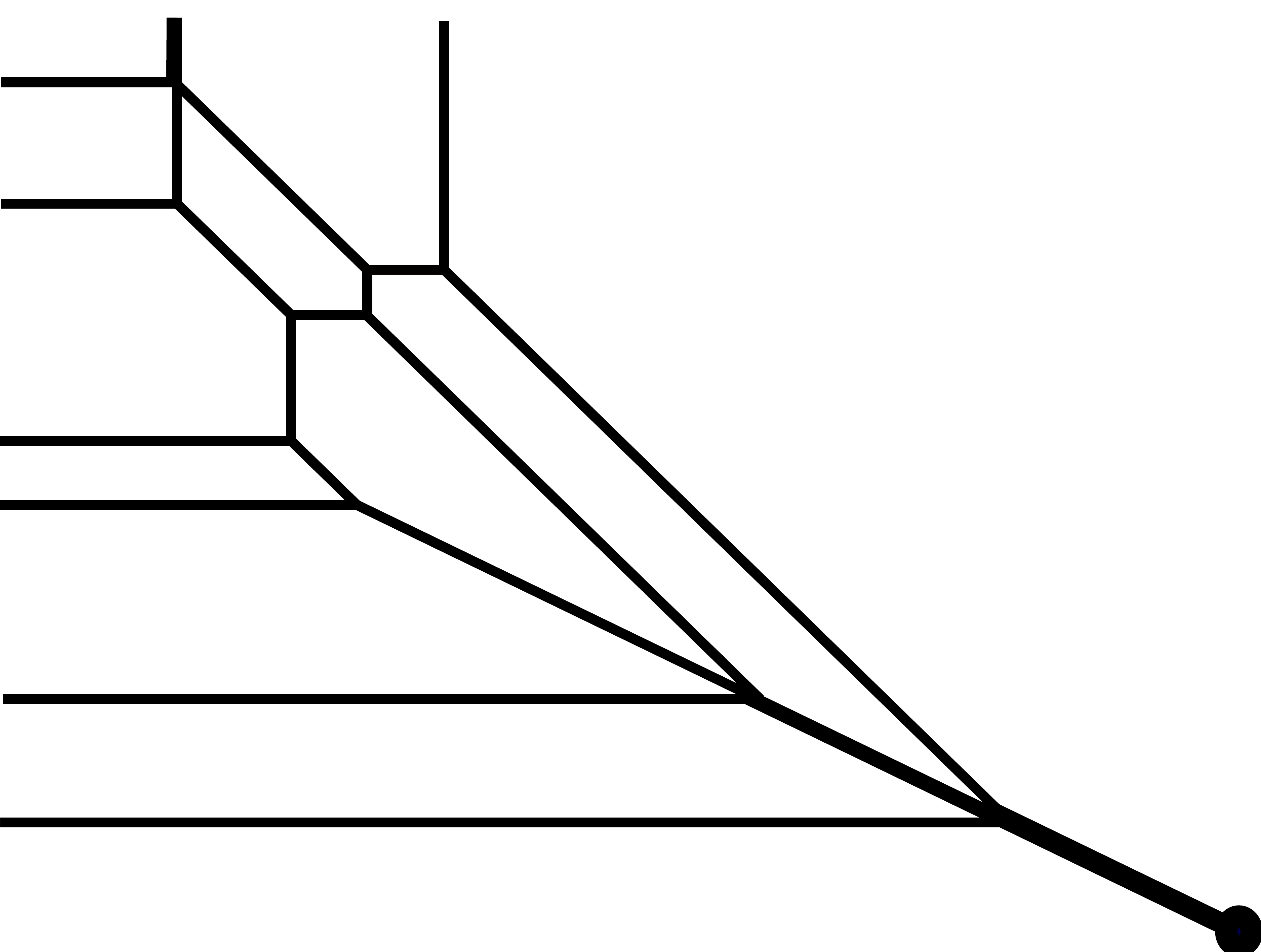}}\hspace{1cm}
\caption{(a): A diagram involving the rank-2 antisymmetric matter for an $SU(6)$ gauge theory. (b): Performing flop transitions to the diagram in Figure \ref{fig:sp3higgs1} and then moving the $(2, -1)$ 7-brane in the direction given by the arrow. (c): The diagram obtained after moving the $(2, -1)$ 7-brane. (d): The diagram obtained after performing flop transitions to the diagram in Figure \ref{fig:sp3higgs3}. 
}
\label{fig:sp3higgs}
\end{figure}

\paragraph{\underline{$Sp(3)+\frac12{\bf TAS}+19/2\bF$}} 
We first consider the Higgsing from the $SU(6)_0$ gauge theory with $N_{\bf TAS} = \frac{1}{2}, N_{\bf AS} = 1, N_{\bf F} = 9$. A diagram for the $SU(6)_0$ gauge theory with $N_{\bf TAS} = \frac{1}{2}, N_{\bf AS} = 1, N_{\bf F} = 9$ is given by Figure \ref{fig:SU6+1over2TAS+1AS+9F}. The Higgsing associated to the antisymmetric matter can be carried out diagrammatically as follows. The essential part involving the antisymmetric matter for an $SU(6)$ gauge theory is depicted in Figure \ref{fig:sp3higgs1}. From the diagram in Figure \ref{fig:sp3higgs1} we first perform flop transitions and move the $(2, -1)$ 7-brane in the direction specified in Figure \ref{fig:sp3higgs2}, which results in the diagram in Figure \ref{fig:sp3higgs3}. In order to perform the Higgsing associated to the antisymmetric matter, we further do flop transitions until we obtain the diagram in Figure \ref{fig:sp3higgs4}. 

\begin{figure}[t]
\centering
\includegraphics[width=12cm]{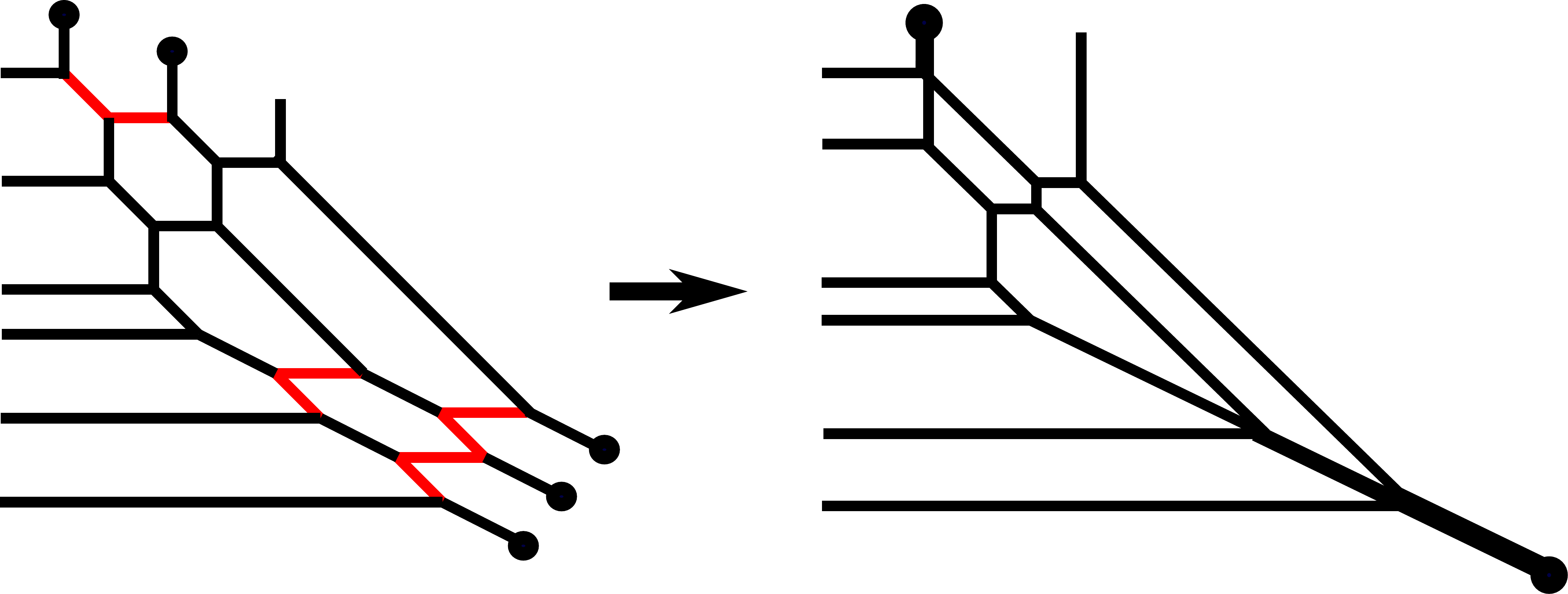}
\caption{Higgsing from a $[SU(6)_{\kappa}] - SU(4)_0 - SU(2)$ quiver theory to $[SU(6)_{\kappa + 1}] - [1{\bf AS}]$}
\label{fig:su6ashiggs}
\end{figure}

Note that the diagram in Figure \ref{fig:sp3higgs4} itself can be also understood from a Higgsing of a quiver theory involving an $SU(6)$ gauge node. The Higgsing is depicted in Figure \ref{fig:su6ashiggs}.  
The left diagram in Figure \ref{fig:su6ashiggs} has an $SU(3) \times SU(3)$ flavor symmetry associated to the external 5-branes. We can then partially Higgs the diagram by shrinking the lines in red of the left diagram in Figure \ref{fig:su6ashiggs}. The tuning opens up a Higgs branch which is related to the space of deformations of pieces of 5-branes between 7-branes. After decoupling the pieces of 5-branes we end up with a diagram at low energies and it is given by the right diagram in Figure \ref{fig:su6ashiggs} which is the same diagram as the one in Figure \ref{fig:sp3higgs4}. Then reading off the gauge theory content from the two diagrams in Figure \ref{fig:su6ashiggs} implies the following relation
\begin{align}
[SU(6)_{\kappa}] - SU(4)_0 - SU(2) \quad \xrightarrow{\text{Higgsing}} \quad [SU(6)_{\kappa + 1}] - \left[1{\bf AS}\right]. \label{HiggstoAS}
\end{align}

\begin{figure}[t]
\centering
\includegraphics[width=12cm]{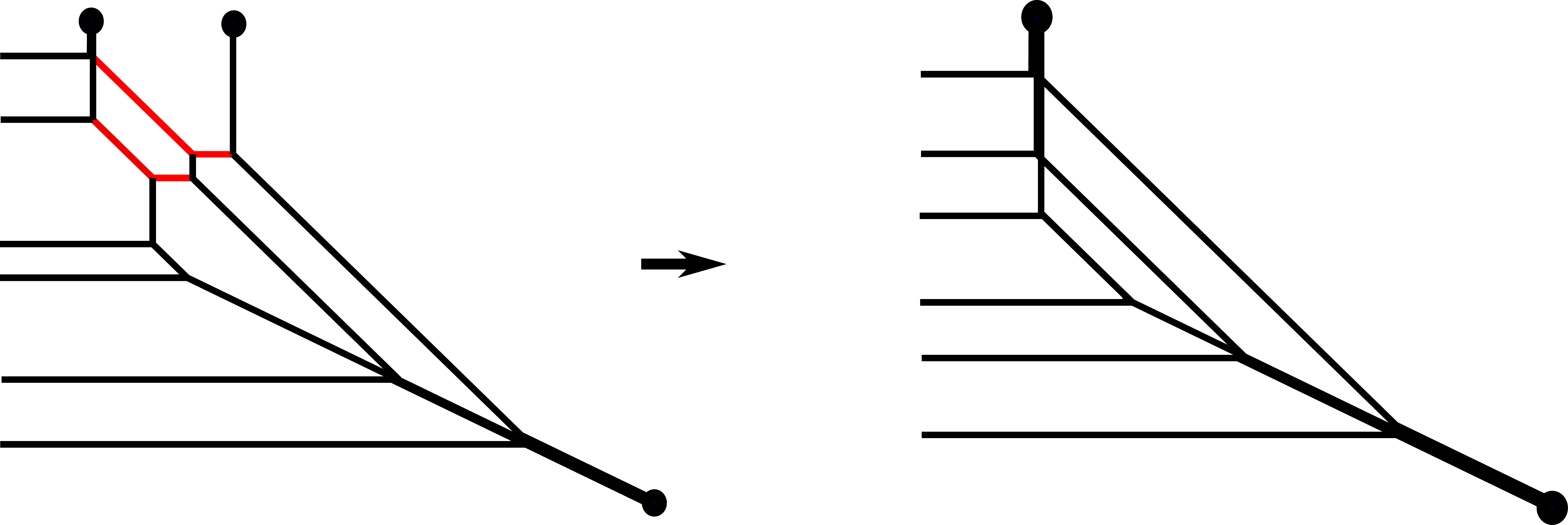}
\caption{Higgsing from $[SU(6)] - 1{\bf AS}$ to $[Sp(3)]$}
\label{fig:sp3higgs5}
\end{figure}
In order to perform the Higgsing associated to the antisymmetric hypermultiplet, we need to further tune the length of lines in Figure \ref{fig:sp3higgs4}. The lines which need to be shrunken are depicted in red in the left diagram in Figure \ref{fig:sp3higgs5}. The tuning opens up a Higgs branch and decoupling pieces of 5-branes yields the right diagram in Figure \ref{fig:sp3higgs5}. Hence, gauging the six horizontal D5-branes in the right diagram in Figure \ref{fig:sp3higgs5} gives rises to an $Sp(3)$ gauge theory.

\begin{figure}[t]
\centering
\subfigure[]{\label{fig:sp3w1htas}
\includegraphics[width=6cm]{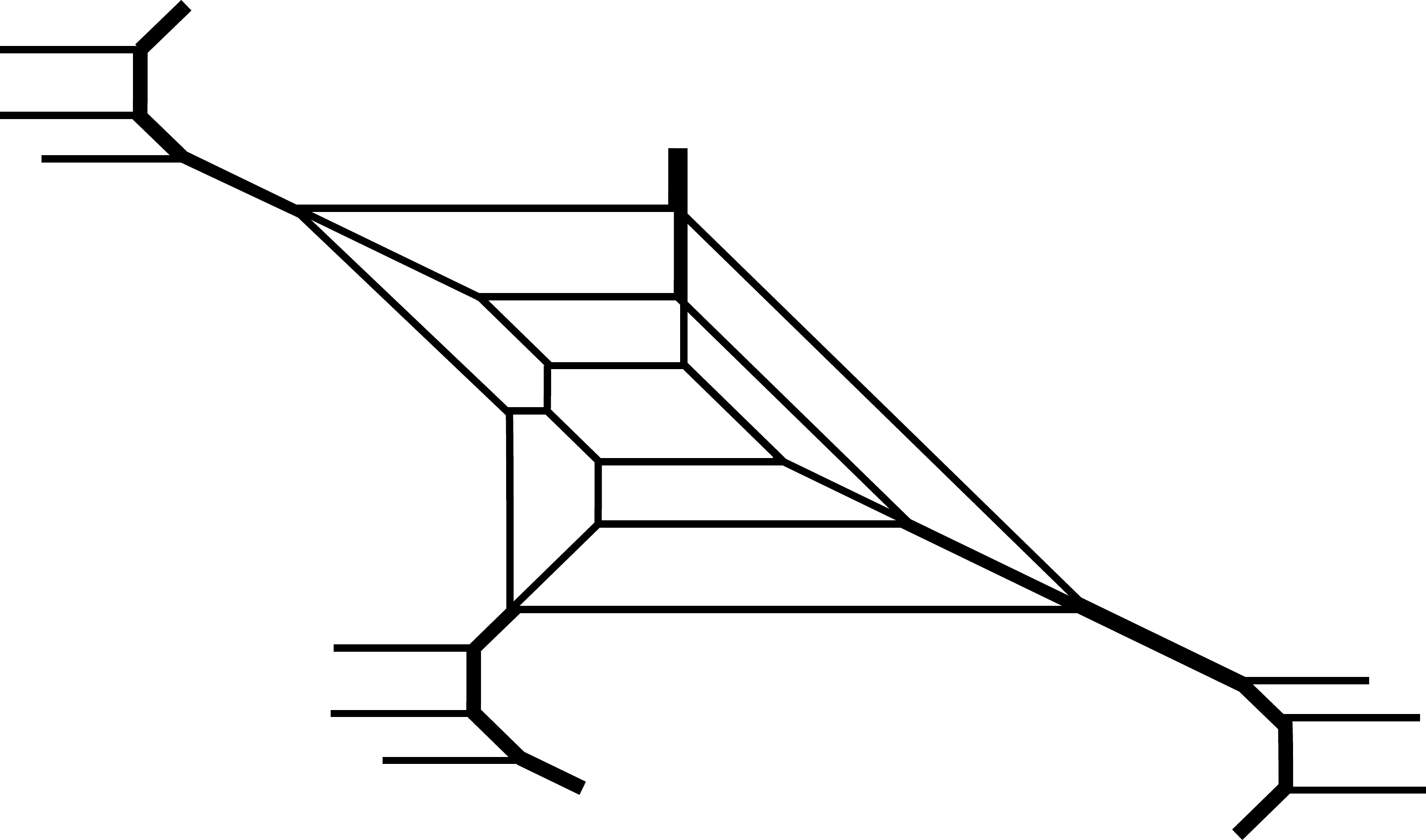}}\hspace{1cm}
\subfigure[]{\label{fig:sp3w1htas2}
\includegraphics[width=6cm]{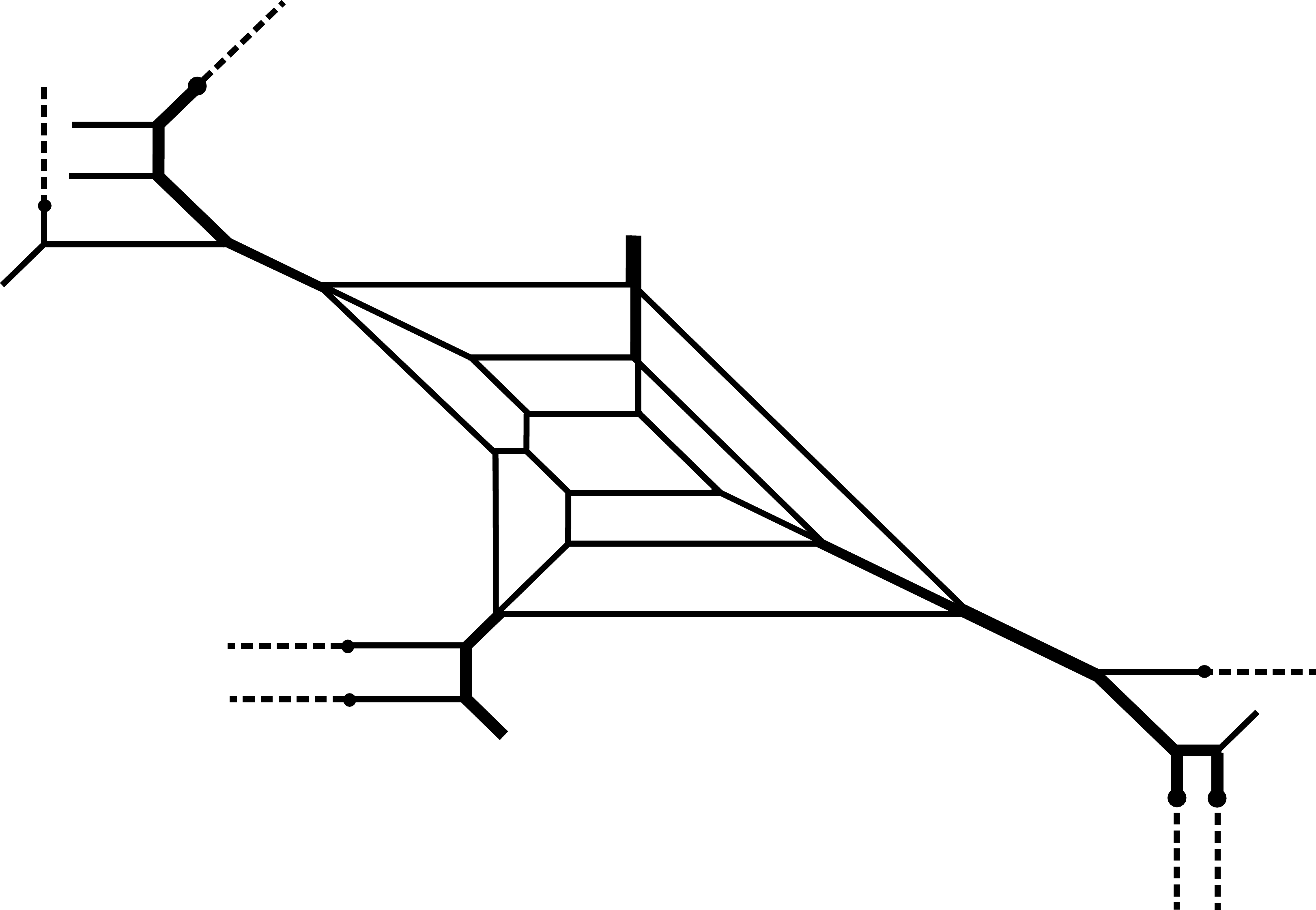}}
\caption{(a): A diagram for the $Sp(3)$ gauge theory with $N_{\bf TAS} = \frac{1}{2}, N_{\bf F} = \frac{19}{2}$ obtained by Higgsings the diagram in Figure \ref{fig:SU6+1over2TAS+1AS+9F}. (b): A Tao diagram from the diagram in Figure \ref{fig:sp3w1htas}.}
\label{fig:sp3w1htas0}
\end{figure}
Therefore, the Higgsing associated to the antisymmetric matter can be achieved diagrammatically by replacing the right part of the diagram in Figure \ref{fig:SU6+1over2TAS+1AS+9F} with the diagram in Figure \ref{fig:sp3higgs5}. Then we obtain a diagram for the $Sp(3)$ gauge theory with $N_{\bf TAS} = \frac{1}{2}, N_{\bf F} = \frac{19}{2}$ and it is given in Figure \ref{fig:sp3w1htas}. By moving 7-branes of the diagram in Figure \ref{fig:sp3w1htas}, it yields the diagram in Figure \ref{fig:sp3w1htas2} and the diagram in Figure \ref{fig:sp3w1htas2} shows that the diagram is a Tao diagram, implying that the the $Sp(3)$ gauge theory with $N_{\bf TAS} = \frac{1}{2}, N_{\bf F} = \frac{19}{2}$ has a 6d uplift. 


\begin{figure}[t]
\centering
\subfigure[]{\label{fig:sp3w1tas}
\includegraphics[width=6cm]{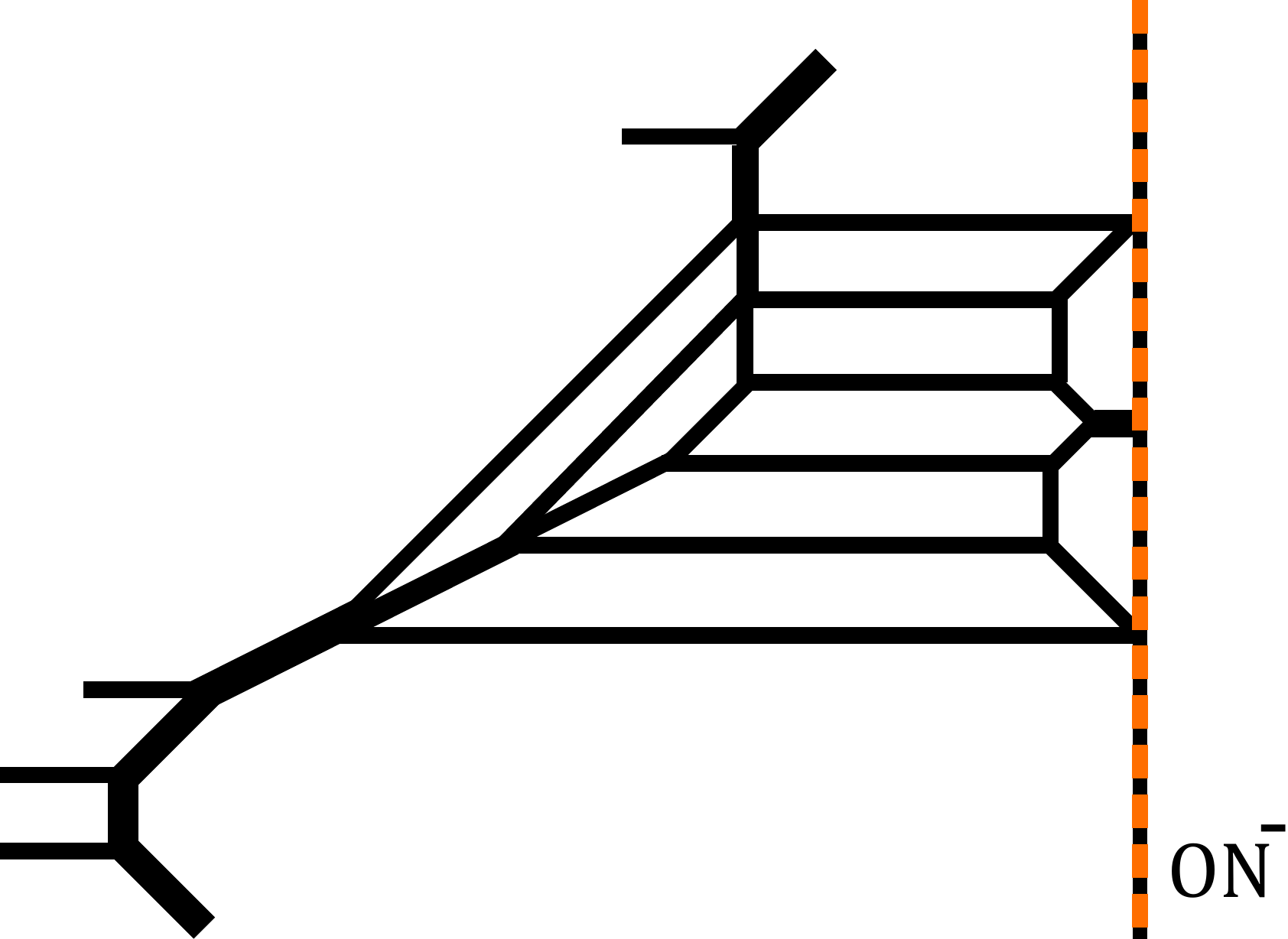}}\hspace{1cm}
\subfigure[]{\label{fig:sp3w1tas2}
\includegraphics[width=6cm]{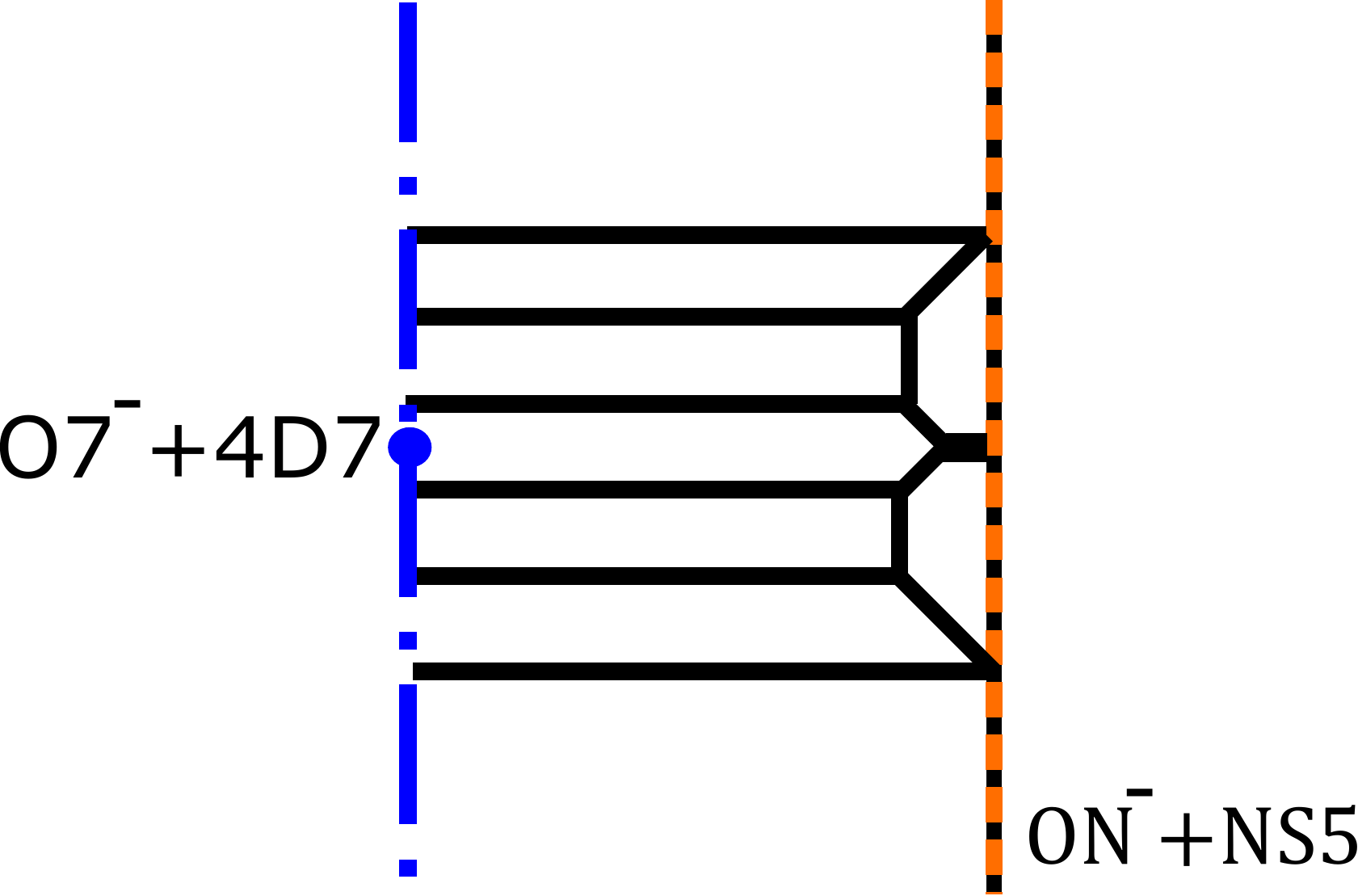}}\hspace{1cm}
\subfigure[]{\label{fig:sp3w1tas3}
\includegraphics[width=6cm]{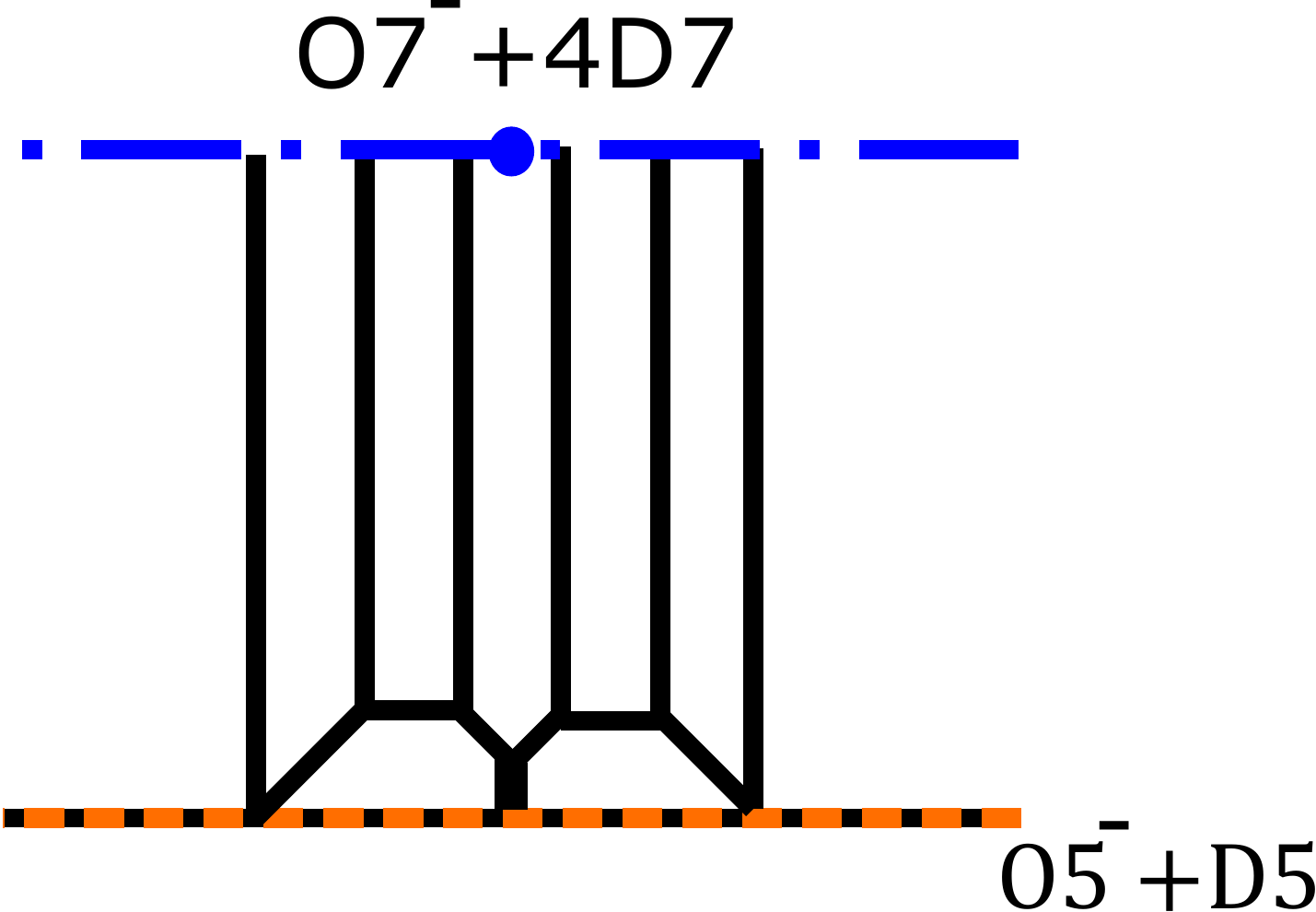}}
\caption{(a): A diagram for the $Sp(3)$ gauge theory with $N_{\bf TAS} = 1, N_{\bf F} = 5$ obtained by Higgsings the diagram in Figure \ref{fig:SU61AS4F}. (b): The diagram after forming an O7$^-$-plane from the diagram in Figure \ref{fig:sp3w1tas}. (c): S-dual of the diagram in Figure \ref{fig:sp3w1tas2}, exhibiting a periodic direction in the vertical direction.}
\label{fig:sp3w1tas0}
\end{figure}
\paragraph{\underline{$Sp(3)+1{\bf TAS}+5\bF$}} 
It is also possible to perform the same Higgsing to the diagram for the $SU(6)_0$ gauge theory with $N_{\bf TAS} = 1, N_{\bf AS} = 1, N_{\bf F} = 4$ in Figure \ref{fig:SU61AS4F}. By replacing the part involving the antisymmetric matter to the part giving the $Sp(3)$ gauge group which is given by the diagram in Figure \ref{fig:sp3higgs5}, we obtain a diagram depicted in Figure \ref{fig:sp3w1tas} for the $Sp(3)$ gauge theory with $N_{\bf TAS} = 1$ and $N_{\bf F} = 5$ where one of the five hypermultiplets in the fundamental representation as well as a hypermultiplet in the rank-3 antisymmetric representation are massless. We can confirm that the $Sp(3)$ gauge theory with $N_{\bf TAS} = 1$ and $N_{\bf F} = 5$ has a 6d UV completion from the diagram in Figure \ref{fig:sp3w1tas}. From the diagram in Figure \ref{fig:sp3w1tas}, we first move the $(1, -1)$ 7-brane and the $(1, 1)$ 7-brane inside the middle 5-brane loop together with the four flavor D7-branes. Then the $(1, -1)$ 7-brane and the $(1, 1)$ 7-brane form an O7$^-$-plane and the diagram becomes the one in Figure \ref{fig:sp3w1tas2}. The combination of an O7$^-$-plane and four D7-branes is S-dual invariant and hence after S-duality we obtain the diagram in Figure \ref{fig:sp3w1tas3} which has a pair of an O7$^-$-plane and an O5$^-$-plane, showing periodicity in the vertical direction. The appearance of the periodicity in the vertical direction implies that the $Sp(3)$ gauge theory with $N_{\bf TAS} = 1$ and $N_{\bf F} = 5$ has a 6d uplift.  

\bigskip
\section{Dualities and 6d uplift of marginal $SU(6)$ gauge theories with rank-3 antisymmetric matter} \label{sec:6d}

Since we have constructed 5-brane web diagrams which imply a 6d UV completion for the realized 5d theory on the web, it is natural to ask what is the 6d theory which completes 5d $SU(6)$ or $Sp(3)$ gauge theories with rank-3 antisymmetric matter at UV. In order to see the 6d uplift explicitly, we need to convert a 5-brane web diagram into some another configuration realizing a 6d theory. One way is to use T-duality and transform a 5-brane web into a brane configuration in type IIA string theory. In fact it is possible to convert the 5-brane web diagrams for the $SU(6)$ gauge theory with $N_{\bf TAS} = 1, N_{\bf F} = 10, \kappa = 0$, the $SU(6)$ gauge theory with $N_{\bf TAS} = \frac{1}{2}, N_{\bf F} = 13, \kappa = 0$ and the $SU(6)$ gauge theory with $N_{\bf TAS} = \frac{1}{2}, N_{\bf Sym} = 1, N_{\bf F} = 1, \kappa = 0$ into type IIA brane system. 

Furthermore, in order to see the 6d uplift of the $SU(6)_0$ gauge theory with $N_{\bf TAS} = \frac{1}{2}, N_{\bf F} = 13$, it will be useful to first go to a dual frame which is given by a 5d quiver theory by moving 7-branes. The 6d uplift of the quiver theory has been known in \cite{Zafrir:2015rga, Hayashi:2015zka, Ohmori:2015tka} and we can make use of the result to see the UV completion of the the $SU(6)_0$ gauge theory with $N_{\bf TAS} = \frac{1}{2}, N_{\bf F} = 13$. Similar deformations by moving 7-branes will yield other dualities from $SU(6)$ gauge theories with a half-hypermultiplet in the rank-3 antisymmetric representation and other matter.

\subsection{6d uplift of $SU(6)_0 + 1{\bf TAS} + 10{\bf F}$}

\label{sec:SU6w1TAS10F}

\begin{figure}[t]
\centering
\subfigure[]{\label{fig:su6w1taso7}
\includegraphics[width=6cm]{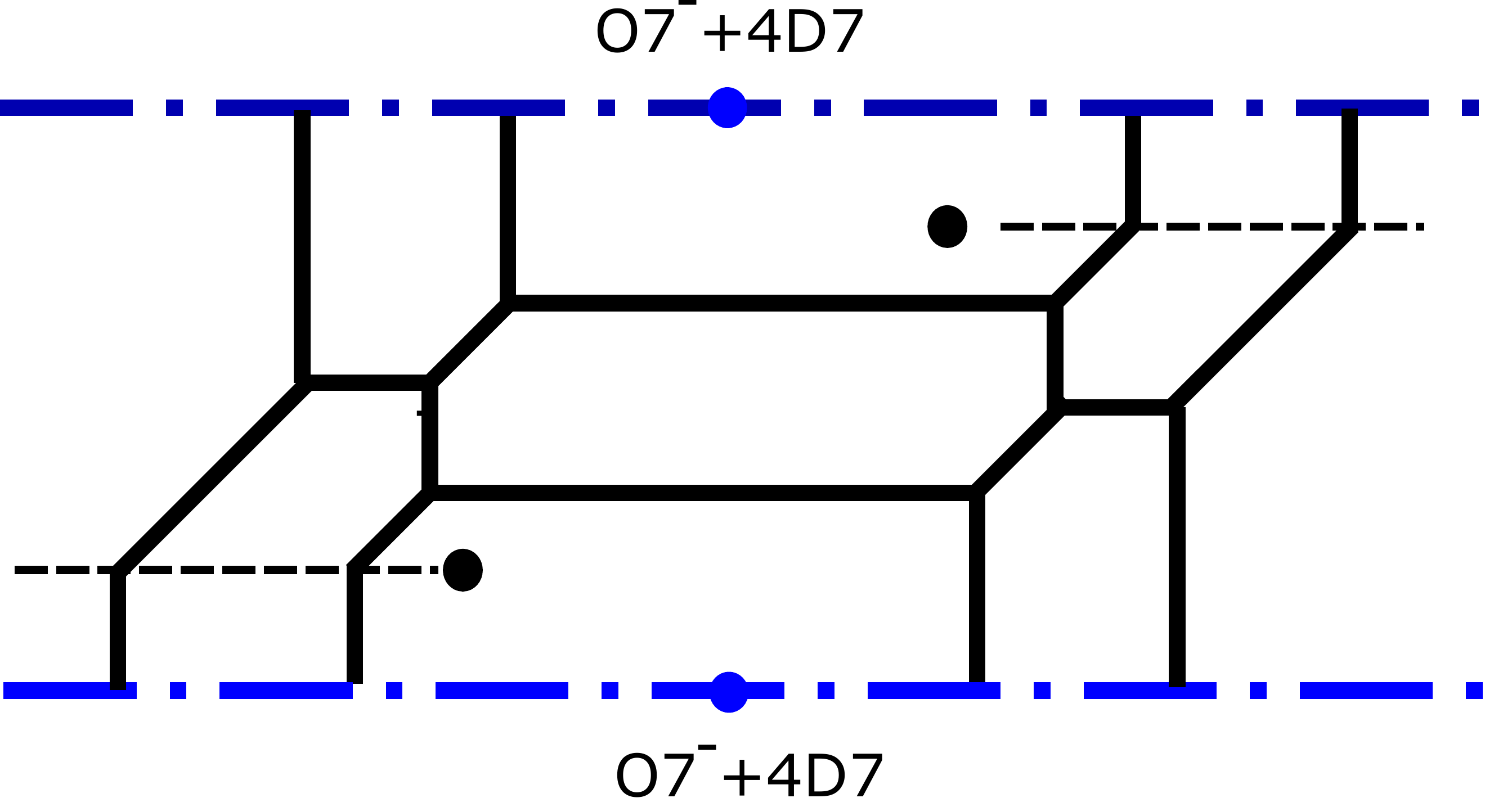}}\hspace{1cm}
\subfigure[]{\label{fig:su6w1tas6d}
\includegraphics[width=6cm]{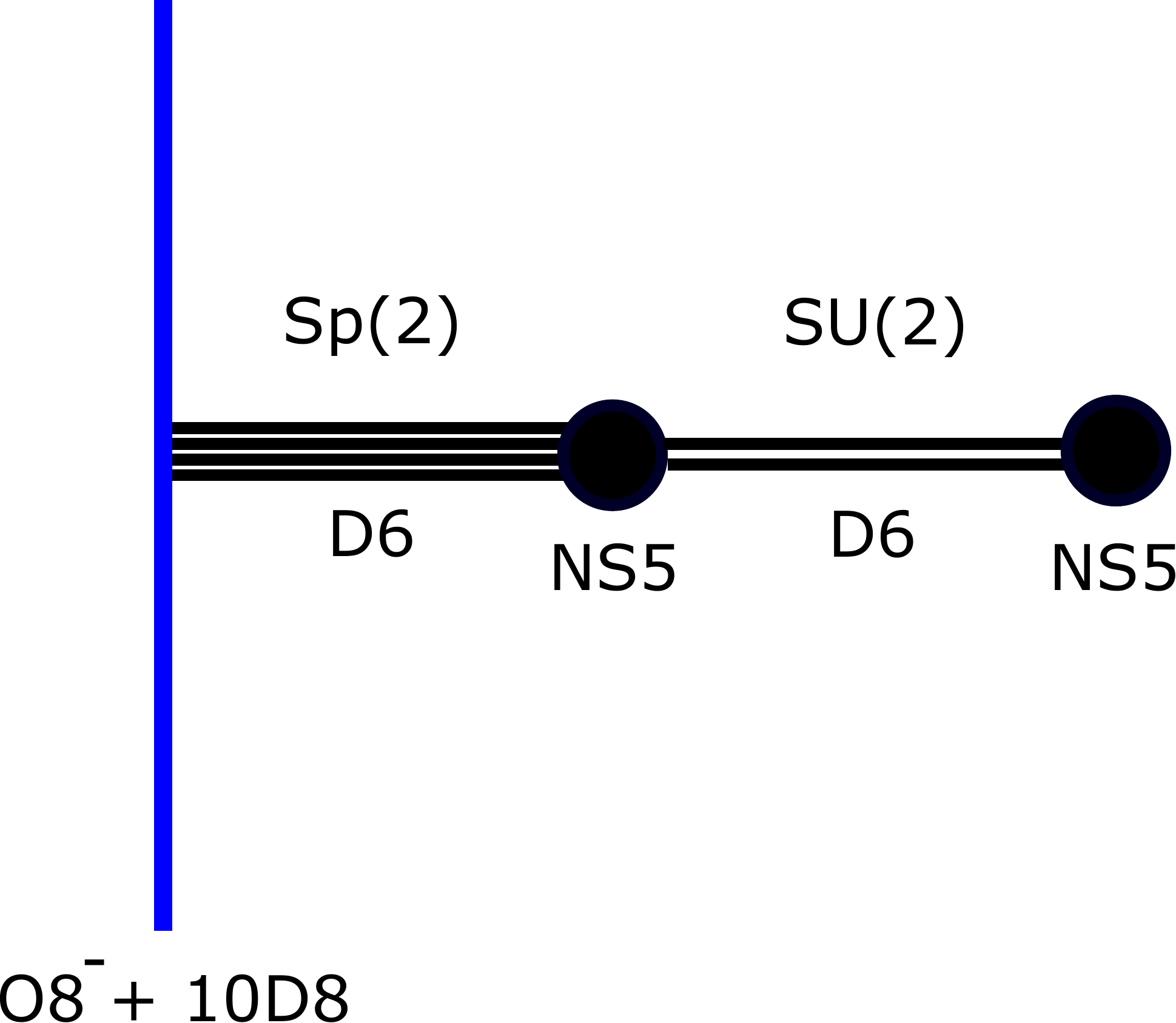}}
\caption{(a): Another diagram for the $SU(6)_0$ aguage theory with $N_{\bf TAS} = 1, N_{\bf F} = 10$ with two O7$^-$-planes. (b): A type IIA brane configuration after applying T-duality along the vertical direction in Figure \ref{fig:su6w1taso7}.}
\label{fig:su6w1tasuplift}
\end{figure}
We start from the 5-brane web diagram in Figure \ref{fig:SU6+1TAS+10F} which realizes the $SU(6)$ gauge theory with $N_{\bf TAS} = 1, N_{\bf F} = 10, \kappa = 0$. In section \ref{sec:SU6w1TAS} we have seen that the $SU(6)_0$ gauge theory with  $N_{\bf TAS} = 1, N_{\bf F} = 10$ has a 6d UV completion since it can be written as a Tao diagram which is given by the right diagram in Figure \ref{fig:SU6+1TAS+10F}. Another way to see that the theory has a 6d UV completion is to form a pair of O7$^-$-planes placed in the vertical direction. For that after flop transitions we move the $(1, 1)$ 7-brane and the $(1, -1)$ 7-brane in the upper part and in the lower part inside 5-brane loops. Then each pair of the $(1, 1)$ 7-brane and the $(1, -1)$ 7-brane are put into the same 5-brane loops and two O7$^-$-planes are formed as in Figure \ref{fig:su6w1taso7}. With the two O7$^-$-planes separated in the vertical direction, we can apply T-duality along the vertical direction which convert a pair of O7$^-$-planes into an O8$^-$-plane in type IIA string theory. Similarly a D5-brane becomes a D6-brane and an NS5-brane still remains to be an NS5-brane in type IIA string theory. Then the 5-brane web diagram in Figure \ref{fig:su6w1taso7} is transformed into the one in Figure \ref{fig:su6w1tas6d}. It is straightforward to read off the gauge theory content from the brane system in Figure \ref{fig:su6w1tas6d} and it is an $Sp(2) - SU(2)$ quiver theory where ten flavors are coupled to the $Sp(2)$ gauge group. Namely the brane configuration implies the following UV completion
\begin{align}
5d\quad [1 {\bf TAS}] - SU(6)_0 - [10 {\bf F}] \quad \xrightarrow{\text{UV completion}} \quad 6d\quad [10 {\bf F}] - Sp(2) - SU(2), \label{1tasUV}
\end{align}
for the $SU(6)_0$ gauge theory with  $N_{\bf TAS} = 1, N_{\bf F} = 10$.

We can further support the 6d uplift \eqref{1tasUV} by counting the number of the parameters from both sides. We compactify the 6d theory on a circle and turn on holonomies for the $SO(20)$ flavor symmetry. Hence we have ten parameters in addition to the radius of the circle, which gives eleven parameters in total. On the other hand the 5d theory has ten mass parameters for the ten flavors and also there is a gauge coupling for the $SU(6)$ gauge theory. Therefore we have also eleven parameters in 5d, which matches with the number of the parameters obtained by a circle compactification of the 6d theory. Note that the rank-3 antisymmetric hypermultiplet of the $SU(6)$ gauge theory is massless and there is no mass parameter for the rank-3 antisymmetric matter. Let us also see the matching of the number of Coulomb branch moduli. After a circle compactification a 6d tensor multiplet becomes a 5d vector multiplet. Hence two tensor multiplets in addition to the Cartan part for the $Sp(2)$ and $SU(2)$ vector multiplets yield $2 + 2 + 1 = 5$ dimensional Coulomb branch moduli space in five dimensions. This agrees with the five Coulomb branch moduli of the $SU(6)$ gauge theory.

\subsection{Dualities and 6d uplift of $SU(6)_0 + \frac{1}{2}{\bf TAS} + 13{\bf F}$}
\label{sec:6dHTAS13F}

\begin{figure}[t]
\centering
\subfigure[]{\label{fig:su6w1htas13f}
\includegraphics[width=6cm]{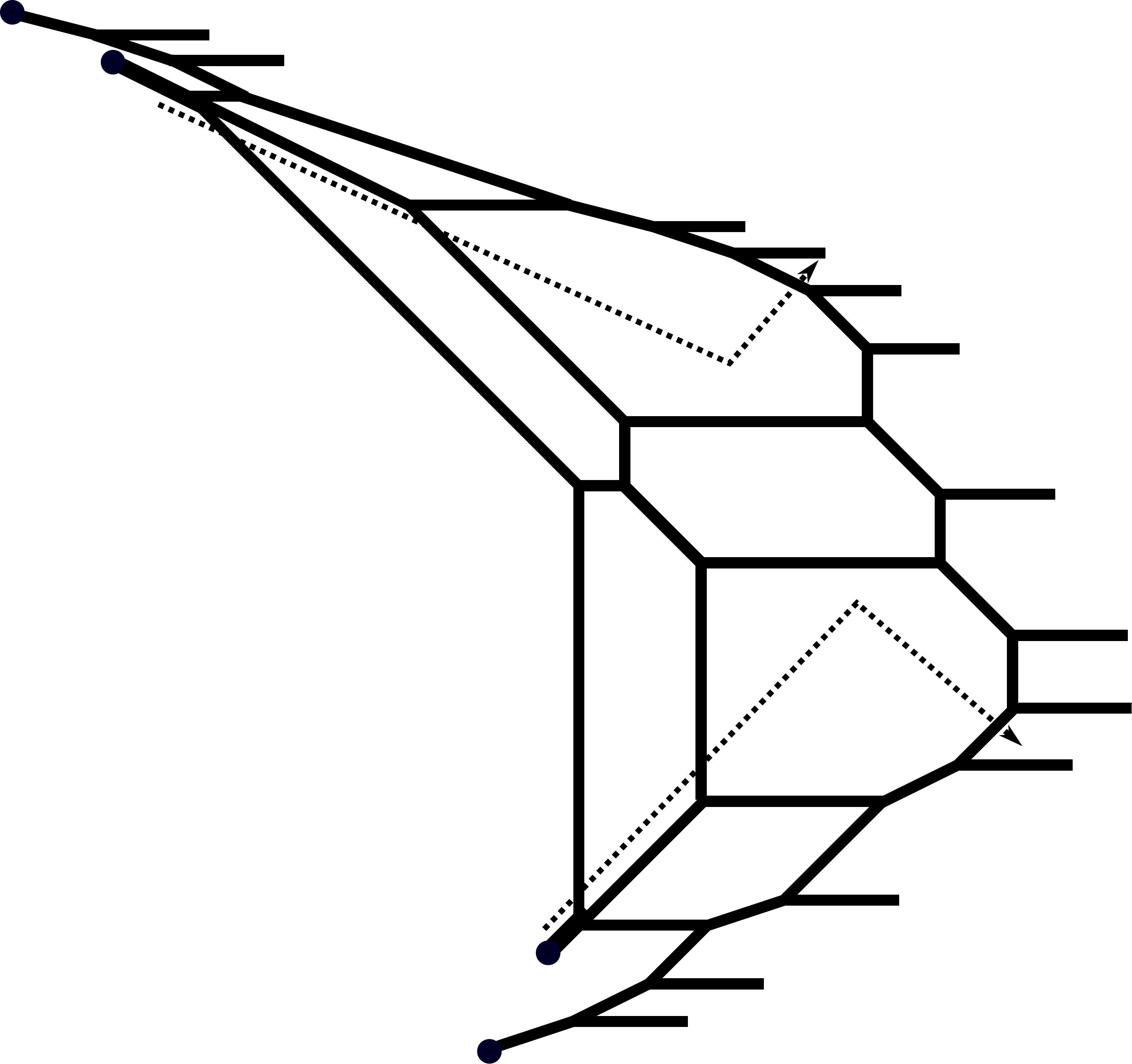}}\hspace{1cm}
\subfigure[]{\label{fig:su6w1htas13f2}
\includegraphics[width=6cm]{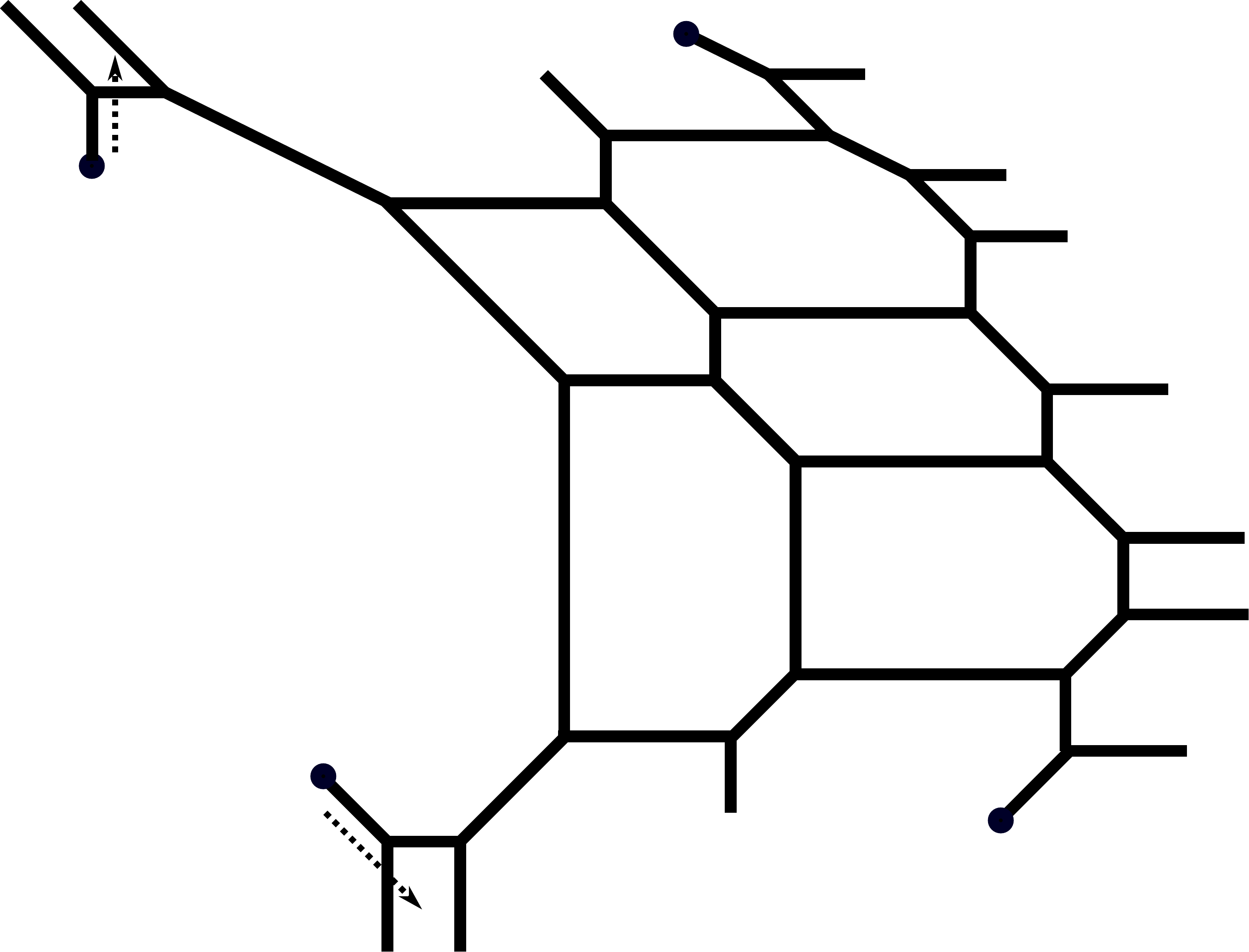}} \hspace{1cm}
\subfigure[]{\label{fig:su6w1htas13f3}
\includegraphics[width=6cm]{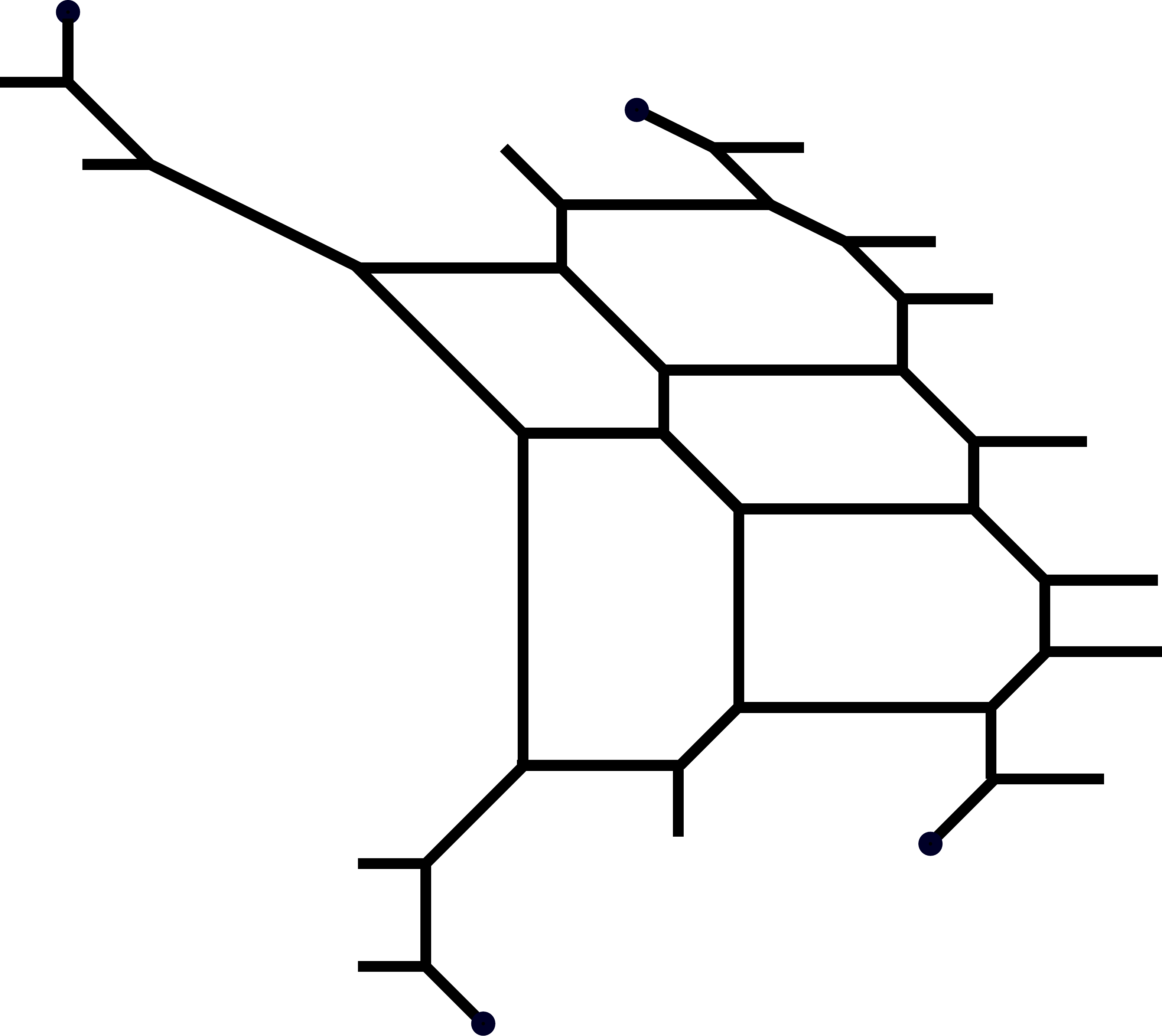}}
\caption{(a): A 5-brane web diagram for the $SU(6)$ gauge theory with $N_{\bf TAS} = \frac{1}{2}, N_{\bf F} = 13, \kappa = 0$. (b): The diagram after moving the $(2, -1)$ 7-brane and the $(1, 1)$ 7-brane along the arrows in Figure \ref{fig:su6w1htas13f}. (c): A 5-brane diagram for the $[4] - SU(3)_0 - SU(4)_0 - [7]$ quiver theory which is obtained from moving the $(0, 1)$ 7-brane and the $(1, -1)$ 7-brane in Figure \ref{fig:su6w1htas13f2}. }
\label{fig:su6w1htasuplift}
\end{figure}
\begin{figure}[t]
\centering
\includegraphics[width=4cm]{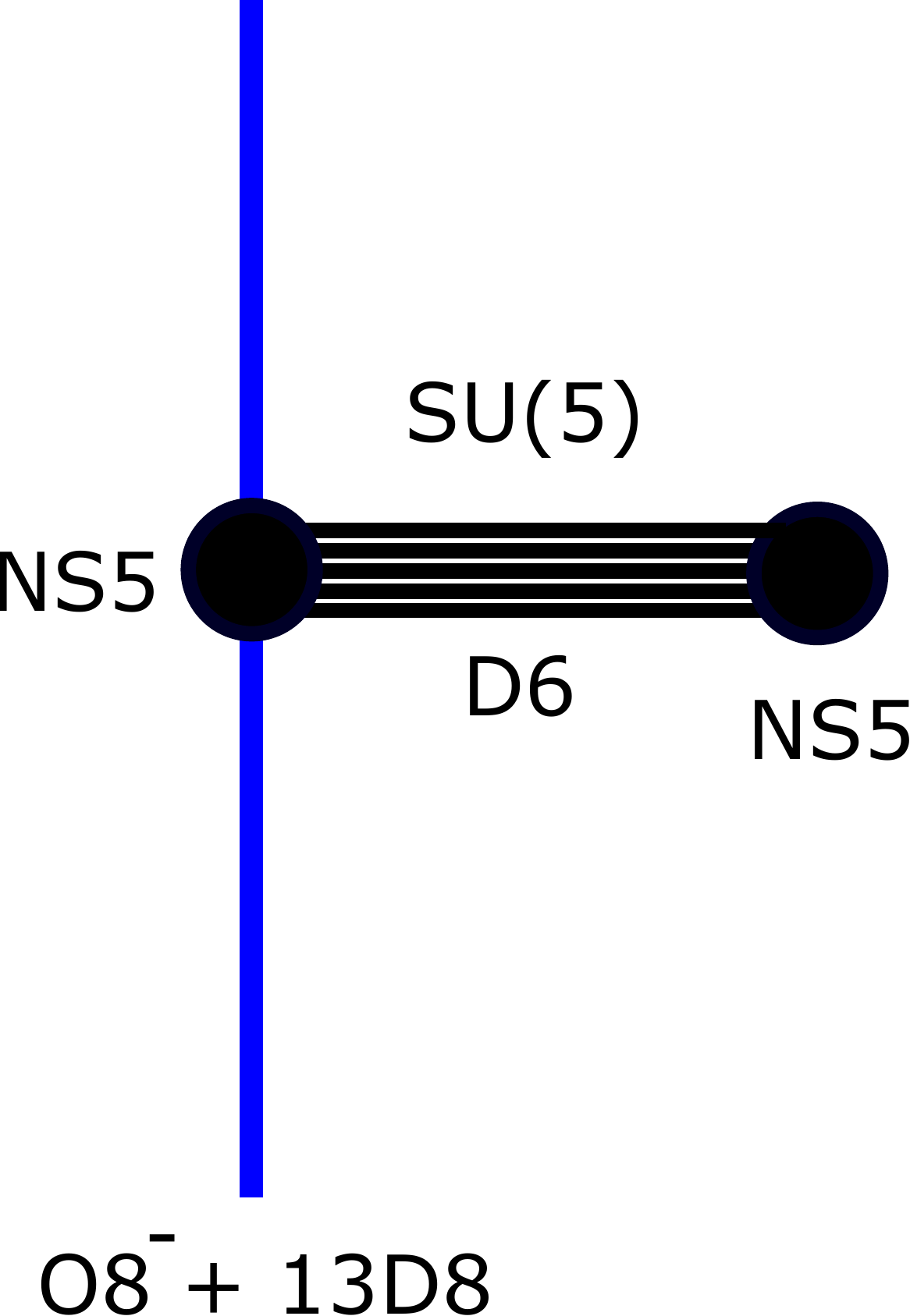}
\caption{Type IIA brane configuration for the $SU(5)$ gauge theory with $N_{\bf AS} = 1, N_{\bf F} = 13$. }
\label{fig:su6w1htas6d}
\end{figure}

We then consider the $SU(6)$ gauge theory with $N_{\bf TAS} = \frac{1}{2}, N_{\bf F} = 13, \kappa = 0$. The 5-brane diagram is given in Figure \ref{fig:SU6+1over2TAS+13F}. From the diagram in Figure \ref{fig:SU6+1over2TAS+13F} we move D7-branes and perform flop transitions to arrive at the diagram in Figure \ref{fig:su6w1htas13f}. From the diagram in Figure \ref{fig:su6w1htas13f}, we move the $(2, 1)$ 7-brane and the $(1, -1)$ 7-brane along the arrows specified in Figure \ref{fig:su6w1htas13f}. Then the diagram becomes the one in Figure \ref{fig:su6w1htas13f2}. We further move the $(0, 1)$ 7-brand and the $(1, 1)$ 7-brane in the diagram along the arrows depicted in Figure \ref{fig:su6w1htas13f2}. The resulting diagram after the movement of the 7-branes is given in Figure \ref{fig:su6w1htas13f3}. Then diagram in Figure \ref{fig:su6w1htas13f3} is nothing but a diagram for the $[4] - SU(3)_0 - SU(4)_0 - [7]$ quiver theory. The 6d UV completion of the quiver theory has been discussed in \cite{Zafrir:2015rga, Hayashi:2015zka, Ohmori:2015tka} from manipulation of the 5-brane web and the 6d uplift is given by the $SU(5)$ gauge theory with $N_{\bf F} = 13, N_{\bf AS} = 1$. We can form a pair of O7$^-$-planes from the diagram in Figure \ref{fig:su6w1htas13f3} and T-duality yields the type IIA brane system in Figure \ref{fig:su6w1htas6d}.
Namely the brane configuration implies the following UV completion
\begin{align}
5d\quad \left[\frac{1}{2} {\bf TAS}\right] - SU(6)_0 - [13 {\bf F}] \quad \xrightarrow{\text{UV completion}} \quad 6d\quad [13 {\bf F}] - SU(5) - [1 {\bf AS}], \label{1htasUV}
\end{align}
for the $SU(6)_0$ gauge theory with  $N_{\bf TAS} = \frac{1}{2}, N_{\bf F} = 13$. We can also see that the number of the mass parameters and the Coulomb branch moduli from a circle compactification of the 6d theory agrees with the number of the mass parameters and the Coulomb branch moduli of the 5d theory. 

Furthermore, since we can deform the diagram of the $SU(6)$ gauge theory with $N_{\bf TAS} = \frac{1}{2}, N_{\bf F} = 13, \kappa = 0$ in Figure \ref{fig:su6w1htas13f} to the diagram of the $[4] - SU(3)_0 - SU(4)_0 - [7]$ quiver theory in Figure \ref{fig:su6w1htas13f3}, the two 5d theories are dual to each other. From the 6d uplift given by \eqref{1htasUV} we can also obtain various other dual 5d theories \cite{Hayashi:2015zka}. Namely the $SU(6)$ gauge theory with $N_{\bf TAS} = \frac{1}{2}, N_{\bf F} = 13, \kappa = 0$ is dual to 
\begin{align}
 [(3n-2){\bf F}]-  SU(n+1)_0 - SU(6-n)_0 - [(13 - 3n){\bf F}], \qquad (n= 1, 2), \label{HTAS13F.quiver1}
\end{align}
and it is also dual to
\begin{align}
[1 {\bf AS}] - SU(6)_0 - [12{\bf F}].
\end{align}

We can also see a relation to the 6d $SU(6)$ gauge theory with $N_{\bf TAS} = \frac{1}{2} , N_{\bf F} = 15$. Note that applying a 5d limit to the 6d $SU(6)$ gauge theory with $N_{\bf TAS} = \frac{1}{2} , N_{\bf F} = 15$ will yield a 5d $SU(6)$ gauge theory with $N_{\bf TAS} = \frac{1}{2}$ and some flavors. A 5d limit may be achieved by decoupling some Coulomb branch moduli of a 5d theory whose UV completion is given by the 6d $SU(6)$ gauge theory with $N_{\bf TAS} = \frac{1}{2} , N_{\bf F} = 15$. On the other hand, it has been proposed that the UV completion of the 5d $[7] - SU(4)_0 - SU(3)_0 - SU(2) - [4]$ quiver theory is the 6d $SU(6)$ gauge theory with $N_{\bf TAS} = \frac{1}{2} , N_{\bf F} = 15$ \cite{Zafrir:2015rga, Hayashi:2015zka}. Then decoupling the Coulomb branch modulus for the $SU(2)$ gauge theory of the 5d $[7] - SU(4)_0 - SU(3)_0 - SU(2) - [4]$ quiver theory gives another quiver given by $[7] - SU(4)_0 - SU(3)_0 -[2]$. We can also obtain the same quiver theory by decoupling the flavors of the $[7] - SU(4)_0 - SU(3)_0 - [4]$ quiver theory which is dual to the 5d $SU(6)_0$ gauge theory with $N_{\bf TAS} = \frac{1}{2} , N_{\bf F} = 13$. By doing a similar deformation to the one through Figure \ref{fig:su6w1htas13f}-\ref{fig:su6w1htas13f3}, it is possible to see that the $[7] - SU(4)_0 - SU(3)_0 -[2]$ quiver theory is dual to the 5d $SU(6)_0$ gauge theory with $N_{\bf TAS} = \frac{1}{2} , N_{\bf F} = 11$. Hence a 5d limit of the 6d $SU(6)$ gauge theory with $N_{\bf TAS} = \frac{1}{2} , N_{\bf F} = 15$ indeed yields the 5d $SU(6)_0$ gauge theory with $N_{\bf TAS} = \frac{1}{2} , N_{\bf F} = 11$. 

\subsection{Other dualities involving marginal $SU(6)$ gauge theories with $N_{\bf TAS} = \frac{1}{2}$}

In section \ref{sec:6dHTAS13F}, we have seen that a deformation of a 5-brane web implies that the $SU(6)$ gauge theory with $N_{\bf TAS} = \frac{1}{2}, N_{\bf F} = 13, \kappa = 0$ is dual to quiver theories given by \eqref{HTAS13F.quiver1}. A similar deformation of 5-brane webs of $SU(6)$ gauge theories with $N_{\bf TAS} = \frac{1}{2}$ can lead to other dualities. 

\paragraph{\underline{$SU(6)_{-3} + \frac{1}{2}{\bf TAS} + 9{\bf F}$}}
\begin{figure}[t]
\centering
\subfigure[]{\label{fig:su6w1htas9f}
\includegraphics[width=6cm]{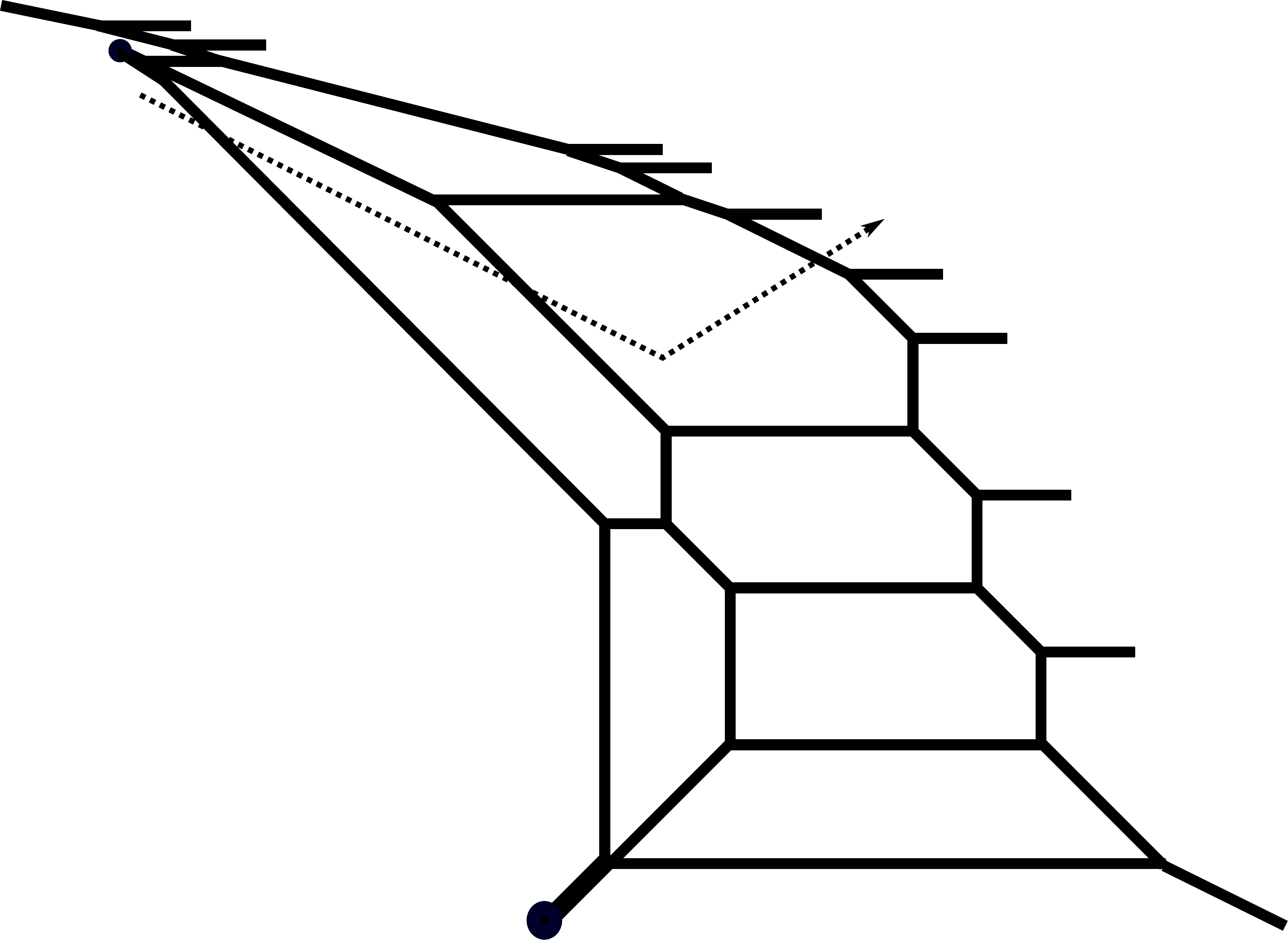}}\hspace{1cm}
\subfigure[]{\label{fig:su6w1htas9f2}
\includegraphics[width=6cm]{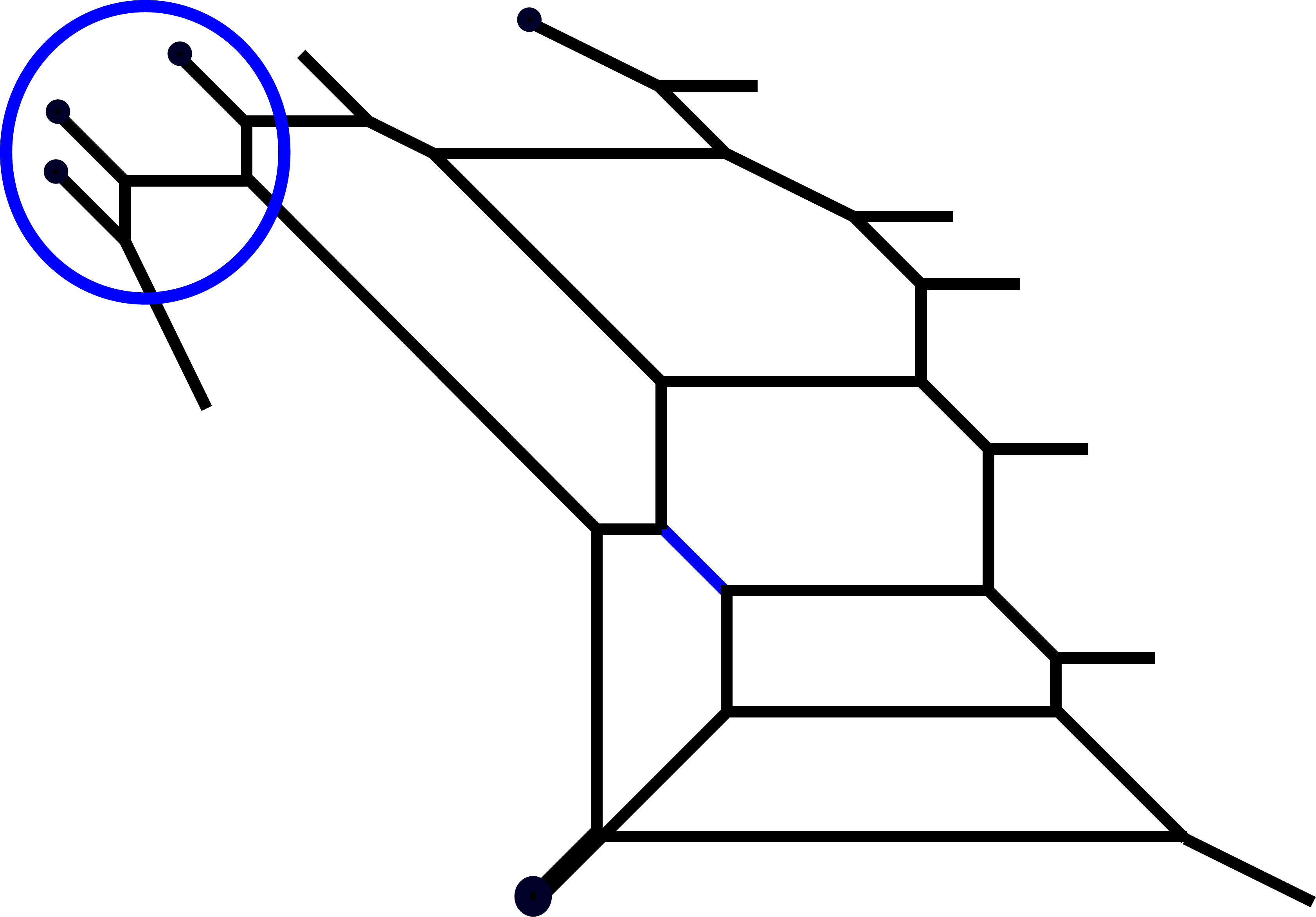}} \hspace{1cm}
\subfigure[]{\label{fig:su6w1htas9f3}
\includegraphics[width=6cm]{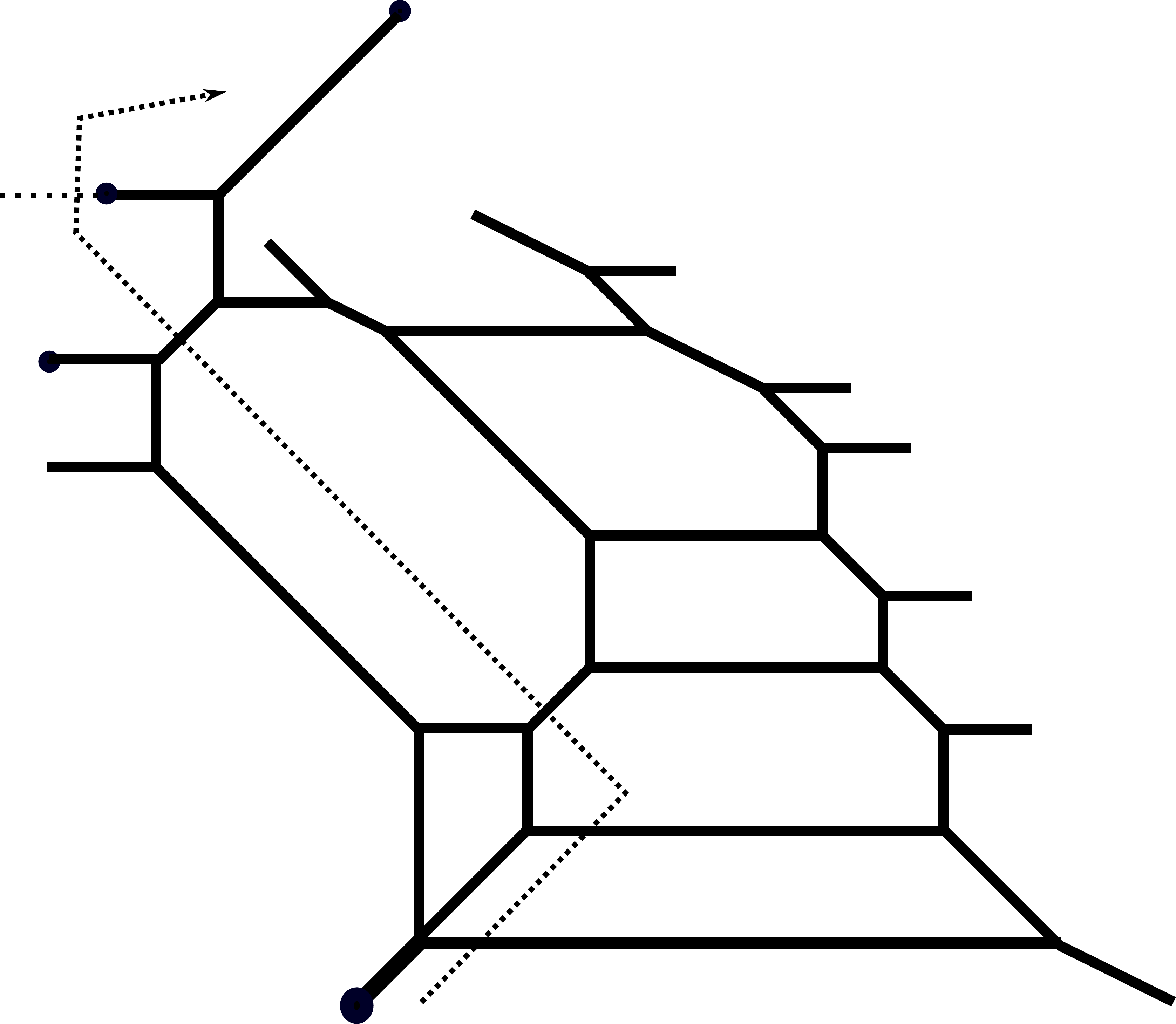}}\hspace{1cm}
\subfigure[]{\label{fig:su6w1htas9f4}
\includegraphics[width=6cm]{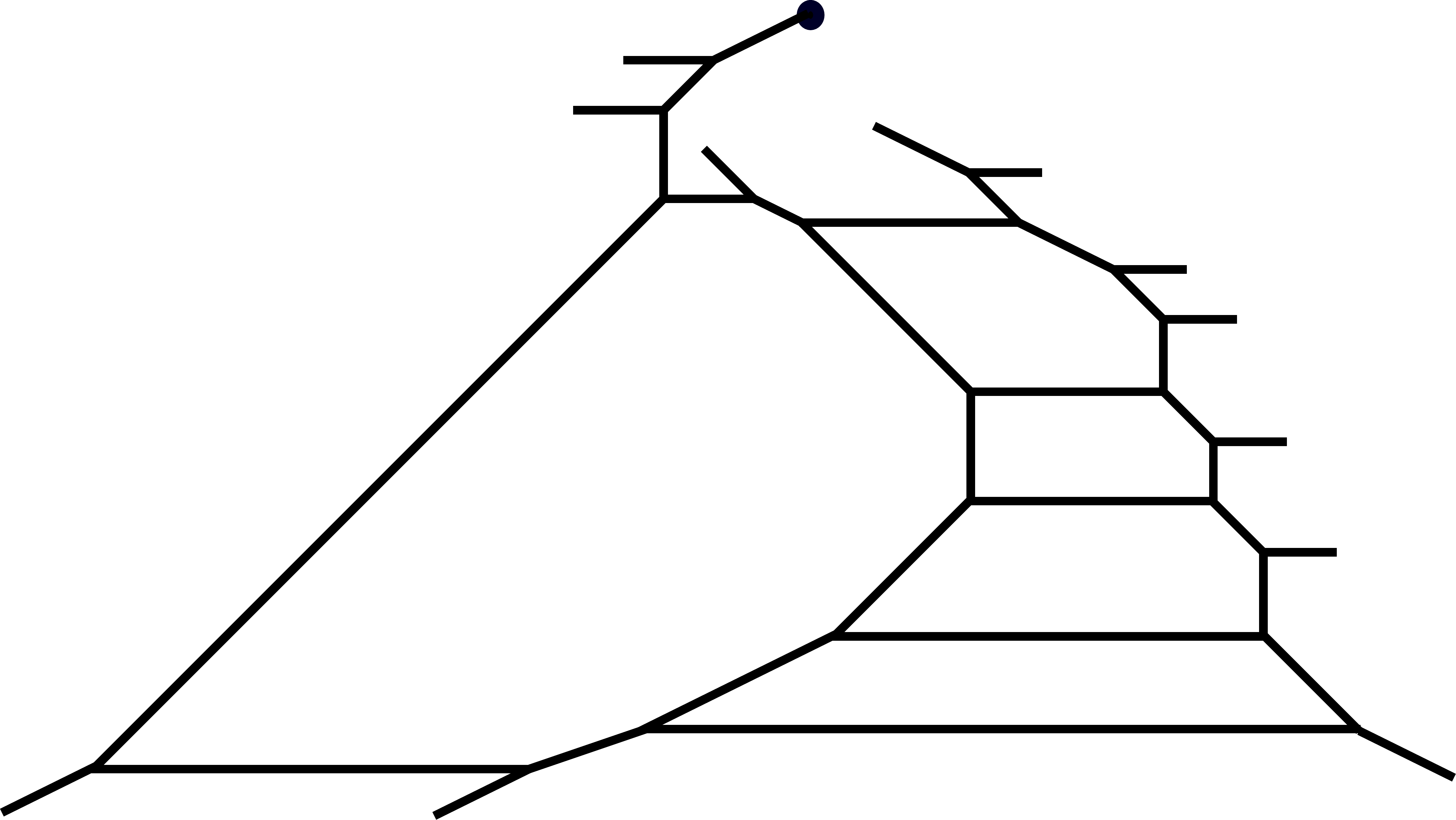}}
\caption{(a): A 5-brane web diagram for the $SU(6)$ gauge theory with $N_{\bf TAS} = \frac{1}{2}, N_{\bf F} = 9, \kappa = -3$. (b): The diagram after moving the $(2, -1)$ 7-brane along the arrow in Figure \ref{fig:su6w1htas9f}. (c): The diagram after we flop the line in blue and also move 7-branes in the blue circle in the diagram in Figure \ref{fig:su6w1htas9f2}. (d): The diagram of the quiver $[2{\bf F}] - SU(2) - SU(5)_{-\frac{5}{2}} - [5{\bf F}]$ obtained by moving the $(1, -1)$ 7-brane along the arrow in Figure \ref{fig:su6w1htas9f3}.}
\label{fig:su6w1htas9fdef}
\end{figure}

We first consider the 5-brane diagram of the $SU(6)_{-3}$ gauge theory with $N_{\bf TAS} = \frac{1}{2}$ and $N_{\bf F} = 9$ which is given in Figure \ref{fig:SU6+1over2TAS+9F}. From the diagram in Figure \ref{fig:SU6+1over2TAS+9F} moving some of the flavor D7-branes lead to the diagram in Figure \ref{fig:su6w1htas9f}. In the diagram in Figure \ref{fig:su6w1htas9f}, we move the $(2, -1)$ 7-brane along the arrow and the diagram becomes the one in Figure \ref{fig:su6w1htas9f2}. The moved $(2, -1)$ 7-brane is now attached at the end of the external $(2, -1)$ 5-brane in the upper right part of the diagram. From the diagram in Figure \ref{fig:su6w1htas9f2}, we flop the line in blue and also move 7-branes in the blue circle to obtain the diagram in Figure \ref{fig:su6w1htas9f3}. Finally, moving the $(1, 1)$ 7-brane along the arrow in Figure \ref{fig:su6w1htas9f3} yields the diagram in Figure \ref{fig:su6w1htas9f4}, which can be interpreted as a diagram of the $[2{\bf F}] - SU(2) - SU(5)_{-\frac{5}{2}} - [5{\bf F}]$ quiver theory. Namely, the $SU(6)_{-3}$ gauge theory with $N_{\bf TAS} = \frac{1}{2}, N_{\bf F} = 9$ is dual to the $[2{\bf F}] - SU(2) - SU(5)_{-\frac{5}{2}} - [5{\bf F}]$. From the quiver theory moving 7-branes can lead to another dual quiver theory such as $[3{\bf F}] - SU(2) - \left[SU(3)_{\frac{1}{2}} - [2{\bf F}]\right] - SU(3)_{-\frac{5}{2}}$. 

\paragraph{\underline{$SU(6)_{0} + \frac{1}{2}{\bf TAS} + 1{\bf AS} + 9{\bf F}$}}
\begin{figure}[t]
\centering
\subfigure[]{\label{fig:su6w1htas1as9f}
\includegraphics[width=6cm]{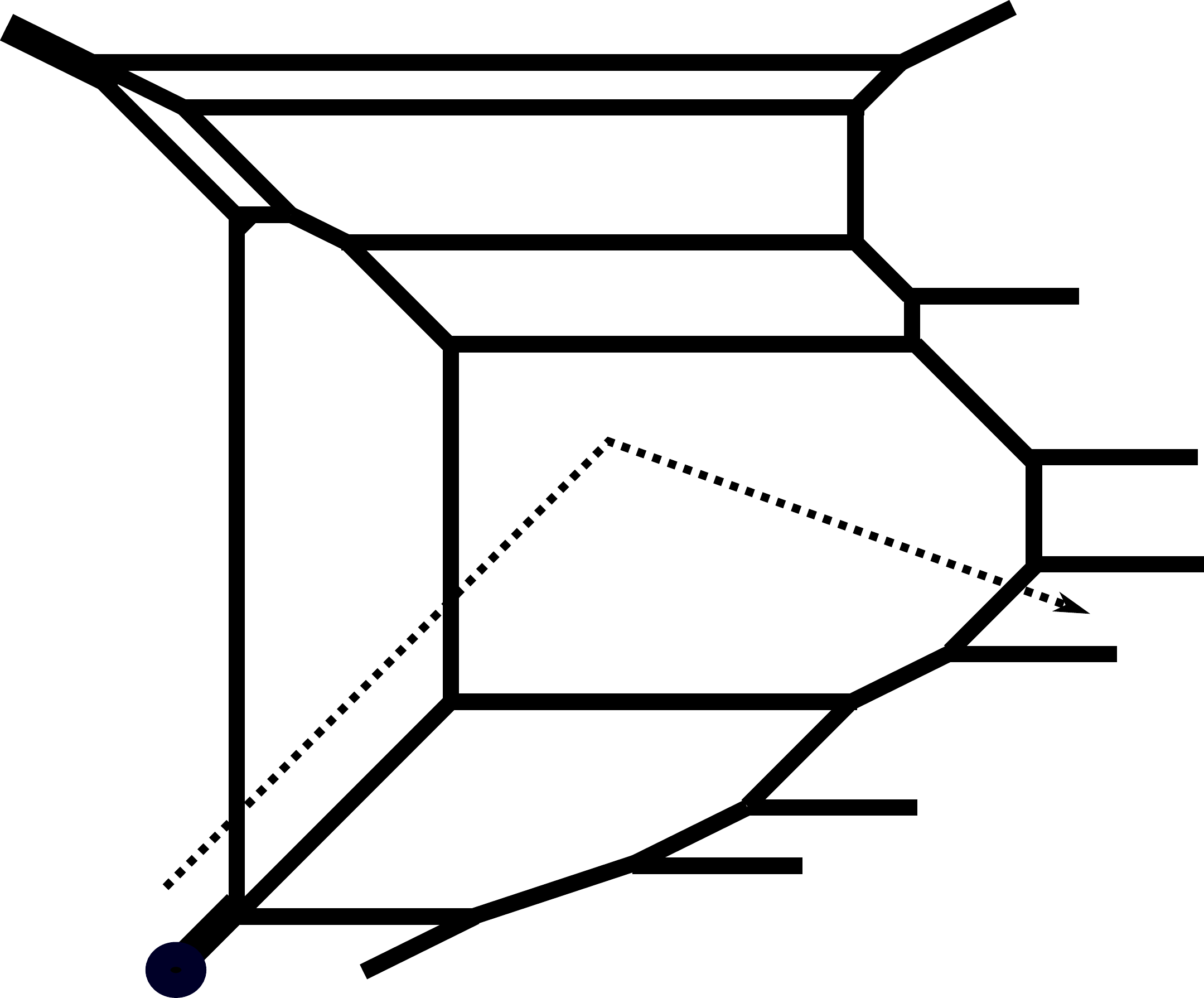}}\hspace{1cm}
\subfigure[]{\label{fig:su6w1htas1as9f2}
\includegraphics[width=6cm]{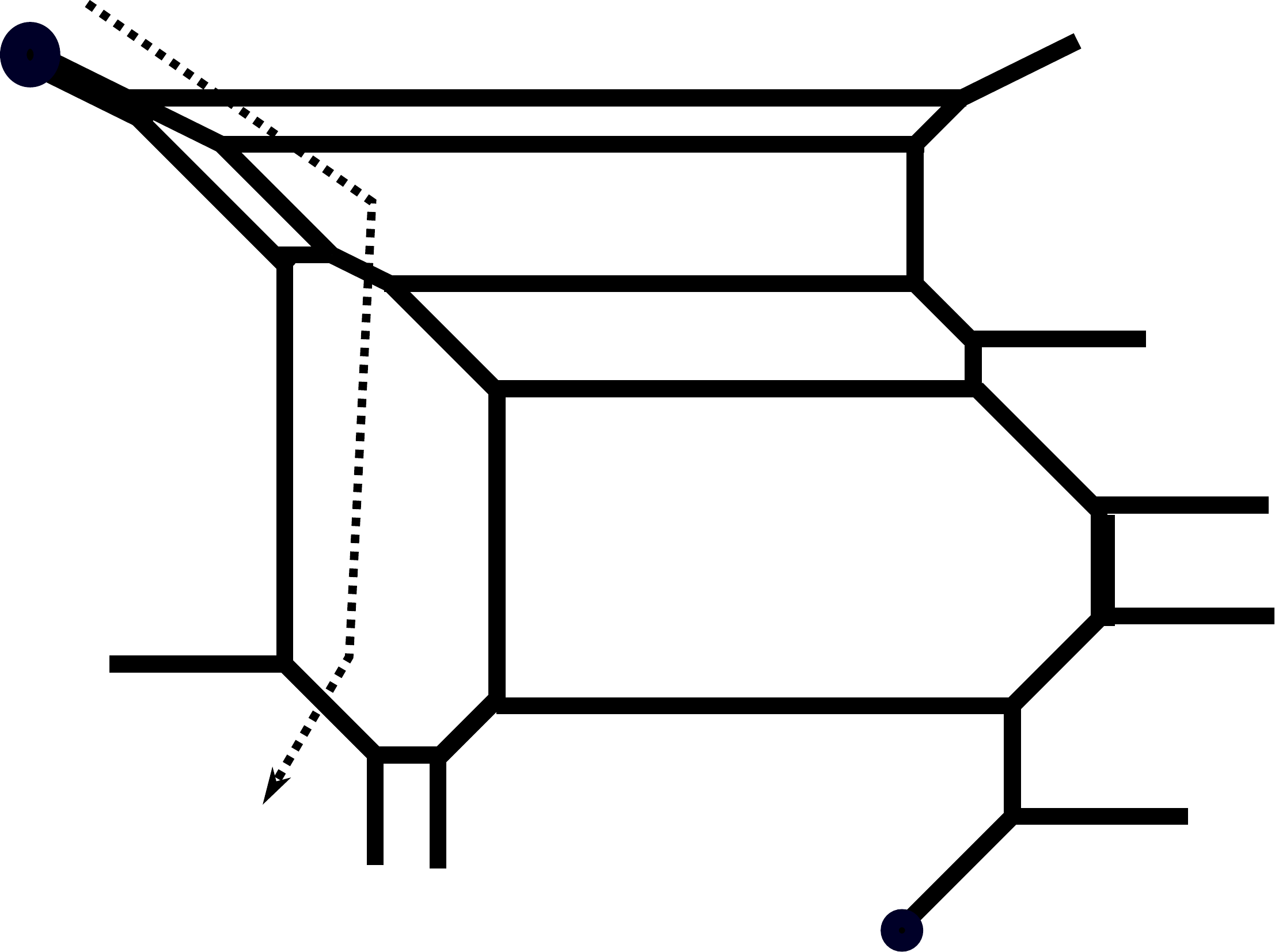}} \hspace{1cm}
\subfigure[]{\label{fig:su6w1htas1as9f3}
\includegraphics[width=6cm]{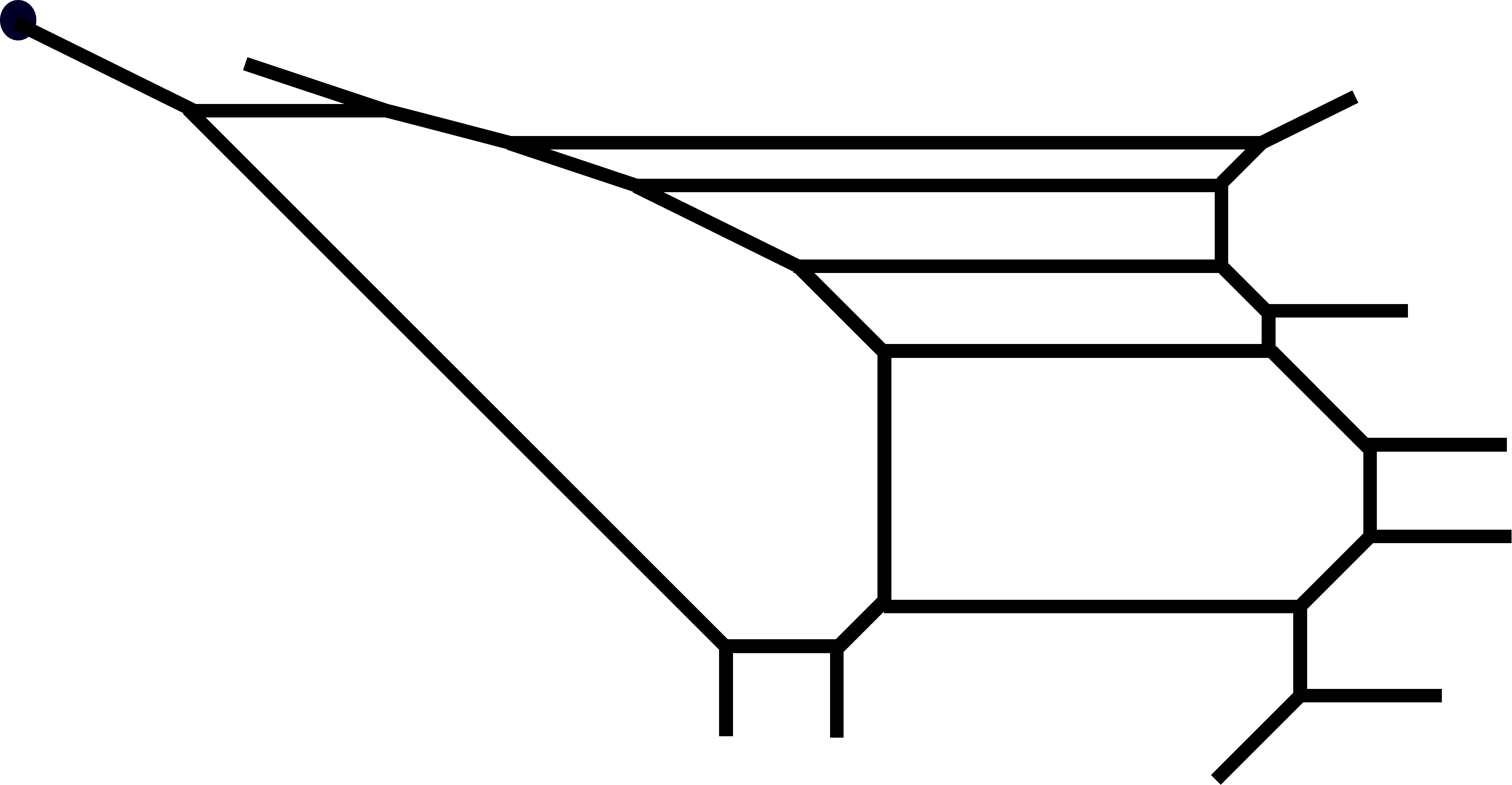}}\hspace{1cm}
\subfigure[]{\label{fig:su6w1htas1as9f4}
\includegraphics[width=6cm]{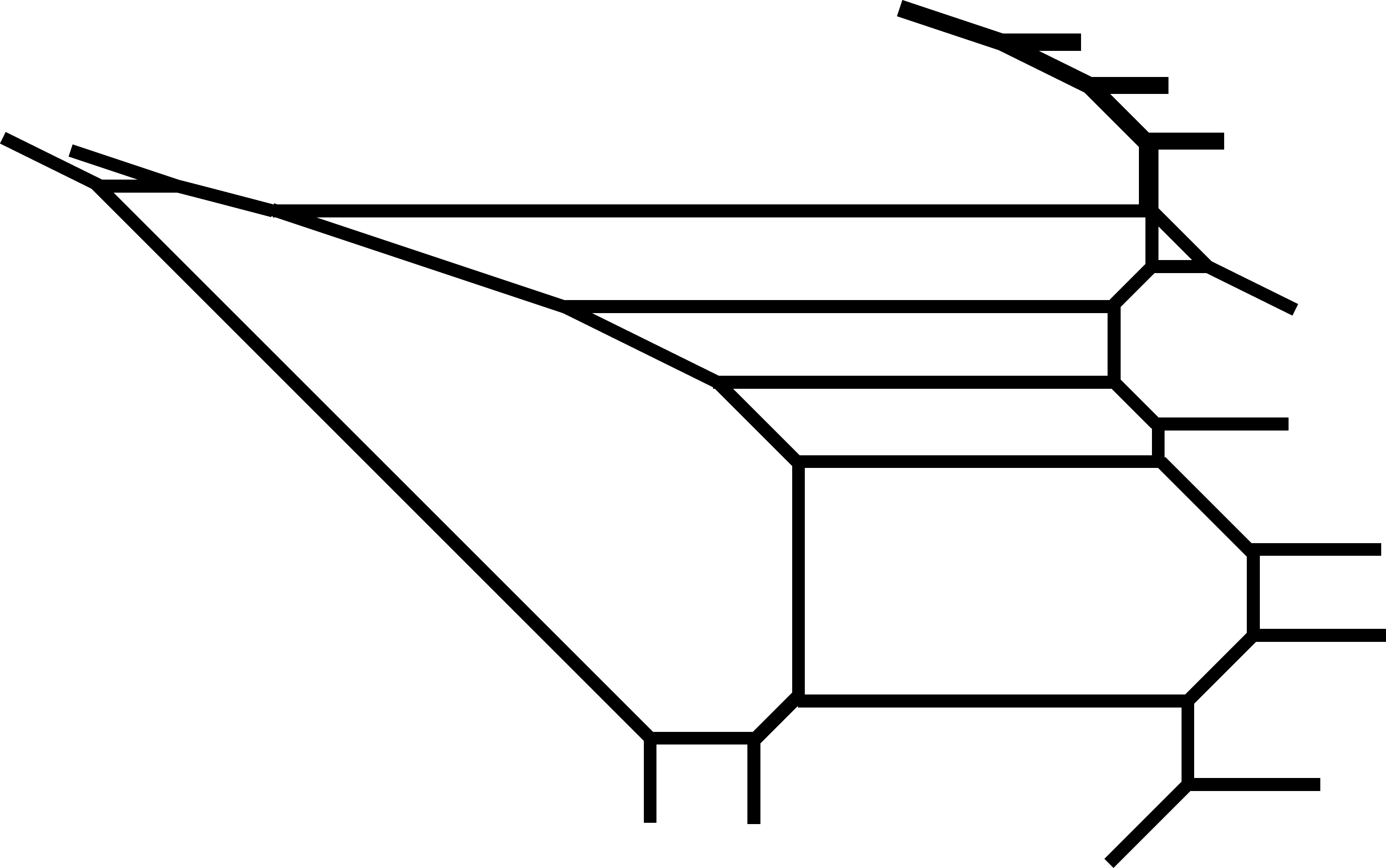}}
\caption{(a): A 5-brane web diagram for the $SU(6)$ gauge theory with $N_{\bf TAS} = \frac{1}{2}, N_{\bf F} = 6, \kappa = \frac{5}{2}$. (b): The diagram after moving the $(1, 1)$ 7-brane along the arrow in Figure \ref{fig:su6w1htas1as9f}. (c): The diagram after moving the $(2, -1)$ 7-brane along the arrow in the diagram in Figure \ref{fig:su6w1htas1as9f2}. It gives the quiver theory $SU(2) - SU(5)_{3} - [4{\bf F}]$. (d): The diagram of the quiver $SU(2) - SU(5)_{1} - [7{\bf F}, 1{\bf AS}]$ by adding three flavors and a hypermultiplet in the antisymmetric representation to the diagram in Figure \ref{fig:su6w1htas1as9f3}.}
\label{fig:su6w1htas1as9fdef}
\end{figure}
In fact, the similar deformation can be applied to the $SU(6)$ gauge theory with a hypermultiplet in the rank-2 antisymmetric representation in addition to $N_{\bf TAS} = \frac{1}{2}$ and flavors. The first example is the $SU(6)$ gauge theory wtih $N_{\bf TAS} = \frac{1}{2}, N_{\bf AS} = 1, N_{\bf F} = 9$ and $\kappa = 0$. The diagram of the theory has been given in Figure \ref{fig:SU6+1over2TAS+1AS+9F}. We then deform the diagram in Figure \ref{fig:SU6+1over2TAS+1AS+9F} to another diagram which can be interpreted as a quiver theory. For that it is enough to focus on a part of the diagram in Figure \ref{fig:SU6+1over2TAS+1AS+9F} which is given in Figure \ref{fig:su6w1htas1as9f}. Compared with the diagram in Figure \ref{fig:SU6+1over2TAS+1AS+9F}, three D7-branes and a $(0, 1)$ 7-brane are decoupled in the upper direction and the diagram yields the $SU(6)$ gauge theory with $N_{\bf TAS} = \frac{1}{2}, N_{\bf F} = 6, \kappa = \frac{5}{2}$. From the diagram in Figure \ref{fig:su6w1htas1as9f}, we first move the $(1, 1)$ 7-brane along the arrow to go to the diagram in Figure \ref{fig:su6w1htas1as9f2} and then move the $(2, -1)$ 7-brane as in Figure \ref{fig:su6w1htas1as9f2}. After the deformation, the resulting theory from the web in Figure \ref{fig:su6w1htas1as9f3} leads to the $SU(2) - SU(5)_{3} - [4{\bf F}]$ quiver theory. Since we start the diagram with three D7-branes and the $(0, 1)$ 7-brane decoupled, we need to reintroduce the 7-branes to the diagram in Figure \ref{fig:su6w1htas1as9f3} which yields the diagram in Figure \ref{fig:su6w1htas1as9f4}. Then reintroducing the $(0, 1)$ 7-brane adds a hypermultiplet in the antisymmetric representation of $SU(5)$ and three D7-branes give three flavors to the $SU(5)$. Hence, the deformations in Figure \ref{fig:su6w1htas1as9fdef} imply that the $SU(6)_0$ gauge theory with $N_{\bf TAS} = \frac{1}{2}, N_{\bf AS} = 1, N_{\bf F} = 9$ is dual to the $SU(2) - SU(5)_{1} - [7{\bf F}, 1{\bf AS}]$ quiver theory. Another deformation by 7-branes may give further dual quiver theory such as $[3{\bf F}] - SU(2) - SU(3)_{\frac{1}{2}} - SU(3)_2 - [2{\bf F}, 1{\bf AS}]$.

\paragraph{\underline{$SU(6)_{\frac{3}{2}} + \frac{1}{2}{\bf TAS} + 1{\bf AS} + 8{\bf F}$}}
\begin{figure}[t]
\centering
\includegraphics[width=7cm]{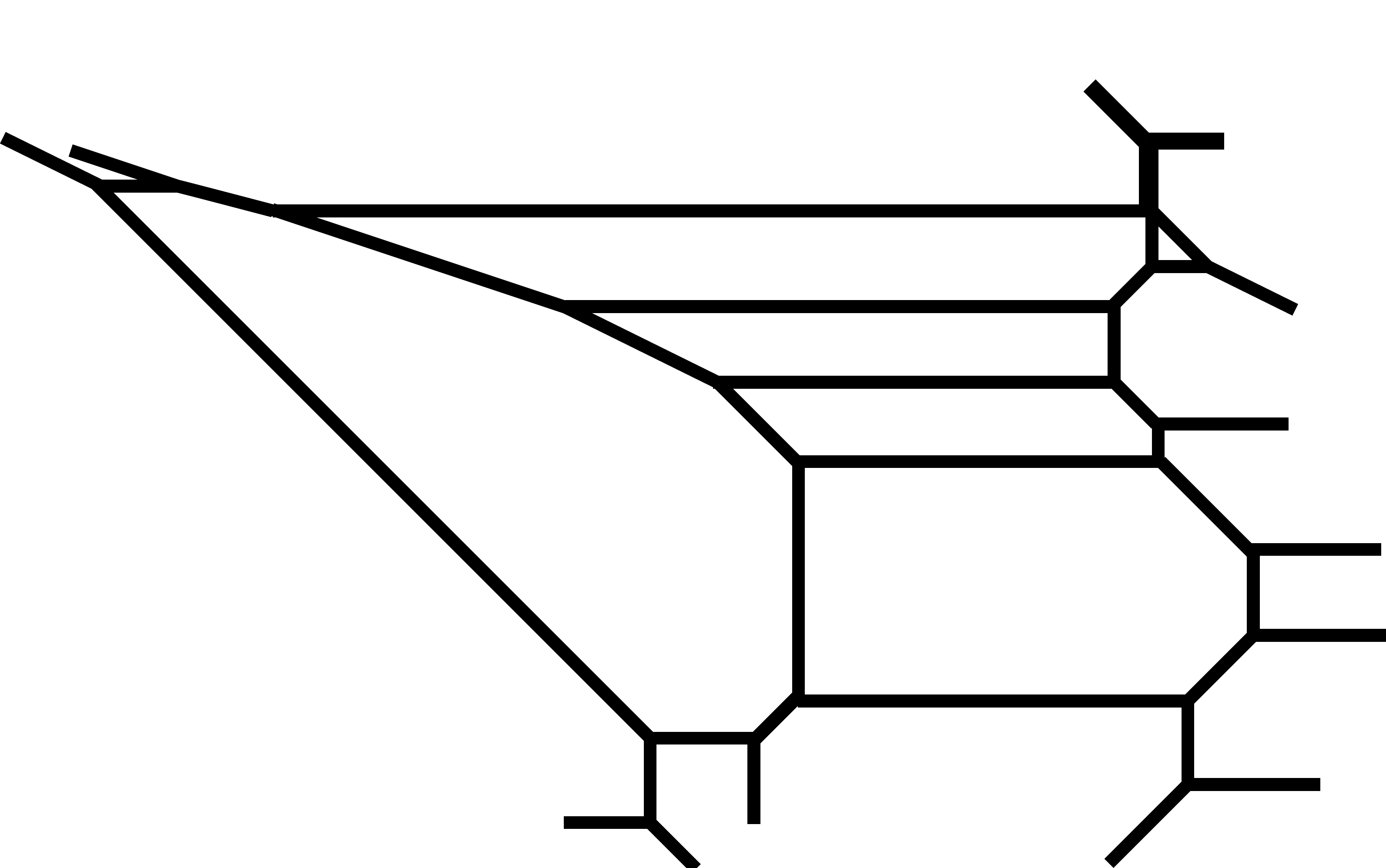}
\caption{The diagram obtained by deforming the diagram in Figure \ref{fig:SU6+1over2TAS+1AS+8F}. The diagram gives the $[1{\bf F}] - SU(2) - SU(5)_{2} - [5{\bf F}, 1{\bf AS}]$ quiver theory. }
\label{fig:su6w1htas1as8f}
\end{figure}

The next example of dualities which involve $SU(6)$ gauge theories with an antisymmetric hypermultiplet in addition to $N_{\bf TAS}  = \frac{1}{2}$ is the $SU(6)$ gauge theory with $N_{\bf TAS} = \frac{1}{2}, N_{\bf AS} = 1$ and $\kappa = \frac{3}{2}$. The diagram has been obtained in Figure \ref{fig:SU6+1over2TAS+1AS+8F}. In order to obtain a dual quiver description we can make use of the deformations from Figure \ref{fig:su6w1htas1as9f} to Figure \ref{fig:su6w1htas1as9f3}. From the diagram in Figure \ref{fig:SU6+1over2TAS+8F}, decoupling one D7-brane and a $(0, 1)$ 7-brane in the upper direction and also remove a D7-brane in the lower direction gives rise to the diagram in Figure \ref{fig:su6w1htas1as9f}. Hence we can reintroduce the 7-branes to the diagram in Figure \ref{fig:su6w1htas1as9f3} for a deformed diagram from the one in Figure \ref{fig:SU6+1over2TAS+1AS+8F}. The final deformed diagram is depicted in Figure \ref{fig:su6w1htas1as8f} and it realizes the $[1{\bf F}] - SU(2) - SU(5)_{2} - [5{\bf F}, 1{\bf AS}]$ quiver theory. Therefore, the $SU(6)_{\frac{3}{2}}$ gauge theory with $N_{\bf TAS} = \frac{1}{2}, N_{\bf AS} = 1, N_{\bf F} = 8$ is dual to the $[1{\bf F}] - SU(2) - SU(5)_{2} - [5{\bf F}, 1{\bf AS}]$ quiver theory. The distribution duality in \cite{Hayashi:2015zka} can yield another dual quiver theory such as $[4{\bf F}] - SU(3)_0 - SU(4)_{\frac{5}{2}} - [2{\bf F}, 1{\bf AS}]$.

\subsection{6d uplift of $SU(6)_0 + \frac{1}{2}{\bf TAS} + 1 {\bf Sym} + 1{\bf F}$}
Here, we discuss  the 5d $SU(6)$ gauge theory with $N_{\bf TAS} = \frac{1}{2}, N_{\bf Sym} = \frac{1}{2}, N_{\bf F} = 1, \kappa = 0$. 
In \cite{Hayashi:2015vhy}, it is discussed that we obtain 
\begin{align}
5d \,\, [1 {\bf Sym}] - SU(N+2k-1) - SU(N+2k-5) - \cdots - SU(N-2k+3) - [N-2k+1 {\bf F}]
\end{align}
by the twisted circle compactification of
\begin{align}
6d \,\, [N {\bf F}] - SU(N) - \cdots - SU(N) - \cdots - SU(N) - [N {\bf F}]
\end{align}
where we have $2k$ $SU(N)$ gauge nodes. Although $N$ was assumed to be even number for simplicity when this was diagrammatically derived in \cite{Hayashi:2015vhy},
we can generalized this relation to the case for odd $N$.
The 5d $SU(6)$ gauge theory with $N_{\bf TAS} = \frac{1}{2}, N_{\bf Sym} = \frac{1}{2}, N_{\bf F} = 1, \kappa = 0$ turns out to be related to the case $N=3$, $k=2$.

\begin{figure}
\centering
\includegraphics[width=10cm]{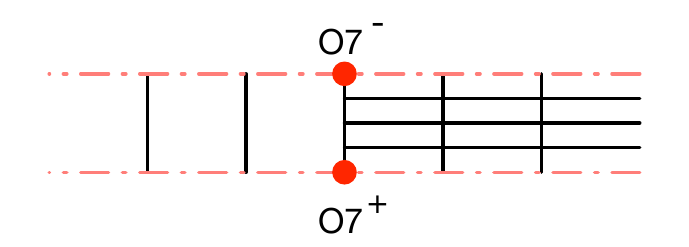}
\caption{Type IIB web diagram corresponding to twisted circle compactification of 
6d SU(3) quiver gauge theory. All the $(p,1)$ 5-branes are written vertically for simplicity.}
\label{fig:Twistedcpt}
\end{figure}

The corresponding web diagram of this theory is given in Figure \ref{fig:Twistedcpt}.
It is straightforward to see that Figure \ref{fig:SU6+1over2TAS+1Sym+1F-1} can be obtained by the Higgsing of this web diagram. From the point of view of the 6d theory, this Higgsing has to be done at the left hand side and the right hand side so that it is compatible with the twist. Therefore, we conclude that
\begin{align}
& 6d \,\, [1 {\bf F}] - SU(2)
- {\overset{\overset{\text{\large [1{\bf F}] }}{\textstyle\vert}}{ SU(3)  }}
- {\overset{\overset{\text{\large [1{\bf F}] }}{\textstyle\vert}}{ SU(3)  }}
 - SU(2) - [1 {\bf F}] 
\cr \cr
& \xrightarrow{\text{Twisted circle compactification}} \quad 
5d \,\, [1 {\bf Sym} ] - SU(6) -   [1/2 {\bf TAS} + 1 {\bf F}] .
\end{align}

\bigskip
\section{Summary and discussion}\label{sec:conclusion}

In this paper, we explicitly constructed 5-brane webs for 5d $SU(6)$ and $Sp(3)$ gauge theories with hypermultiplets in the rank-3 antisymmetric representation. For an $SU(6)$ gauge theory with a half-hypermultiplet in the rank-3 antisymmetric representation, we started from a 5-brane diagram for the $SO(12)$ gauge theory with a half-hypermultiplet in the conjugate spinor representation, and utilized the embedding $SO(12)\supset SU(6)\times U(1)$ where the rank-3 antisymmetric representation of the $SU(6)$ is not charged under the $U(1)$. Then decoupling the degree of freedom associated with the $U(1)$ yields a 5-brane configuration for the $SU(6)$ gauge theory of the Chern-Simons level $\kappa=\frac{5}{2}$ with a half-hypermultiplet in the rank-3 antisymmetric representation. 
We also confirmed the validity of the brane diagram by computing the monopole string tension. 
Using the topological vertex, we computed the Nekrasov partition function for the $SU(6)_\frac52+\frac{1}{2}{\bf TAS}$ theory up to two instanton orders and confirmed that the instanton part correctly captures the Chern-Simons level, which also supports our 5-brane construction of the $SU(6)_\frac52+ \frac{1}{2}{\bf TAS}$ theory. 
This would be the first quantitative result for the partition function for the $SU(6)_\frac52$ gauge theory with a half-hypermultiplet in the rank-3 antisymmetric representation. 

By increasing the number of half-hypermultiplets in the rank-3 antisymmetric representation and also adding various hypermultiplets in other representations, we constructed 5-brane diagrams for 5d marginal $SU(6)$ gauge theories with $N_{\bf TAS}=\frac12,1, 2$ hypermultiplets in the rank-3 antisymmetric representation, which are summarized in Table \ref{Tab:su6marginal}. The matter content of what we found for the 5d marginal $SU(6)$ gauge theories with rank-3 antisymmetric hypermultiplets is in agreement with those classified in \cite{Jefferson:2017ahm}.
Moreover, global symmetries that one can read off from the 5-brane webs also support our 5-brane construction for the marginal $SU(6)$ theories with half-hypermultiplets in the rank-3 antisymmetric representation. The 5-brane web diagrams also imply 6d uplifts or dualities for some of the marginal theories. Interestingly, as discussed in section \ref{sec:6d}, some $SU(6)$ gauge theories with half-hypermultiplets in the rank-3 antisymmetric representation are dual to quiver theories. 

For marginal $Sp(3)$ gauge theories with half-hypermultiplets in the rank-3 antisymmetric representation, we used a Higgsing of marginal $SU(6)$ gauge theories involving a hypermultiplet in the rank-2 antisymmetric representation. 
Possible Higgsings are discussed in \eqref{HiggstoSp3v1}-\eqref{HiggstoSp3v3}. Since we know two of the UV $SU(6)$ theories of the three Higgsings,  we explicitly realized 5-brane configuration for the $Sp(3)+\frac{1}{2}{\bf TAS}+ \frac{19}{2}{\bf F}$ theory in Figure \ref{fig:sp3w1htas0}, and for the $Sp(3)+1{\bf TAS}+ 5{\bf F}$ theory in Figure \ref{fig:sp3w1tas0}. 

\vskip 0.5cm
Although we have constructed 5-brane webs for many of the marginal $SU(6)$ gauge theories with rank-3 antisymmetric matter in Table \ref{Tab:su6marginal} which are classified in \cite{Jefferson:2017ahm}, there are some marginal theories that we did not find their 5-brane web configurations. We note that not having a 5-brane web for a marginal theory does not imply that 5-brane webs for its descendent theories are not constructed. For instance, consider 
$SU(6)_0+3/2 {\bf TAS}+ 5{\bf F}$ or $SU(6)_2+3/2 {\bf TAS}+ 3{\bf F}$. As discussed in section  \ref{sec:su6+2TAS}, a 5-brane web for the $SU(6)_\frac12+3/2 {\bf TAS}$ theory is given in Figure \ref{fig:su6w3htsa}. In fact, one can introduce flavors to the 5-brane web diagram properly to make the CS level to be that of the marginal theories of interest.  For $SU(6)_0+3/2 {\bf TAS}+ 5{\bf F}$, one may find a configuration with 3 D7-branes added above and 2 D7-branes below, so that the resulting configuration has the CS level 0. It is also not so difficult to find that a little manipulation of the 7-branes allows a pair of 7-branes which can be converted to an O7$^-$-plane, and hence together with an ON$^-$-plane, it yields a configuration with two orientifolds horizontally separated. On the other hand, we were not able to make this 5-brane web to be a conventional 5-brane configuration with O7$^-$-plane and O5-plane vertically apart, after performing the S-duality. Hence, we did not include such 5-brane configuration in Table \ref{Tab:su6marginal}.  

It is also worth noting that there may be some intrinsic issues on 5-brane realization of marginal theories. For example, our construction utilizes 5-brane web diagrams and there will be some restriction to theories which 5-brane web on a plane can realize. Second, the marginal theories in \cite{Jefferson:2017ahm} were classified based on 
only necessary conditions and hence it may be still possible that some of them may not have a UV completion.  We here make some comments on marginal $SU(6)$ gauge theories which we have not constructed from 5-brane webs in this paper. One class of such theories is $SU(6)$ gauge theories with $N_{\bf TAS} = \frac{1}{2}, N_{\bf AS} = 2$ and flavors. Note that a half-hypermultiplet in the rank-3 antisymmetric representation can arise from a Higgsing from $[SU(6)] - SU(3)_0$ whereas a hypermultiplet in the rank-2 antisymmetric representation can be realized by a Higgsing from $[SU(6)] - SU(4)_0 - SU(2)$. Hence an $SU(6)$ gauge theory with $N_{\bf TAS} = \frac{1}{2}, N_{\bf AS} = 2$ may be obtained by considering the Higgsing from a quiver
\begin{align}
SU(2) - SU(4)_0 - {\overset{\overset{\text{\large$SU(3)_0$}}{\textstyle\vert}}{SU(6)}} - SU(4)_0 - SU(2),\label{E6Dynkin}
\end{align}
where the CS level for the $SU(6)$ gauge node needs to be chosen so that the quiver theory has a UV completion. The quiver theory \eqref{E6Dynkin} has a shape of the $E_6$ Dynkin diagram and it is difficult to realize the quiver theory from 5-brane web diagrams on a plane. Therefore the Higgsed theories, which are $SU(6)$ gauge theories with $N_{\bf TAS} = \frac{1}{2}, N_{\bf AS} = 2$, would be also difficult to be obtained by 5-brane webs on a plane.  We will need trivalent gauging for web diagrams considered in \cite{Hayashi:2017jze}. 

There are also marginal $SU(6)$ gauge theories which only have the rank-2 antisymmetric representation as matter. The $SU(6)$ gauge theories have three hypermultiplets in the rank-2 antisymmetric representation and the CS level can be $|\kappa| = 0, 1, 2, 3$ \cite{Jefferson:2017ahm}. Since a hypermultiplet in the rank-2 antisymmetric representation arises from a Higgsing from $[SU(6)] - SU(4)_0 - SU(2)$, an $SU(6)$ gauge theory with $N_{\bf AS} = 3$ can be realized by considering the Higgsing from a quiver
\begin{align}
SU(2) - SU(4)_0 - {\overset{\overset{\overset{\overset{\text{\large$SU(2)$}}{\textstyle\vert}}{\text{\large$SU(4)_0$}}}{\textstyle\vert}}{SU(6)}} - SU(4)_0 - SU(2)\label{E6affineDynkin}
\end{align}
The quiver theory \eqref{E6affineDynkin} has a shape of the affine $E_6$ Dynkin diagram which might have a 6d UV completion for some specific CS level for the $SU(6)$ gauge node. In particular when the CS level for the $SU(6)$ gauge node is zero, the Higgsed theory becomes the $SU(6)$ gauge theory with $N_{\AS} = 3, \kappa = 3$ since each Higgsing which gives the rank-2 antisymmetric matter increases the CS level by one as in \eqref{HiggstoAS}. The $SU(6)$ theory is exactly one of the marginal $SU(6)$ theories classified in \cite{Jefferson:2017ahm}. Since this quiver is an affine $E_6$ Dynkin quiver it is difficult to realize it by a 5-brane web on a plane, implying that the Higgsed theories would be also difficult to be obtained by a brane web on a plane. In order to construct the affine $E_6$ quiver theory, we will need trivalent gauging for web diagrams again. 

5-brane webs we found for marginal theories show the periodicity either as a Tao diagram \cite{Kim:2015jba} or as a 5-brane configuration with two orientifolds. It clearly suggests that they are a realization of 6d theory on a circle with/without some twists. For some marginal theories, we discussed their 6d uplifts. It would be interesting to find  6d uplifts for other marginal theories as well as possible dual quiver descriptions. 

In this paper we have focused on half-hypermultiplets in the rank-3 antisymmetric representation for $SU(6)$ and $Sp(3)$ gauge theories. It would be also interesting to generalize the result to rank-3 antisymmetric matter for other gauge theories such as $SU(7)$ or $Sp(4)$ gauge theories which were also discussed in \cite{Jefferson:2017ahm}. 







\acknowledgments
We would like to thank Gabi Zafrir for discussions.
SSK and KL would like to gratefully acknowledge APCTP for hosting the Focus program ``Strings, Branes and Gauge Theories 2018'' and thank Fudan University for hospitality for their visit. SKK thanks YMSC at Tsinghua university for hosting ``Tsinghua Summer Workshop on Geometry and Physics 2018'' and also KIAS and POSTECH for hospitality for his visit. The work of HH is supported in part by JSPS KAKENHI Grant Number JP18K13543. SSK is supported by the UESTC Research Grant A03017023801317.  KL is supported in part by the National Research Foundation of Korea Grant NRF-2017R1D1A1B06034369. FY is supported by the Fundamental Research Funds for the Central Universities A0920502051904-48, by Start-up research grant A1920502051907-2-046, in part by NSFC grant No. 11501470 and No. 11671328, and by Recruiting Foreign Experts Program No. T2018050 granted by SAFEA.


\bigskip
\appendix 


\bigskip

\bibliographystyle{JHEP}
\bibliography{ref}

\providecommand{\href}[2]{#2}\begingroup\raggedright\begin{thebibliography}{10}

\bibitem{Aharony:1997ju}
O.~Aharony and A.~Hanany, {\it {Branes, superpotentials and superconformal
  fixed points}},  {\em Nucl.Phys.} {\bf B504} (1997) 239--271,
  [\href{http://arxiv.org/abs/hep-th/9704170}{{\tt hep-th/9704170}}].

\bibitem{Aharony:1997bh}
O.~Aharony, A.~Hanany, and B.~Kol, {\it {Webs of (p,q) five-branes,
  five-dimensional field theories and grid diagrams}},  {\em JHEP} {\bf 9801}
  (1998) 002, [\href{http://arxiv.org/abs/hep-th/9710116}{{\tt
  hep-th/9710116}}].

\bibitem{Aganagic:2003db}
M.~Aganagic, A.~Klemm, M.~Marino, and C.~Vafa, {\it {The Topological vertex}},
  {\em Commun.Math.Phys.} {\bf 254} (2005) 425--478,
  [\href{http://arxiv.org/abs/hep-th/0305132}{{\tt hep-th/0305132}}].

\bibitem{Iqbal:2007ii}
A.~Iqbal, C.~Kozcaz, and C.~Vafa, {\it {The Refined topological vertex}},  {\em
  JHEP} {\bf 0910} (2009) 069, [\href{http://arxiv.org/abs/hep-th/0701156}{{\tt
  hep-th/0701156}}].

\bibitem{Leung:1997tw}
N.~C. Leung and C.~Vafa, {\it {Branes and toric geometry}},  {\em
  Adv.Theor.Math.Phys.} {\bf 2} (1998) 91--118,
  [\href{http://arxiv.org/abs/hep-th/9711013}{{\tt hep-th/9711013}}].

\bibitem{Sen:1996vd}
A.~Sen, {\it {F theory and orientifolds}},  {\em Nucl.Phys.} {\bf B475} (1996)
  562--578, [\href{http://arxiv.org/abs/hep-th/9605150}{{\tt hep-th/9605150}}].

\bibitem{Brunner:1997gk}
I.~Brunner and A.~Karch, {\it {Branes and six-dimensional fixed points}},  {\em
  Phys. Lett.} {\bf B409} (1997) 109--116,
  [\href{http://arxiv.org/abs/hep-th/9705022}{{\tt hep-th/9705022}}].

\bibitem{Bergman:2015dpa}
O.~Bergman and G.~Zafrir, {\it {5d fixed points from brane webs and
  O7-planes}},  {\em JHEP} {\bf 12} (2015) 163,
  [\href{http://arxiv.org/abs/1507.03860}{{\tt arXiv:1507.03860}}].

\bibitem{Benini:2009gi}
F.~Benini, S.~Benvenuti, and Y.~Tachikawa, {\it {Webs of five-branes and N=2
  superconformal field theories}},  {\em JHEP} {\bf 0909} (2009) 052,
  [\href{http://arxiv.org/abs/0906.0359}{{\tt arXiv:0906.0359}}].

\bibitem{Bergman:2013aca}
O.~Bergman, D.~Rodr\'iguez-G\'omez, and G.~Zafrir, {\it {5-Brane Webs, Symmetry
  Enhancement, and Duality in 5d Supersymmetric Gauge Theory}},  {\em JHEP}
  {\bf 1403} (2014) 112, [\href{http://arxiv.org/abs/1311.4199}{{\tt
  arXiv:1311.4199}}].

\bibitem{Bergman:2014kza}
O.~Bergman and G.~Zafrir, {\it {Lifting 4d dualities to 5d}},  {\em JHEP} {\bf
  1504} (2015) 141, [\href{http://arxiv.org/abs/1410.2806}{{\tt
  arXiv:1410.2806}}].

\bibitem{Zafrir:2015ftn}
G.~Zafrir, {\it {Brane webs and $O5$-planes}},  {\em JHEP} {\bf 03} (2016) 109,
  [\href{http://arxiv.org/abs/1512.08114}{{\tt arXiv:1512.08114}}].

\bibitem{Hayashi:2018bkd}
H.~Hayashi, S.-S. Kim, K.~Lee, and F.~Yagi, {\it {5-brane webs for 5d $
  \mathcal{N} $ = 1 G$_{2}$ gauge theories}},  {\em JHEP} {\bf 03} (2018) 125,
  [\href{http://arxiv.org/abs/1801.03916}{{\tt arXiv:1801.03916}}].

\bibitem{Hayashi:2018lyv}
H.~Hayashi, S.-S. Kim, K.~Lee, and F.~Yagi, {\it {Dualities and 5-brane webs
  for 5d rank 2 SCFTs}},  {\em JHEP} {\bf 12} (2018) 016,
  [\href{http://arxiv.org/abs/1806.10569}{{\tt arXiv:1806.10569}}].

\bibitem{Jefferson:2018irk}
P.~Jefferson, S.~Katz, H.-C. Kim, and C.~Vafa, {\it {On Geometric
  Classification of 5d SCFTs}},  {\em JHEP} {\bf 04} (2018) 103,
  [\href{http://arxiv.org/abs/1801.04036}{{\tt arXiv:1801.04036}}].

\bibitem{Tachikawa:2011yr}
Y.~Tachikawa and S.~Terashima, {\it {Seiberg-Witten Geometries Revisited}},
  {\em JHEP} {\bf 09} (2011) 010, [\href{http://arxiv.org/abs/1108.2315}{{\tt
  arXiv:1108.2315}}].

\bibitem{Ohmori:2018ona}
K.~Ohmori, Y.~Tachikawa, and G.~Zafrir, {\it {Compactifications of 6d $N = (1,
  0)$ SCFTs with non-trivial Stiefel-Whitney classes}},  {\em JHEP} {\bf 04}
  (2019) 006, [\href{http://arxiv.org/abs/1812.04637}{{\tt arXiv:1812.04637}}].

\bibitem{Jefferson:2017ahm}
P.~Jefferson, H.-C. Kim, C.~Vafa, and G.~Zafrir, {\it {Towards Classification
  of 5d SCFTs: Single Gauge Node}},
  \href{http://arxiv.org/abs/1705.05836}{{\tt arXiv:1705.05836}}.

\bibitem{Hayashi:2017btw}
H.~Hayashi, S.-S. Kim, K.~Lee, and F.~Yagi, {\it {Discrete theta angle from an
  O5-plane}},  {\em JHEP} {\bf 11} (2017) 041,
  [\href{http://arxiv.org/abs/1707.07181}{{\tt arXiv:1707.07181}}].

\bibitem{Seiberg:1996bd}
N.~Seiberg, {\it {Five-dimensional SUSY field theories, nontrivial fixed points
  and string dynamics}},  {\em Phys.Lett.} {\bf B388} (1996) 753--760,
  [\href{http://arxiv.org/abs/hep-th/9608111}{{\tt hep-th/9608111}}].

\bibitem{Morrison:1996xf}
D.~R. Morrison and N.~Seiberg, {\it {Extremal transitions and five-dimensional
  supersymmetric field theories}},  {\em Nucl.Phys.} {\bf B483} (1997)
  229--247, [\href{http://arxiv.org/abs/hep-th/9609070}{{\tt hep-th/9609070}}].

\bibitem{Intriligator:1997pq}
K.~A. Intriligator, D.~R. Morrison, and N.~Seiberg, {\it {Five-dimensional
  supersymmetric gauge theories and degenerations of Calabi-Yau spaces}},  {\em
  Nucl.Phys.} {\bf B497} (1997) 56--100,
  [\href{http://arxiv.org/abs/hep-th/9702198}{{\tt hep-th/9702198}}].

\bibitem{Closset:2018bjz}
C.~Closset, M.~Del~Zotto, and V.~Saxena, {\it {Five-dimensional SCFTs and gauge
  theory phases: an M-theory/type IIA perspective}},
  \href{http://arxiv.org/abs/1812.10451}{{\tt arXiv:1812.10451}}.

\bibitem{Hayashi:2013qwa}
H.~Hayashi, H.-C. Kim, and T.~Nishinaka, {\it {Topological strings and 5d $T_N$
  partition functions}},  {\em JHEP} {\bf 1406} (2014) 014,
  [\href{http://arxiv.org/abs/1310.3854}{{\tt arXiv:1310.3854}}].

\bibitem{Hayashi:2014wfa}
H.~Hayashi and G.~Zoccarato, {\it {Exact partition functions of Higgsed 5d
  $T_N$ theories}},  {\em JHEP} {\bf 1501} (2015) 093,
  [\href{http://arxiv.org/abs/1409.0571}{{\tt arXiv:1409.0571}}].

\bibitem{Kim:2015jba}
S.-S. Kim, M.~Taki, and F.~Yagi, {\it {Tao Probing the End of the World}},
  {\em PTEP} {\bf 2015} (2015), no.~8 083B02,
  [\href{http://arxiv.org/abs/1504.03672}{{\tt arXiv:1504.03672}}].

\bibitem{Hayashi:2015xla}
H.~Hayashi and G.~Zoccarato, {\it {Topological vertex for Higgsed 5d T$_{N}$
  theories}},  {\em JHEP} {\bf 09} (2015) 023,
  [\href{http://arxiv.org/abs/1505.00260}{{\tt arXiv:1505.00260}}].

\bibitem{Kim:2017jqn}
S.-S. Kim and F.~Yagi, {\it {Topological vertex formalism with O5-plane}},
  {\em Phys. Rev.} {\bf D97} (2018) 026011,
  [\href{http://arxiv.org/abs/1709.01928}{{\tt arXiv:1709.01928}}].

\bibitem{Hanany:1996ie}
A.~Hanany and E.~Witten, {\it {Type IIB superstrings, BPS monopoles, and
  three-dimensional gauge dynamics}},  {\em Nucl.Phys.} {\bf B492} (1997)
  152--190, [\href{http://arxiv.org/abs/hep-th/9611230}{{\tt hep-th/9611230}}].

\bibitem{Kutasov:1995te}
D.~Kutasov, {\it {Orbifolds and solitons}},  {\em Phys. Lett.} {\bf B383}
  (1996) 48--53, [\href{http://arxiv.org/abs/hep-th/9512145}{{\tt
  hep-th/9512145}}].

\bibitem{Sen:1996na}
A.~Sen, {\it {Duality and orbifolds}},  {\em Nucl. Phys.} {\bf B474} (1996)
  361--378, [\href{http://arxiv.org/abs/hep-th/9604070}{{\tt hep-th/9604070}}].

\bibitem{Kapustin:1998fa}
A.~Kapustin, {\it {D(n) quivers from branes}},  {\em JHEP} {\bf 12} (1998) 015,
  [\href{http://arxiv.org/abs/hep-th/9806238}{{\tt hep-th/9806238}}].

\bibitem{Hanany:1999sj}
A.~Hanany and A.~Zaffaroni, {\it {Issues on orientifolds: On the brane
  construction of gauge theories with SO(2n) global symmetry}},  {\em JHEP}
  {\bf 07} (1999) 009, [\href{http://arxiv.org/abs/hep-th/9903242}{{\tt
  hep-th/9903242}}].

\bibitem{Hayashi:2015fsa}
H.~Hayashi, S.-S. Kim, K.~Lee, M.~Taki, and F.~Yagi, {\it {A new 5d description
  of 6d D-type minimal conformal matter}},  {\em JHEP} {\bf 08} (2015) 097,
  [\href{http://arxiv.org/abs/1505.04439}{{\tt arXiv:1505.04439}}].

\bibitem{Hayashi:2015zka}
H.~Hayashi, S.-S. Kim, K.~Lee, and F.~Yagi, {\it {6d SCFTs, 5d Dualities and
  Tao Web Diagrams}},  \href{http://arxiv.org/abs/1509.03300}{{\tt
  arXiv:1509.03300}}.

\bibitem{Hayashi:2015vhy}
H.~Hayashi, S.-S. Kim, K.~Lee, M.~Taki, and F.~Yagi, {\it {More on 5d
  descriptions of 6d SCFTs}},  {\em JHEP} {\bf 10} (2016) 126,
  [\href{http://arxiv.org/abs/1512.08239}{{\tt arXiv:1512.08239}}].

\bibitem{Gaberdiel:1997ud}
M.~R. Gaberdiel and B.~Zwiebach, {\it {Exceptional groups from open strings}},
  {\em Nucl.Phys.} {\bf B518} (1998) 151--172,
  [\href{http://arxiv.org/abs/hep-th/9709013}{{\tt hep-th/9709013}}].

\bibitem{Gaberdiel:1998mv}
M.~R. Gaberdiel, T.~Hauer, and B.~Zwiebach, {\it {Open string-string junction
  transitions}},  {\em Nucl.Phys.} {\bf B525} (1998) 117--145,
  [\href{http://arxiv.org/abs/hep-th/9801205}{{\tt hep-th/9801205}}].

\bibitem{DeWolfe:1999hj}
O.~DeWolfe, A.~Hanany, A.~Iqbal, and E.~Katz, {\it {Five-branes, seven-branes
  and five-dimensional E(n) field theories}},  {\em JHEP} {\bf 03} (1999) 006,
  [\href{http://arxiv.org/abs/hep-th/9902179}{{\tt hep-th/9902179}}].

\bibitem{Hayashi:2016abm}
H.~Hayashi, S.-S. Kim, K.~Lee, and F.~Yagi, {\it {Equivalence of several
  descriptions for 6d SCFT}},  {\em JHEP} {\bf 01} (2017) 093,
  [\href{http://arxiv.org/abs/1607.07786}{{\tt arXiv:1607.07786}}].

\bibitem{Zafrir:2015rga}
G.~Zafrir, {\it {Brane webs, $5d$ gauge theories and $6d$ $\mathcal{N}$$=(1,0)$
  SCFT's}},  {\em JHEP} {\bf 12} (2015) 157,
  [\href{http://arxiv.org/abs/1509.02016}{{\tt arXiv:1509.02016}}].

\bibitem{DeWolfe:1998pr}
O.~DeWolfe, T.~Hauer, A.~Iqbal, and B.~Zwiebach, {\it {Uncovering infinite
  symmetries on [p, q] 7-branes: Kac-Moody algebras and beyond}},  {\em
  Adv.Theor.Math.Phys.} {\bf 3} (1999) 1835--1891,
  [\href{http://arxiv.org/abs/hep-th/9812209}{{\tt hep-th/9812209}}].

\bibitem{Ohmori:2015tka}
K.~Ohmori and H.~Shimizu, {\it {$S^1/T^2$ Compactifications of 6d
  $\mathcal{N}=(1,0)$ Theories and Brane Webs}},
  \href{http://arxiv.org/abs/1509.03195}{{\tt arXiv:1509.03195}}.

\bibitem{Hayashi:2017jze}
H.~Hayashi and K.~Ohmori, {\it {5d/6d DE instantons from trivalent gluing of
  web diagrams}},  {\em JHEP} {\bf 06} (2017) 078,
  [\href{http://arxiv.org/abs/1702.07263}{{\tt arXiv:1702.07263}}].

\end{thebibliography}\endgroup
\end{document}